\documentclass[10pt,journal,twocolumn]{IEEEtran}

\usepackage{amsmath,amsthm,amssymb,bbm,enumerate}
\usepackage{graphicx,float,etoolbox}
\usepackage{xcolor,svg}
\usepackage{pgfplots,balance,microtype}
\usepackage{lipsum}
\newtoggle{paper}
\toggletrue{paper}
\usepackage{cite}
\usepackage{url}
\newcommand{\mc}{\mathcal}
\newcommand{\mrm}{\mathrm}

\newcommand{\norm}[1]{\left\| #1 \right\|}

\newcommand{\DSBS}{\mathsf{DSBS}}
\newcommand{\Rach}{\mc{R}_{\mrm{ach}}}

\newcommand{\RNO}{R^{\mrm{NO-SR}}_{\mrm{opt}}}
\togglefalse{paper}
\def \arxive {1}
\title{Coordination Through Shared Randomness}
\DeclareMathOperator*{\argmin}{arg\,min}
\theoremstyle{plain} \newtheorem{thm}{Theorem}
\theoremstyle{plain} \newtheorem{lemma}{Lemma}
\theoremstyle{plain} 
\theoremstyle{plain} \newtheorem{defn}{Definition}
\theoremstyle{definition} \newtheorem{exmp}{Example}
\theoremstyle{plain} 
\theoremstyle{definition} \newtheorem{remark}{Remark}
\pgfplotsset{compat=1.14}
\begin{document}
\author{Gowtham R. Kurri, \IEEEmembership{Member,~IEEE}, Vinod M. Prabhakaran, \IEEEmembership{Member,~IEEE}, and Anand D. Sarwate, \IEEEmembership{Senior Member,~IEEE}\thanks{This work was supported by the Department of Atomic Energy, Government of India, under project no. RTI4001. The work of Gowtham R. Kurri was supported in part by a Travel Fellowship from the Sarojini Damodaran Foundation. This work was done while Gowtham R. Kurri was at the Tata Institute of Fundamental Research, India. This article was presented in part at the 2018 IEEE International Symposium on Information Theory (ISIT)  and 2019 Information Theory Workshop (ITW). 

Gowtham R. Kurri is with the School of Electrical, Computer and Energy Engineering, Arizona State University, Tempe, AZ 85287 USA
(email: \protect\url{gowthamkurri@gmail.com}). 

Vinod M. Prabhakaran is with the School of Technology and Computer Science, Tata Institute of Fundamental Research, Mumbai 400005, India 
(email: \protect\url{vinodmp@tifr.res.in}).

Anand D. Sarwate is with the Department of Electrical and Computer Engineering, Rutgers, The State University of New Jersey, Piscataway, NJ 08854 USA
 (email: \protect\url{asarwate@ece.rutgers.edu}).
 
Copyright \copyright \ 2021 IEEE. Personal use of this material is permitted.  However, permission to use this material for any other purposes must be obtained from the IEEE by sending a request to pubs-permissions@ieee.org. 
 }}

\maketitle

\begin{abstract}
We study a distributed sampling problem where a set of processors want to output (approximately) independent and identically distributed samples from a given joint distribution with the help of a common message from a coordinator. 
Each processor has access to a subset of sources from a set of independent sources  of  ``shared'' randomness. We consider two cases -- in the ``omniscient coordinator setting'', the coordinator has access to all these sources of shared randomness, while in the ``oblivious coordinator setting", it has access to none. 
In addition, all processors and the coordinator may privately randomize. In the omniscient coordinator setting, when the subsets at the processors are disjoint (individually shared randomness model), we characterize the rate of communication required from the coordinator to the processors over a multicast link. For the two-processor case, the optimal rate matches a special case of relaxed Wyner's common information proposed by Gastpar and Sula~(2019), thereby providing an operational meaning to the latter. We also give an upper bound on the communication rate for the ``randomness-on-the-forehead'' model where each processor observes all but one source of randomness and present an achievable strategy for the general case where the processors have access to arbitrary subsets of sources of randomness.
Also, we consider a more general model where the processors observe components of correlated sources (with the coordinator observing all the components), where we characterize the communication rate when all the processors wish to output the same random sequence. In the oblivious coordinator setting, we completely characterize the trade-off region between the communication and shared randomness rates for the general case where the processors have access to arbitrary subsets of sources of randomness. 
\end{abstract}%

\section{Introduction}
In a coordination problem~\cite{CuffPC10}, a set of users communicate over a network to ensure that their outputs follow a joint behaviour specified by a prescribed joint distribution of outputs. A fundamental question here is to characterize the optimal rate of communication among the users. There are many different ways to mathematically formalize this problem depending on the information available to the users, the nature of the communication, and the type of behaviour sought by the system designer.

\begin{figure}[t]
\centering
\includegraphics{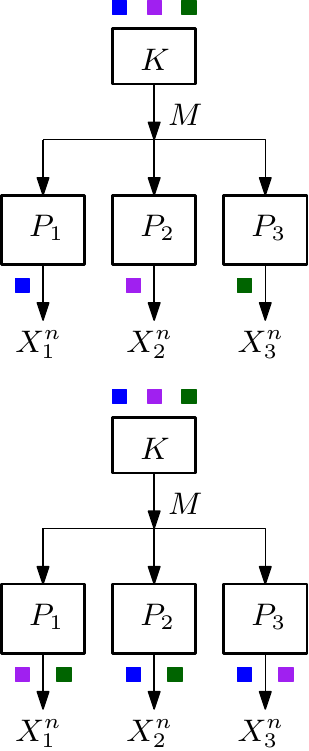}
\caption{Omniscient coordinator setting: (Top) Individually shared randomness model. (Bottom) Randomness-on-the-forehead model. Blue, purple and green colors represent shared random variables $W_1, W_2$ and $W_3$, respectively. Coordinator $K$ sends a common message $M$ to the processors $P_1,P_2,$ and $P_3$ so that they may output $X_1^n,X_2^n,$ and $X_3^n$, respectively, where $(X_{1i},X_{2i},X_{3i})$, $i=1,\dots,n$, are (approximately)  i.i.d. with $q_{X_1X_2X_3}$.}
\label{fig:model}
\end{figure}

An early work of this kind is due to Wyner~\cite{Wyner75} who characterized the minimum rate of common randomness required by two processors to produce (approximately) independent and identically distributed (i.i.d.) samples from a given joint distribution $q_{XY}$; this rate is known as Wyner's common information.
Bennett et al.~\cite{Bennet02}, Winter~\cite{winter2002compression}, Cuff~\cite{Cuff13}, Bennett et al.~\cite{BennettDHSW14}, and Wilde et al.~\cite{wilde2012information} studied a processor observing i.i.d. $X^n$ that sends a message over a noiseless link to another processor to approximate a noisy channel $q_{Y|X}$ between them. The non-asymptotic version of this problem was studied by Harsha et al.~\cite{Harsha2010}. Satpathy and Cuff~\cite{SaketC13}, and Vellambi et al.~\cite{VellambiKB18} studied the cascade network with more than two processors. Cuff et al.~\cite{CuffPC10} studied several two-node and three-node networks in which the nodes try to produce correlated random variables. Non-interactive distributed sampling relying on correlated sources was studied by Kamath and Anantharam~\cite{KamathA2016}. Exact distributed sampling was studied by Anantharam and Borkar \cite{Anantharam2007}, Kumar et al.~\cite{KumarLG2014}, and Vellambi and Kliewer~\cite{VellambiK2016,VellambiK2018}. Coordination in the finite-length regime  was studied by Cervia et al.~\cite{CerviaOS19}. The form of coordination discussed above has been called \emph{strong coordination} to contrast with a weaker form called \emph{empirical coordination}. In strong coordination, the distribution of the sequence of samples need to be close to that of independent and identically distributed (i.i.d.) copies of the desired distribution, whereas in empirical coordination only the empirical distribution of the sequence of samples is required to be close to the desired distribution~\cite{CuffPC10}. Empirical coordination has also been extensively investigated~\cite{CuffPC10,Raginsky2013,Treust2014,Treust2015,Treust2015a,TreustB2016,CerviaLT2016,MylonakisSS19,MylonakisSS19ITW}.  

Non-interactive common randomness (CR) generation was first studied by G\'{a}cs and K\"{o}rner~\cite{GacsK73} and a companion result was later shown by Witsenhausen~\cite{Wit75}. See Mossel et al.~\cite{Mosseletal2006}, Yang~\cite{Yang2007}, and Bogdanov and Mossel~\cite{BogdanovM11} for more recent works on this. CR generation with interactive communication was studied by Ahlswede and Csisz\'{a}r~\cite{AhlswedeC98}. CR generation with a helper was studied by Csisz\'{a}r and Narayan~\cite{CsiszarN00}. Generation of CR by keeping it secret from an eavesdropper, i.e., \emph{secret key agreement}, has been studied by several authors~\cite{AhlswedeC93,Maurer93,CsiszarN00,Tyagi13,LiuCV15,MukherjeeKS16,ChanMKZ18}.
 Privacy amplification, where two users extract a secret key from a common random variable about which an eavesdropper has partial information, has also been studied~\cite{BennettBR88,BennettBCM95,MaurerW97,CachinM97}.
 
A source of common randomness available to all the users is a potentially useful resource for coordination. It is known that common randomness (significantly) helps reduce the amount of communication required for {strong coordination}, but not for empirical coordination~\cite{CuffPC10}. The focus of the present paper is on strong coordination. In general, a common source of randomness may not always be available to all users, e.g., in a decentralized network. However, some subsets of users may share randomness. We shall call this form of randomness ``shared randomness'' in contrast to common randomness which is accessible to all the users. Also, the communication links may be available only between certain users or may be shared (e.g., in a wireless network). Motivated by this, we study the settings below which model these aspects while being simple enough to be tractable. We note here an earlier work~\cite{KadampotB17} on coordination using shared randomness and incomplete communication graph which studied a three-user cascade network where only two users share randomness.

  Consider $t$ processors, a coordinator $K$, and a rate limited communication link from the coordinator to the processors. First, we study the \emph{omniscient coordinator setting}. Here, the coordinator has access to $h$ independent random variables $W_1,W_2,\dots,W_h$ and each processor has access to a subset of these random variables. In addition, all processors and the coordinator can privately randomize. The processors want to generate approximately (in the sense of asymptotically vanishing total variation distance) i.i.d. samples from a given joint distribution $q_{X_1 X_2\dots X_t}$. Notice that, in the absence of shared random variables $W_1,W_2,\dots,W_h$, this setting reduces to Wyner's common information problem~\cite{Wyner75}, whose multi-user generalization, among other things, was studied by Xu et al.~\cite{Liu2010}.
  
In the general problem, the structure of the collection of subsets of the variables $W_1, W_2, \ldots, W_h$ is arbitrary. However, we can identify two extreme cases which are interesting and provide some insight into the types of achievability strategies that may be effective. In particular, we give special attention to two models in which $h = t$: $(i)$ the \emph{individually shared randomness model} where processor $P_i$ has access to random variable $W_i$, $i\in[1:t]$ and $(ii)$ the \emph{randomness-on-the-forehead model}\footnote{The metaphor is that each processor $P_i$ is like a person at a party who is wearing a hat labeled with $W_i$. They can see all hats except the one they are wearing.} where processor $P_i$ has access to all random variables except $W_i$, $i\in[1:t]$. Figure~\ref{fig:model} shows these two models for $t=3$. 
 
 In the omniscient coordinator setting with $t = 2$ processors, note that  the individually shared randomness model and the randomness-on-the-forehead model are equivalent. It is easy to infer from the literature~\cite{Wyner75,Bennet02,winter2002compression} that a rate of $\min\{0.5C(X_1;X_2)$, $I(X_1;X_2)\}$ is achievable under {unlimited shared randomness}, where
\begin{align}\label{eqn:wyner_discussion}C(X_1;X_2):=\underset{X_1-U-X_2}{\min}I(X_1,X_2;U)\end{align} is Wyner's common information~\cite{Wyner75}. Firstly, note that shared randomness can be converted to common randomness using a simple network coding technique, In particular, the coordinator can send the \texttt{XOR} of two individually shared random strings producing $2$ bits of common randomness for every bit sent. Then, Wyner's result~\cite{Wyner75} shows that $0.5C(X_1;X_2)$ is achievable (see Figure~\ref{wynerscheme}). On the other hand, note that using their shared randomness, coordinator and processor $P_1$ can sample $X_1^n$ i.i.d. with distribution $q_{X_1}$. We can treat coordinator and processor $P_1$ as a single entity (encoder) having an input i.i.d. $X_1^n$, which sends a message $M$ to processor $P_2$ (decoder), which has to produce $X_2^n$ according to the desired distribution, implying that $I(X_1;X_2)$ is achievable using channel simulation~\cite{Bennet02,winter2002compression} (see Figure~\ref{channelsim}). 

These ideas illustrate different aspects relevant to our problem. However, it turns out that neither of these ideas are optimal, in general. The novelty of our optimal achievable scheme is that it builds on these ideas treating them as guideposts while strictly improving over them. It uses shared randomness in two different ways: some part is turned into common randomness using network coding and the remaining part is used jointly between the coordinator the respective processor.  Our scheme builds upon a non-trivial synthesis of the above two ideas making optimal use of the shared randomness. Please refer Section~\ref{section:two_proc} for details. The former and the latter schemes arise as extreme cases in our scheme bridging the gap between these two schemes. Our proofs for the multi-processor setting generalize the ideas from the proofs of the omniscient coordinator setting with two processors.

The optimal rate of communication from the coordinator to the two processors who want to output approximately i.i.d. samples from a given joint distribution $q_{X_1X_2}$ under unlimited individually shared randomness is given by (Theorem~\ref{theorem:unlimited_shared_randomness})
\begin{align}
\min_{p_{U|X_1,X_2}}\max\left\{I(X_1;X_2|U),I(X_1,X_2;U)\right\}\label{eqn:expresn}.
\end{align}
A more general form of the optimization problem in \eqref{eqn:expresn} was studied independently by Wang et al.~\cite{WangLG2016} and Gastpar and Sula~\cite{GastparS19}, which they defined as the \emph{relaxed Wyner's common information} (see Remark~\ref{gastpar} for details). Our result (Theorem~\ref{theorem:unlimited_shared_randomness}) can be thought of as giving an alternative operational interpretation to the optimization problem in \eqref{eqn:expresn}.

\begin{figure}[htbp]
\centering
\includegraphics{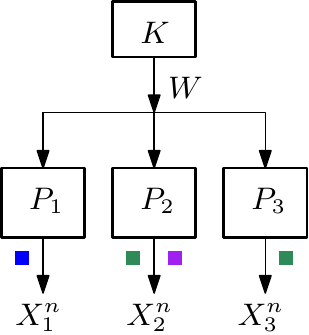}
\caption{Oblivious coordinator setting. Blue, purple, and green colors represent shared random variables $W_1, W_2$, and $W_3$, respectively. Coordinator $K$ sends common randomness $W$ to the processors $P_1,P_2,$ and $P_3$ so that they may output $X_1^n,X_2^n,$ and $X_3^n$, respectively, where $(X_{1i},X_{2i},X_{3i})$, $i=1,\dots,n$, are (approximately) i.i.d. with $q_{X_1X_2X_3}$.}
\label{fig:modelobv}
\end{figure}

We also study the \emph{oblivious coordinator setting} which is similar to the omniscient coordinator setting except that the coordinator does not have access to any of the shared random variables $W_1,W_2,\dots,W_h$. Figure~\ref{fig:modelobv} shows the oblivious coordinator setting for $t=3$ and a specific shared randomness structure. One extreme in the problem space is when the random variables $W_1, W_2, \ldots, W_h$ are not present. In this case, the oblivious coordinator setting also reduces to Wyner's common information problem~\cite{Wyner75},\cite{Liu2010}. The oblivious coordinator setting is similar to Wyner's common information problem~\cite{Wyner75,Liu2010}. Coordinators in both Wyner's common information problem and the oblivious coordinator setting send a uniformly distributed common random message to all the processors in order to produce approximately i.i.d. samples. However, because the processors have access to subsets of the shared random variables, the communication rate required by the coordinator is potentially smaller in the oblivious coordinator setting. In that sense, the oblivious coordinator setting can be seen as an extension of Wyner's common information problem~\cite{Wyner75,Liu2010}. In fact, our results recover the multi-user generalization of Wyner's common information~\cite{Liu2010}.

The oblivious coordinator setting is closely related to channel resolvability~\cite{Wyner75} in the sense that the coordinator sends uniformly distributed common random message to all the processors as in Wyner's common information problem where an application of channel resolvability is inherent. However, we additionally have shared random variables here. Taking this shared randomness also into account, our proof technique builds on the use of versions of channel resolvability in recent works~\cite{Hayashi}, \cite{Cuff13},~\cite{BlochK13}.

In brief, the main contributions of this work are as follows:
\begin{itemize}
\item In the omniscient coordinator setting, for the individually shared randomness model, we characterize the optimal transmission rate under unlimited shared randomness (Theorem~\ref{theorem:generalize}). Our characterization is in terms of a notion of multivariate mutual information (namely, Watanabe's total correlation~\cite{Watanabe60}).
\item In the omniscient coordinator setting, for the randomness-on-the-forehead model, we give an upper bound on the optimal transmission rate under unlimited shared randomness, which turns out to be tight for some special cases (Theorem~\ref{theorem:t=3}). Our upper bound is in terms of another notion of multivariate mutual information (Han's dual total correlation~\cite{Han75}). We also give an achievable strategy in the omniscient coordinator setting for the general case where the processors have access to arbitrary subsets of sources of randomness.
\item In the omniscient coordinator setting, for the general case where the processors have access to arbitrary subsets of sources of randomness, we characterize the trade-off region between the communication and shared randomness rates when all the processors wish to output the same random sequence (Theorems~\ref{theorem_greedy}, \ref{theorem:foreheadequal} and \ref{theorem:equal}).  Indeed, we consider a more general model, i.e., omniscient coordinator with correlated shared randomness model, where the processors observe components of correlated sources (with the coordinator observing all the components), and characterize the optimal transmission rate when all the processors wish to output the same random sequence (Theorem~\ref{theorem:assisted}).
\item In the oblivious coordinator setting, we completely characterize the trade-off region between the communication and shared randomness rates for the general case where the processors have access to arbitrary subsets of sources of randomness (Theorem~\ref{theorem_oblv_general}).
\end{itemize}
 
The remainder of this paper is organized as follows. We present our problem definition in Section~\ref{prob_def}. The two-processor setting with an omniscient coordinator is presented in Section~\ref{section:two_proc}, the multi-processor setting (including the individually shared randomness model and the randomness-on-the-forehead model) is presented in Section~\ref{section:multi_proc}, and the omniscient coordinator with correlated shared randomness model is presented in Section~\ref{section:assisted}. The Oblivious coordinator setting is presented in Section~\ref{section:oblivious}.

% !TEX root = Generalization.tex

\section{Problem Definition}\label{prob_def}
\emph{Notation:} We use a capital letter (e.g.,\ $P_X$) to denote a random p.m.f.~(see, e.g.,~\cite{YassaeeAG14}, \cite{Cuff13}) and a lower-case letter (like $p_X$) to denote a non-random p.m.f. %
For any two sequences of random p.m.f.'s $\{ P_{X^{(n)}} : n \in \mathbb{N} \}$ and $\{ Q_{X^{(n)}} : n \in \mathbb{N} \}$ on a sequence of sets $\{ \mathcal{X}^{(n)} : n \in \mathbb{N} \}$ (where $\mathcal{X}^{(n)}$ is arbitrary and can differ from the Cartesian product $\mathcal{X}^n$), we write $P_{X^{(n)}}\approx Q_{X^{(n)}}$ if $\lim_{n\rightarrow \infty}\mathbbm{E} \norm{ P_{X^{(n)}}-Q_{X^{(n)}} }_1=0$.  

We present definitions for the \emph{omniscient coordinator setting} here. Similar definitions for the \emph{oblivious coordinator setting} and the \emph{omniscient coordinator with correlated shared randomness model} can be written down analogously (see Section~\ref{section:assisted} and Section~\ref{section:oblivious}, respectively, for details).  Our model consists of a coordinator $K$, processors $P_1,\dots,P_t$. The coordinator has $h$ independent sources of randomness $W_1,\dots,W_h$ where each $W_j$ is uniformly distributed in $[1:2^{nR_j}]$ for $j \in [1:h]$ and each processor has access to a subset of these random variables. Let $\mathcal{V}_i$ denote the shared randomness accessible to $P_i$, i.e., $\mathcal{V}_i=\{ j: W_j\ \text{is accessible to } P_i\}$, and $\mathcal{V}:=\left(\mathcal{V}_i \right)_{i\in[1:t]}$. Let $\mathcal{X}_i$ be a finite alphabet for each  $i\in[1:t]$. The goal is to produce $(X^n_1,X_2^n,\dots,X^n_t) \in \mathcal{X}_1^n \times \mathcal{X}_2^n \times \cdots \times \mathcal{X}_t^n$ such that they are approximately (in the sense of asymptotically vanishing total variation distance) distributed according to $q_{X_1\dots X_t}^{(n)}:=\prod_{i=1}^n q_{X_1\dots X_t}(x_{1i},\dots,x_{ti})$. When $h=t$, and for all $i\in[1:t]$, $\mathcal{V}_i=\{i \}$, we call this the \emph{individually shared randomness model}. When $h=t$, and for all $i\in[1:t]$, $\mathcal{V}_i=[1:t]\setminus\{i \}$, we call this the \emph{randomness-on-the-forehead model}. On observing $W_1,\dots,W_h$, the coordinator $K$ produces a message $M\in[1:2^{nR}]$ according to $p(m|w_{[1:h]})$ (with $w_{\mathcal{S}}:=\{w_j:j\in\mathcal{S}\}$) and sends it over a common communication link to $t$ processors. Processor $P_i$ produces $X_i^n\in\mathcal{X}^n_i$ according to a random map $p(x_i^n|m,w_{\mathcal{V}_i})$, $i\in[1:t]$.

\begin{defn}\label{defn1}
An $(n,2^{nR},2^{nR_1},\dots,2^{nR_h})$ \emph{simulation code} consists of $\left(p(m|w_{[1:t]}),p(x_1^n|m,w_{\mathcal{V}_1}),\dots,p(x_t^n|m,w_{\mathcal{V}_t})\right)$, where $m\in[1:2^{nR}], w_i\in[1:2^{nR_i}]$, $i\in[1:h]$.  
\end{defn}
The joint distribution of $(W_{[1:h]},M,X_1^n,\dots,X_t^n)$ and induced distribution on $(X_1^n\dots,X_t^n)$ are given by
\begin{align*}
p(w_{[1:h]},m,x_1^n,\dots,x_t^n)&=\frac{p(m|w_{[1:h]})\prod_{i=1}^tp(x_i^n|m,w_{\mathcal{V}_i})}{2^{n(R_1+\dots+R_h)}},\\
p(x_1^n,\dots,x_t^n)&=\sum_{w_{[1:h]},m}p(w_{[1:h]},m,x_1^n,\dots,x_t^n).
\end{align*}
\begin{defn}\label{defn2}
A rate tuple $(R,R_1,\dots,R_h)$ is said to be \emph{achievable for a p.m.f. $q_{X_1,\dots,X_t}$} if there exists a sequence of $(n,2^{nR},2^{nR_1},\dots$,$2^{nR_h})$ simulation codes such that
\begin{align}
\lim_{n\rightarrow \infty} \lVert p_{X_1^n,\dots,X_t^n}^{\mrm{(induced)}}-q_{X_1,\dots,X_t}^{(n)}\rVert_1=0\label{eqn:correctness}.
\end{align}
\end{defn}
The \emph{simulation rate region} $\mathcal{R}(\mathcal{V})$ is the closure of the set of all achievable rate tuples $(R,R_1,\dots,R_h)$. Let $\mathcal{R}^{\mrm{Indv}}$ and $\mathcal{R}^{\mrm{Forehead}}$ denote the simulation rate regions for the individually shared randomness model and the randomness-on-the-forehead model, respectively.

\begin{defn}
The \emph{optimal transmission rate} $R_{\mrm{opt}}(\mathcal{V})$ is the infimum of all the rates $R$ such that there exists $R_1,\dots,R_h$ so that 
$(R,R_1,\dots,R_h)\in\mathcal{R}(\mathcal{V})$. Let $R^{\mrm{Indv}}_{\mrm{opt}}$ and 
$R^{\mrm{Forehead}}_{\mrm{opt}}$ denote the respective infima for the individually shared randomness model and the randomness-on-the-forehead model.
\end{defn}

% !TEX root = Generalization.tex
\section{Omniscient Coordinator Setting: Two Processors}\label{section:two_proc}
We start with the simplest setting: an omniscient coordinator with two processors. This case will present the proof techniques clearly and later we will discuss how the techniques can be generalized to multiple-processor scenario.
Recall that for $t = 2$ processors, the individually shared randomness model and the randomness-on-the-forehead model are identical. We state results for the individually shared randomness model and the randomness-on-the-forehead results follow by switching $R_1$ and $R_2$. Without loss of generality then we drop the superscripts and refer to the simulation rate region $\mc{R}$ and optimal transmission rate $R_{\mrm{opt}}$. To simplify the subscripts we define $X = X_1$ and $Y = X_2$.
% !TEX root = Generalization.tex

Let $q_{X,Y}=q_{X_1,X_2}$, and $\Rach$ be the set of all non-negative rate triplets $(R,R_1,R_2)$ such that
\begin{align} \label{eqn:1}
R+R_1&\geq I(X,Y;U,U_1),\nonumber\\
R+R_2&\geq I(X,Y;U,U_2),\nonumber\\
R&\geq I(U_1;U_2|U),\nonumber\\
R+R_1+R_2&\geq I(U_1;U_2|U)+I(X,Y;U,U_1,U_2),\nonumber\\
2R+R_1+R_2&\geq I(U_1;U_2|U)+I(X,Y;U)\nonumber\\
&\hspace{1cm}+I(X,Y;U,U_1,U_2),\nonumber\\
2R&\geq I(U_1;U_2|U)+I(X,Y;U),
\end{align}
for some p.m.f. $p(x,y,u,u_1,u_2)=q(x,y)p(u,u_1,u_2|x,y)$ s.t. $X-(U,U_1)-(U,U_2)-Y$.
\begin{thm}\label{theorem:achievability}
For the two-processor simulation problem with an omniscient coordinator, the set of rates $\Rach$ is achievable: $\Rach \subseteq \mc{R}$.
\end{thm}
The proofs of this theorem and the subsequent theorems are presented in Section~\ref{section:proofs}. We show the above result to be tight in some settings. When the shared randomness rates $R_1$ and $R_2$ are sufficiently large, we can characterize the optimal transmission rate.
\begin{thm}\label{theorem:unlimited_shared_randomness}
The optimal transmission rate for the omniscient coordinator setting with two processors is given by the following expression:
\begin{align}
&R_{\mrm{opt}}=\min\max\big\{I(X;Y|U),I(X,Y;U)\big\}\nonumber\\
&=\min \max \left\{I(X;Y|U), \frac{1}{2}\big(I(X,Y;U)+I(X;Y|U)\big)\right\}\nonumber,
\end{align}
where the minimum is over all probability mass functions 
	\begin{align*}
	p(x,y,u)=q(x,y)p(u|x,y)
	\end{align*}
such that
	\begin{align*}
	|\mathcal{U}|\leq |\mathcal{X}||\mathcal{Y}|+2.
	\end{align*}
\end{thm}
\begin{remark}\label{gastpar}
An optimization problem closely related to the first expression of $R_{\mrm{opt}}$ in Theorem~\ref{theorem:unlimited_shared_randomness} was studied in the context of information-theoretic caching~\cite{WangLG2016,GastparS19}. In particular, Gastpar and Sula~\cite{GastparS19} defined relaxed Wyner's common information as  $$C_\gamma(X;Y):=\min\limits_{p_{U|XY}: I(X;Y|U)\leq \gamma} I(X,Y;U), \ \gamma>0.$$
The optimal transmission rate $R_{\mrm{opt}}$ of Theorem~\ref{theorem:unlimited_shared_randomness} can be expressed in terms of $C_{\gamma}:=C_\gamma(X;Y)$ as follows.
\begin{align*}
R_{\mrm{opt}}&=\min_\gamma\max\{\gamma,C_\gamma\}\\
&=C_{\gamma^*},
\end{align*} 
where $\gamma^*$ is the fixed point of the function $C_\gamma$, i.e., the solution to $C_\gamma=\gamma$.
\end{remark}

\begin{figure}[htbp]
\centering
\includegraphics[scale=0.88]{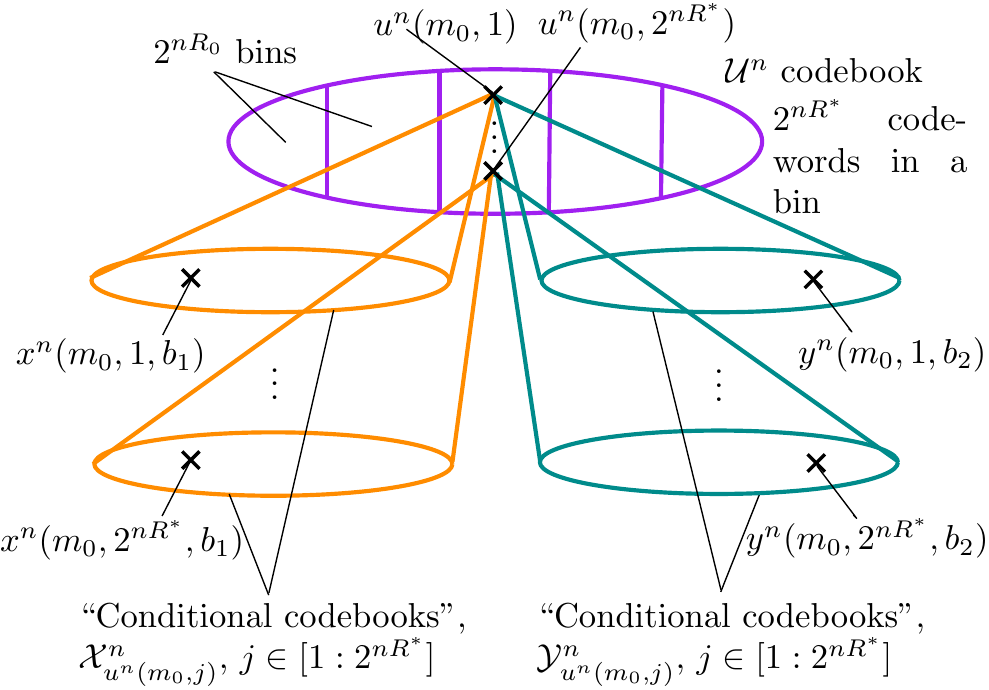}
\caption{A schematic diagram of the coding scheme for the achievability part of Theorem~\ref{theorem:unlimited_shared_randomness} (when the shared randomness rates are large enough): Indices $m_0,b_1$ and $b_2$ are determined by the shared randomness in the following way: Index $m_0$ which is uniformly distributed on $[1:2^{nR_0}]$ is a concatenation of two $\frac{nR_0}{2}$ length bit strings $m_{01}$ and $m_{02}$, where $m_{0i}$ is obtained from shared randomness $w_i$, for $i=1,2$. Index $b_i$ which is independent of $m_0$ and uniformly distributed on $[1:2^{n\tilde{R}_i}]$ is also obtained from shared randomness $w_i$, for $i=1,2$. Note that $m_0,b_1,b_2$ are mutually independent of each other. The coordinator finds an $m^*$ inside the bin indexed by $m_0$, such that $\left(u^n(m_0,m^*),x^n(m_0,m^*,b_1),y^n(m_0,m^*,b_2)\right)$ is consistent with high probability. Loosely, $R^*>I(X;Y|U)$ ensures that there exists such an $m^*$. The coordinator then sends $(m_{01}\oplus m_{02},m^*)$ as a common message to the processors at a rate $R=\frac{R_0}{2}+R^*$. Note that $P_i$ has access to $m_{0i}$ and recovers $m_0$. The processors $P_1$ and $P_2$ output $x^n(m_0,m^*,b_1)$ and $y^n(m_0,m^*,b_2)$, respectively. Roughly, $R_0+R^*>I(X,Y;U)$ ensures that the output is according to the desired distribution. Since $R=\frac{R_0}{2}+R^*$, the above rate constraints imply that $\max \left\{I(X;Y|U),\frac{1}{2}\big(I(X;Y|U)+I(U;X,Y)\big)\right\}$ is achievable when the shared randomness rates are large enough.} 
\label{fig:picture}
%\vspace{-18pt}
\end{figure}

\begin{figure*}[htbp]
\begin{center}
\includegraphics{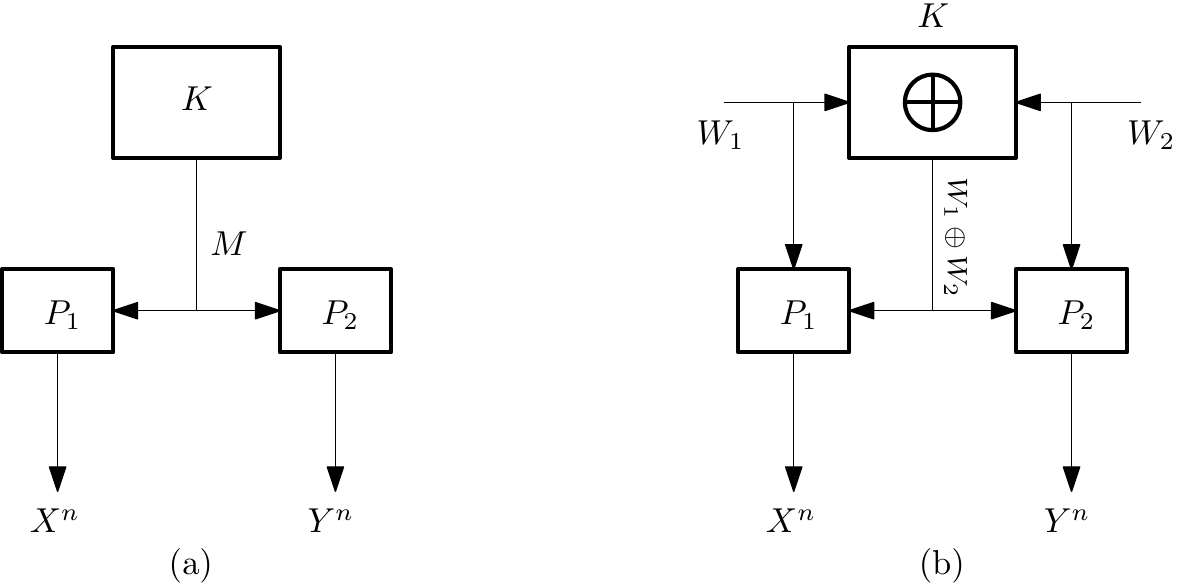}
\end{center}
\caption[A rate of $0.5C(X;Y)$ is achievable.]{A rate of $0.5C(X;Y)$ is achievable. (a) The model on the left is the setup for Wyner's common information problem~\cite{Wyner75}.  The coordinator sends a uniformly distributed common message to both the processors who may output i.i.d. samples from a given joint distribution.  The infimum of achievable common randomness rates is called Wyner's common information $C(X;Y)$. (b) The model on the right is for our problem. Consider $W_1,W_2$ each to be uniformly distributed on $[1:2^{nR}]$. We treat $W_1$ and $W_2$ each as an $nR$-length bit string. Let the coordinator transmit the bit string $M=W_1 \oplus W_2$ $(`\oplus$' denotes bit-wise $XOR)$ over common communication link to both the processors. Note that rate of transmission is $R$. From this both the processors can recover $(W_1,W_2)$ which is a common random variable uniformly distributed on $[1:2^{n(2R)}]$. Then, Wyner's result~\cite{Wyner75} shows that $2R\geq C(X;Y)$ is achievable, i.e., $R\geq 0.5C(X;Y)$ is achievable.}\label{wynerscheme}
\end{figure*}

\begin{figure*}[htbp]
\begin{center}
\includegraphics{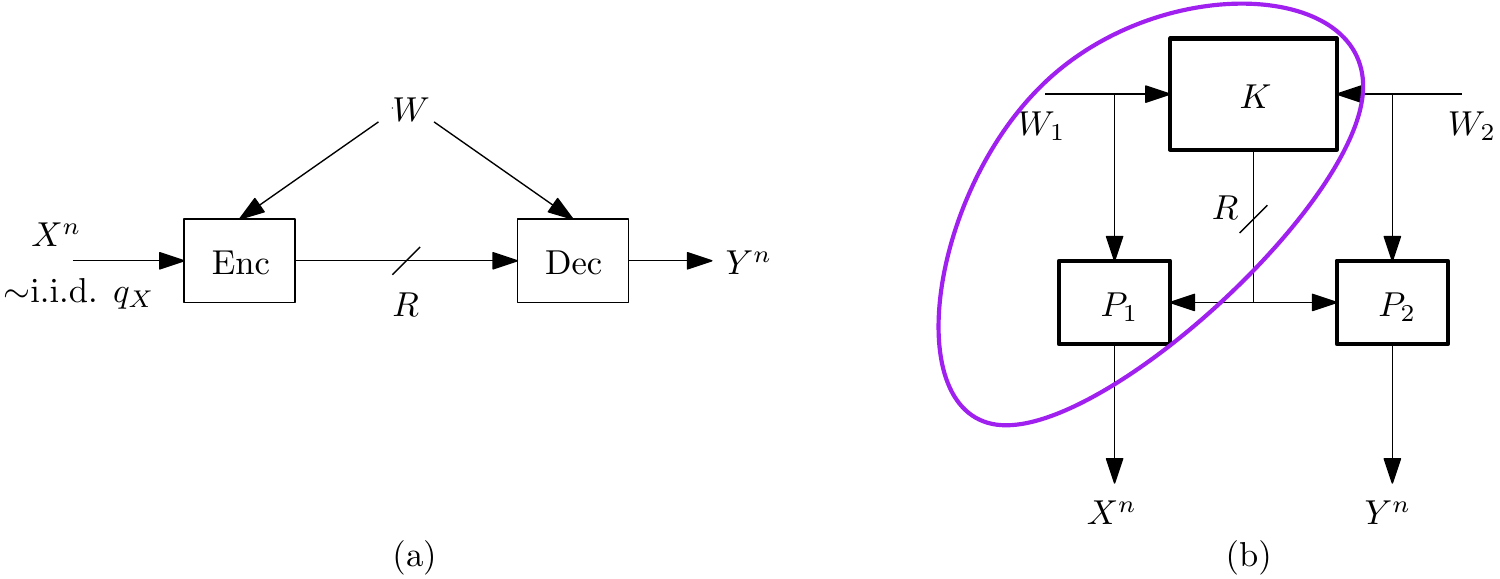}
\end{center}
\caption[A rate of $I(X;Y)$ is achievable.]{A rate of $I(X;Y)$ is achievable. (a) The model on the left is the setup for the channel simulation problem of Bennett et al.~\cite{Bennet02} and Winter~\cite{winter2002compression}. An encoder observing i.i.d. $X^n$ sends a message of rate $R$ to decoder to approximate a noisy channel $q_{Y|X}$ between them. Common randomness $W$ assists them in this. A rate of $R=I(X;Y)$ is achievable in this model under unlimited shared randomness~\cite{Bennet02,winter2002compression}. (b) The model on the right is our problem. Using their shared randomness $W_1$, the coordinator $K$ and the processor $P_1$ sample $X^n$ i.i.d. with distribution $q_{X}$. We can treat the coordinator $K$ and the processor $P_1$ as a single entity (encoder) which treats the sampled $X^n$ as an input and  sends a message $M$ of rate $R$ to the processor $P_2$ (decoder). The encoder and decoder share randomness $W_2$. The decoder $P_2$ produces $Y^n$ according to the desired distribution, implying that $I(X;Y)$ is achievable~\cite{Bennet02,Cuff13}.}\label{channelsim}
\end{figure*}

A discussion of the intuition behind our achievable scheme by focusing on Theorem~\ref{theorem:unlimited_shared_randomness} is in order. 
%\subsection{Discussion} \label{section:discussion}
%In this section, we discuss the intuition behind our achievability part by focusing on Theorem~\ref{theorem:unlimited_shared_randomness}. 
%Some of the discussion here will be informal. See Section~\ref{section:proofs} for precise details. 
%\RRR{As we discuss below, results from the literature~\cite{Wyner75,Bennet02,Cuff13} imply that when the shared randomness rates are large enough, a rate $R$ of $\min\left\{0.5C(X;Y),I(X;Y)\right\}$ is achievable, where $C(X;Y)$ is Wyner's common information. Our achievability scheme builds on the ideas behind this and we show by an example that our results strictly improve over $\min\left\{0.5C(X;Y),I(X;Y)\right\}$.}
 Based on the literature~\cite{Wyner75,Bennet02,Cuff13}, we first make some quick observations. In Figure~\ref{wynerscheme}, we may use network coding to turn shared randomness into common randomness and then employ Wyner's scheme~\cite{Wyner75} for coordination when common randomness is available. This gives an achievable rate of $0.5C(X;Y)$ for our problem, where $C(X;Y)$ is Wyner's common information~\ref{eqn:wyner_discussion}. In Figure~\ref{channelsim}, we use the channel simulation problem~\cite{Bennet02,Cuff13} and argue that a rate of $I(X;Y)$ is achievable for our problem. 
% \RRR {Our achievability scheme builds on the ideas behind this and we show by an example that our results strictly improve over $\min\left\{0.5C(X;Y),I(X;Y)\right\}$.}
  Our achievable scheme builds on the ideas behind these. While the complete technical details are in Section \ref{section:proofs}, an intuitive explanation is given in Figure~\ref{fig:picture}. 

The main new ingredient in proving the converse is an upper bound on $I(X^n;Y^n|M)$, the conditional mutual information of the outputs of the processors conditioned on the message from the coordinator. In the absence of any shared randomness (i.e., the setup for Wyner's common information problem), this quantity is zero, i.e., the outputs are conditionally independent conditioned on the message from the coordinator. However, in our setup, this no longer need be true due to the presence of shared randomness. Nevertheless, we show that $I(X^n;Y^n|M)$ cannot be arbitrarily large, and in particular, is upper bounded by the size of common message.
We have the following upper and lower bounds on $R_{\mrm{opt}}$.
\begin{thm}\label{theorem:bounds}
\begin{align}
0.5I(X;Y)\leq R_{\mrm{opt}}\leq \min\{0.5C(X;Y),I(X;Y)\}\label{eqn:bounds}.
\end{align}
Furthermore, the lower bound is tight if and only if there exists a $P_{U|XY}$ such that $U-X-Y$ and $U-Y-X$ are Markov chains and $I(X;Y)\leq I(X;U)+I(Y;U)$. The upper bound $R_{\mrm{opt}}\leq I(X;Y)$ is tight if and only if $X$ is independent of $Y$. The upper bound $R_{\mrm{opt}}\leq 0.5{C(X;Y)}$ is tight if $X=(X^\prime,V)$ and $Y=(Y^\prime,V)$, where $X^\prime$ and $Y^\prime$ are conditionally independent given $V$.
\end{thm}
%\begin{proof}
%Consider the second expression for $R_{\mrm{opt}}$ in Theorem~\ref{theorem:unlimited_shared_randomness}. To see the lower bound, notice that $I(X,Y;U)+I(X;Y|U)\geq I(X;Y)$. For the upper bound, choosing $U$ to be a minimizer in \eqref{eqn:wyner_discussion} gives us $R_{\mrm{opt}}\leq 0.5C(X;Y)$. Choosing $U=\emptyset$ gives us $R_{\mrm{opt}} \leq I(X;Y)$.
%\end{proof}
%It is easy to see that
%\begin{align}
%R_{\mrm{opt}}\leq\min\left\{0.5C(X;Y),I(X;Y)\right\}\label{eqn:discussion_lessthan_CInI}.
%\end{align}
%To see this, consider the second expression for $R_{\mrm{opt}}$ in Theorem~\ref{theorem:unlimited_shared_randomness}. Choosing $U$ to be a minimizer in \eqref{eqn:wyner_discussion} gives us $R_{\mrm{opt}}\leq 0.5C(X;Y)$. Choosing $U=\emptyset$ gives us $R_{\mrm{opt}} \leq I(X;Y)$. 

Next, we present an example where the upper bound in \eqref{eqn:bounds} is strict.

\begin{exmp}\label{open}
Consider a doubly symmetric binary source $\DSBS(a)$ on $\{0,1\}^2$ with joint distribution 
	\begin{align*}
	q(x,y) =  \begin{bmatrix} \frac{a}{2} & \frac{1 - a}{2} \\
		\frac{1 - a}{2} & \frac{a}{2} 
		\end{bmatrix}
	\end{align*}
where $a\in[0,0.5]$.
%$q(x,y) = 0.5(1-a)\delta_{xy}+0.5a(1-\delta_{xy})$, $a\in[0,0.5]$ and $x,y\in\{0,1\}$. 
For this distribution $I(X;Y)=1-h(a)$, where $h(\cdot)$ is the binary entropy function defined by $h(t):=-t\log{t}-(1-t)\log{(1-t)}$. Define $p^t(u|x,y) := tp^\bot(u|x,y)+(1-t)p^*(u|x,y), t\in[0,1],$ where
\begin{align*}
p^\bot(0|x,y)&=0.5=p^\bot(1|x,y)\ , \forall\  x,y,\\
p^*(u|x,y)&=\argmin_{p(u|x,y): X-U-Y}I(X,Y;U).
\end{align*}
The distribution $p^*(u|x,y)$ was found by Wyner~\cite{Wyner75}:
\begin{alignat*}{3}
p^*(0|0,1)&=p^*(1|1,0)  &&=0.5,\\
p^*(0|1,1)&=p^*(1|0,0) &&= 
b^2/(1-a),
%\frac{b^2}{(1-a)},
\end{alignat*}
%\begin{align*}
%p^*(0|0,1)&=p^*(1|1,0)&&=0.5\\
%p^*(0|1,1)&=p^*(1|0,0)&&=\frac{a_1^2}{(1-a_0)},
%\end{align*}
where $b=\frac{1}{2}\left(1- \sqrt{1-2a} \right)$ and the common information $C(X;Y) = I_{p^*}(X,Y;U)=1+h(a)-2h(b)$. Let $f(t)=\max\left\{I_{p^t}(X;Y|U),\frac{1}{2}\big(I_{p^t}(X,Y;U)+I_{p^t}(X;Y|U)\big)\right\}$, where $I_{p^t}(X;Y|U)$ and $I_{p^t}(X,Y;U)$ are calculated under $p^t(u|x,y)$:
\begin{align*}
&I_{p^t}(X,Y;U)=1+h(a)-h\left( \alpha,\frac{a}{2},\frac{a}{2},1-a-\alpha \right)\\
&I_{p^t}(X;Y|U)=2 h \left(\alpha+\frac{a}{2} \right)-h \left(\alpha,\frac{a}{2},\frac{a}{2},1-a-\alpha \right),
\end{align*}
where $\alpha=\left(1-t\right)b^2+\frac{t}{2}\left(1-a\right)$.
We can verify that the two endpoints are $f(0)=0.5C(X;Y)$ and $f(1)=I(X;Y)$ (See Appendix~\ref{appreviewers1}).
We find a $t^*$ such that $I_{p^{t^*}}(X,Y;U)=I_{p^{t^*}}(X;Y|U)$, i.e., $t^*$ such that
 \begin{align*}
 1+h(a)&=2h\left(\left(1-t^*\right)b^2+\frac{t^*}{2}\left(1-a\right)+\frac{a}{2}\right)\\
 \Rightarrow \quad t^*&=\frac{1}{\left(\frac{1-a}{2}-b^2\right)}\left(h^{-1}\left(\frac{1+h(a)}{2}\right)-\frac{a}{2}-b^2\right).
 \end{align*}
For any $a\in(0,0.5)$, we can numerically see that $f(t^*)<\min\left\{f(0),f(1)\right\}=\min\left\{0.5C(X;Y),I(X;Y)\right\}$ (Figure~\ref{graphs} illustrates this fact for $a=0.1$ and $a=0.2$) implying that $R_{\mrm{opt}}<\min\left\{0.5C(X;Y),I(X;Y)\right\}$ since $R_{\mrm{opt}} \leq f(t^*)$. Moreover, we conjecture that $p^{t^*}(u|x,y)$ (with $t^*$ as identified above) is an optimizer for the expressions of $R_{\mrm{opt}}$ in Theorem~\ref{theorem:unlimited_shared_randomness}. The conjecture is supported by the fact that, it can be numerically checked that $p^{t^*}(u|x,y)$ is a minimizer among all the conditional p.m.f.'s $p(u|x,y)$ with $|\mathcal{U}|=2$. 
%\ads{probably want a comment as to why the conjecture is hard to prove / we didn't spend the time to prove it.}
\begin{figure}[htbp]
\centering
\newlength\fheight 
    \newlength\fwidth 
    \setlength\fheight{4.5cm} 
    \setlength\fwidth{7.6cm}
\definecolor{mycolor1}{rgb}{0.00000,0.44700,0.74100}%
\definecolor{mycolor2}{rgb}{1.00000,0.00000,1.00000}%
\begin{tikzpicture}

\begin{axis}[%
width=0.951\fwidth,
height=\fheight,
at={(0\fwidth,0\fheight)},
scale only axis,
xmin=0,
xmax=1,
xlabel style={font=\color{white!15!black}},
xlabel={$t$},
ymin=0.15,
ymax=0.55,
ylabel style={font=\color{white!15!black}},
ylabel={$f(t)$},
axis background/.style={fill=white},
legend style={at={(0.97,0.5)}, anchor=east, legend cell align=left, align=left, draw=white!15!black}
]
\addplot [color=mycolor1, line width=1.0pt]
  table[row sep=crcr]{%
0	0.436380283400076\\
0.00279999999999991	0.431499055555295\\
0.00600000000000001	0.426557884692081\\
0.00950000000000006	0.421713981411722\\
0.0134000000000001	0.4168360930299\\
0.0177	0.411951459591998\\
0.0224	0.407082762497141\\
0.0275000000000001	0.40224869478021\\
0.0329999999999999	0.39746451880138\\
0.0388999999999999	0.392742573837561\\
0.0452999999999999	0.388021828003421\\
0.0521	0.383391652121826\\
0.0593999999999999	0.37879591957208\\
0.0671999999999999	0.374251203473264\\
0.0756000000000001	0.369718728770988\\
0.0845	0.365268373707793\\
0.0940000000000001	0.360862568888059\\
0.1042	0.356474776169439\\
0.1151	0.352125858796536\\
0.1267	0.347832816323742\\
0.1391	0.343576307207822\\
0.1523	0.339374487447541\\
0.1663	0.33524216255484\\
0.1813	0.331138614803407\\
0.1972	0.327109445067336\\
0.2142	0.323121673675278\\
0.2323	0.319195303007056\\
0.2515	0.315347013931011\\
0.2719	0.311572886227332\\
0.2936	0.307872200330448\\
0.3167	0.304246959760662\\
0.3412	0.300715313268914\\
0.3436	0.300527573378146\\
0.3608	0.312924083055196\\
0.378	0.32494217372874\\
0.3952	0.336585280718171\\
0.4124	0.347856966449423\\
0.4296	0.358760853917631\\
0.4468	0.369300576421957\\
0.464	0.379479739618213\\
0.4812	0.389301892950686\\
0.4984	0.398770508260472\\
0.5156	0.407888963906333\\
0.5328	0.416660533132085\\
0.55	0.425088375711256\\
0.5672	0.433175532122551\\
0.5844	0.440924919678386\\
0.6016	0.448339330157219\\
0.6188	0.45542142858892\\
0.636	0.462173752918356\\
0.6532	0.468598714331232\\
0.6705	0.474733116308751\\
0.6878	0.480540847546466\\
0.7051	0.486023979838117\\
0.7224	0.491184460397482\\
0.7397	0.496024113604635\\
0.757	0.500544642801195\\
0.7743	0.504747632091284\\
0.7916	0.508634548114012\\
0.8089	0.512206741760539\\
0.8262	0.515465449814522\\
0.8435	0.518411796499324\\
0.8608	0.521046794918987\\
0.8781	0.523371348382871\\
0.8954	0.525386251606146\\
0.9127	0.5270921917801\\
0.93	0.528489749507709\\
0.9473	0.529579399600975\\
0.9646	0.530361511737473\\
0.9819	0.530836350974203\\
1	0.531004406410719\\
};
\addlegendentry{$a=0.1$}

\addplot [color=mycolor2, line width=1.0pt]
  table[row sep=crcr]{%
0	0.352952450491633\\
0.00570000000000004	0.346612136545404\\
0.0119	0.340182625832955\\
0.0185	0.333790024675116\\
0.0255000000000001	0.327444257460282\\
0.0328999999999999	0.321154344760102\\
0.0407	0.314928388317052\\
0.0489999999999999	0.308700802941808\\
0.0577000000000001	0.302559241628198\\
0.0669	0.29644380631286\\
0.0766	0.290369631440877\\
0.0868	0.284349818431383\\
0.0974999999999999	0.278395695856848\\
0.1087	0.272517046729866\\
0.1205	0.26667419966715\\
0.1328	0.26092792792823\\
0.1457	0.255241148140532\\
0.1592	0.249625991853822\\
0.1734	0.244054936128429\\
0.1882	0.23857965686195\\
0.2037	0.233173980752178\\
0.2199	0.227850460366026\\
0.2369	0.222590150843885\\
0.2546	0.217436417826332\\
0.2731	0.212371111851642\\
0.2924	0.207406515164156\\
0.3125	0.202553633843804\\
0.3335	0.19780045854464\\
0.3554	0.193160168361872\\
0.3782	0.188644769263606\\
0.4019	0.184265264727702\\
0.4265	0.180031818015786\\
0.4427	0.177497550666439\\
0.4659	0.185639679534829\\
0.4888	0.193353744760011\\
0.5116	0.2007111749631\\
0.5342	0.207682616998486\\
0.5567	0.214302586315447\\
0.5791	0.220573208483627\\
0.6014	0.226497077596756\\
0.6236	0.232077161518353\\
0.6457	0.237316725103683\\
0.6677	0.24221926776353\\
0.6897	0.246808623997168\\
0.7116	0.251065634178802\\
0.7335	0.255011557787713\\
0.7554	0.258646179287083\\
0.7772	0.261954891546168\\
0.799	0.264954916931875\\
0.8208	0.267646216875044\\
0.8426	0.270028784279182\\
0.8644	0.272102633732726\\
0.8862	0.273867793345498\\
0.9079	0.275318321415127\\
0.9296	0.276463055436915\\
0.9513	0.277302028131565\\
0.973	0.277835265984609\\
1	0.278071905112638\\
};
\addlegendentry{$a=0.2$}

\end{axis}
\end{tikzpicture}%
\caption{In both the plots $(t^*,f(t^*))$ is the minimum point which illustrates that $f(t^*)<\min\{f(0),f(1)\}$, where $(0,f(0))$ and $(1,f(1))$ are the respective corner points. (Top) Case when $f(0)=0.5C(X;Y)<I(X;Y)=f(1)$. $t^*=0.343436$ for $a=0.1$. (Bottom) Case when $f(0)=0.5C(X;Y)>I(X;Y)=f(1)$. $t^*=0.442523$ for $a=0.2$.}
\label{graphs}
\end{figure}
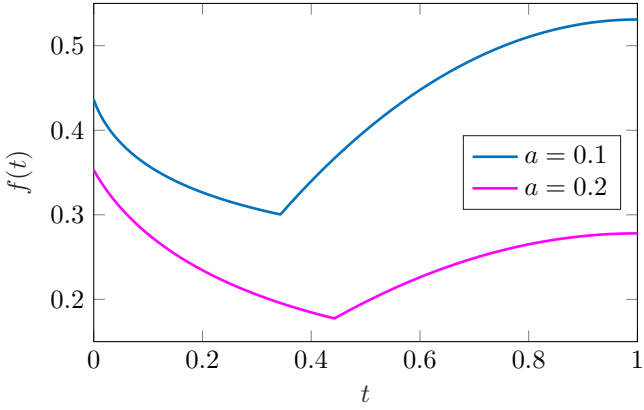
\end{exmp}
\begin{remark}\label{Gastpar:remark}
Independently, Wang et al.\cite{WangLG2016}, and Sula and Gastpar~\cite{SulaG19} addressed the problem of computing the relaxed Wyner common information, $C_\gamma$, of $\DSBS$. Unbeknownst to us, the same choice of auxiliary random variable was proposed in \cite{WangLG2016} which preceded our work, but it was expressed in a different form. The optimality of this choice remains open. In \cite{SulaG19}, the authors also conjecture that it is indeed optimal.
\end{remark}

As expected, when the shared randomness rates approach zero, the optimal transmission rate is equal to Wyner's common information, $C(X;Y)$ as stated in the following theorem. Note that this will not directly follow from \cite{Wyner75} due to the presence of rate triples with shared randomness rate that are non-zero but approaching zero. 
\begin{thm}\label{theorem:noshared}
Let $\RNO = \inf\{ R : (R,0,0) \in \mc{R} \}$ be the smallest transmission rate with no shared randomness. Then $\RNO=C(X;Y).$
\end{thm}
For the case when $X$ and $Y$ are equal, we can completely characterize the simulation rate region as follows.
\begin{thm}\label{theorem:X=Y=Z}
Suppose the output random variables $X$ and $Y$ are identical: $X = Y$ almost surely. Then the simulation rate region is given by the set of all non-negative rate triplets $(R,R_1,R_2)$ such that
\begin{align*}
R+\min\left\{R_1,R_2\right\}&\geq H(X),\\\nonumber
R&\geq \frac{H(X)}{2}.
\end{align*}
\end{thm}

% !TEX root = Generalization.tex

\subsection{Proofs} \label{section:proofs}
\begin{IEEEproof}[Proof of Theorem~\ref{theorem:achievability}]
The proof employs the Output Statistics of Random Binning (OSRB) framework developed by Yassaee et al.~\cite{YassaeeAG14}. 
{We first consider a random binning scheme as follows} (this is along the lines of what Yassaee et al.~\cite{YassaeeAG14} call the ``source coding side'' of the problem). Let $(U^n,U_1^n,U_2^n,X^n,Y^n)$ be i.i.d. with distribution $p(u,u_1,u_2,x,y)=q(x,y)p(u,u_1,u_2|x,y)$ such that $X-(U,U_1)-(U,U_2)-Y$ is a Markov chain. Now, we consider the following random binning:
\begin{itemize}
\item To each $u^n$, assign uniformly and independently three random bin indices $m_0\in[1:2^{nR_0}], f\in[1:2^{n\hat{R}_0}]$ and $m^*\in [1:2^{nR^*}]$.
\item To each pair $(u^n,u_i^n)$, assign uniformly and independently two random bin indices $f_i\in[1:2^{n\hat{R}_i}]$ and $b_i\in[1:2^{n\tilde{R}_i}]$ for $i=1,2$.
\end{itemize}
Further, we use Slepian-Wolf decoders to estimate $(u^n,u_i^n)$ as $(\hat{u}^n_{(i)},\hat{u}_i)$ from $f_i,b_i,m_0,f,m^*$ for $i=1,2$. We denote a Slepian-Wolf decoder by $P^{SW}(\hat{x}^n_{[1,T]}|z^n,b_{[1:T]})$, which equals $1$ if $\hat{x}^n_{[1,T]}$ is the only jointly typical sequence \cite[Chapter 2]{GamalK12} with $z^n$ in the bin $b_{[1:T]}$, where $b_i$ is the bin index corresponding to random binning of $\mathcal{X}_i^n$, $i\in[1:T]$. Otherwise, $\hat{x}^n_{[1,T]}$ is taken to be a fixed arbitrary sequence.
\noindent Then the random p.m.f. induced by the binning can be expressed as follows: 
%\ads{What is $v$ marked in red below? Seems to be from an old version. Also I think the term in green is a repeat and should be removed?}
\begin{align}
P(u^n,&u_1^n,u_2^n,x^n,y^n,m_0,f,m^*,f_1,b_1,f_2,b_2,\nonumber\\
&\hat{u}_1^n,\hat{u}_2^n,\hat{u}^n_{(1)},\hat{u}^n_{(2)})\nonumber\\
&\hspace{-1cm}=p(u^n,u_1^n,u_2^n)p(x^n|u^n,u_1^n)p(y^n|u^n,u_2^n){P(m_0,f|u^n)}\nonumber\\
&\times P(m^*|u^n)P(f_1,b_1|u^n,u_1^n)P(f_2,b_2|u^n,u_2^n)\nonumber\\
&\times P^{SW}(\hat{u}^n_{(1)},\hat{u}_1^n|f_1,b_1,m_0,f,m^*)\nonumber\\
&\times P^{SW}(\hat{u}^n_{(2)},\hat{u}_2^n|f_2,b_2,m_0,f,m^*)\label{eqn:theorem:achievability_protA_simp_1}\\
&\hspace{-1cm}=p(u^n,u_1^n,u_2^n)P(m_0,f,f_1,b_1,f_2,b_2|u^n,u_1^n,u_2^n) \nonumber\\
&\times P(m^*|u^n)P^{SW}(\hat{u}^n_{(1)},\hat{u}_1^n|f_1,b_1,m_0,f,m^*)\nonumber\\
&\times P^{SW}(\hat{u}^n_{(2)},\hat{u}_2^n|f_2,b_2,m_0,f,m^*)\nonumber\\
&\times p(x^n|u^n,u_1^n)p(y^n|u^n,u_2^n)\label{eqn:theorem:achievability_protA_simp_2}\\
&\hspace{-1cm}= P(b_1,b_2,f_1,f_2,m_0,f)P(u^n,u_1^n,u_2^n|b_1,b_2,f_1,f_2,m_0,f)\nonumber\\
&\times P(m^*|u^n)P^{SW}(\hat{u}^n_{(1)},\hat{u}_1^n|f_1,b_1,m_0,f,m^*)\nonumber\\
&\times P^{SW}(\hat{u}^n_{(2)},\hat{u}_2^n|f_2,b_2,m_0,f,m^*)\nonumber\\
&\times p(x^n|u^n,u_1^n)p(y^n|u^n,u_2^n)\label{eqn1:protA},
\end{align}
where \eqref{eqn:theorem:achievability_protA_simp_1} follows from Markov chain $X-(U,U_1)-(U,U_2)-Y$ and binning, \eqref{eqn:theorem:achievability_protA_simp_2} follows from binning.

Now consider a random coding scheme as follows (analogous to what Yassaee et al.~\cite{YassaeeAG14} call the ``main problem assisted with extra shared randomness''). We generate $b_1,b_2,f_1,f_2,m_0,f$ independently and uniformly from the sets $[1:2^{n\tilde{R}_1}],[1:2^{n\tilde{R}_2}],[1:2^{n\hat{R}_1}],[1:2^{n\hat{R}_2}],[1:2^{nR_0}]$ and $[1:2^{n\hat{R}_0}]$ respectively. We treat `$m_0$' as an $nR$-length string of bits i.e., a concatenation of two messages $m_{01},m_{02}$, each consisting of $\frac{nR_0}{2}$ bits. For $i=1,2$, we treat $m_{0i}$ and $b_i$ together as the shared randomness $w_i$ that is shared between the coordinator and processor $P_i$. In addition, we have extra shared randomness $f,f_1$ and $f_2$ which we will eliminate later, where $f$ is shared among coordinator and both the processors, $f_i$ is shared between coordinator and processor $P_i$, for $i=1,2$. The coordinator on observing $b_1,b_2,f_1,f_2,m_0,f$ produces $u^n,u_1^n,u_2^n$ according to $P(u^n,u_1^n,u_2^n|b_1,b_2,f_1,f_2,m_0,f)$ of \eqref{eqn1:protA} and sends $(m_{01}\oplus m_{02},m^*(u^n))$ as a common message $m$ to both the processors, where $m^*(u^n)$ is produced according to $P(m^*|u^n)$ of \eqref{eqn1:protA}. Thus, both the processors can recover `$m_0$' exactly since $P_1$ already has $m_{01}$ and finds $m_{02}=(m_{01}\oplus m_{02})\oplus m_{01}$ and similarly does $P_2$. Then processor $P_1$ uses Slepian-Wolf decoder $P^{SW}(\hat{u}^n_{(1)},\hat{u}_1^n|f_1,b_1,m_0,f,m^*)$ of \eqref{eqn1:protA} to obtain $(\hat{u}^n_{(1)},\hat{u}_1^n)$ as an estimate of $(u^n,u_1^n)$ and produces $x^n$ according to $p(x^n|\hat{u}^n_{(1)},\hat{u}_1^n)$. Similarly, processor $P_2$ uses $P^{SW}(\hat{u}^n_{(2)},\hat{u}_2^n|f_2,b_2,m_0,f,m^*)$ of \eqref{eqn1:protA} and produces $y^n$ according to $p(y^n|\hat{u}^n_{(2)},\hat{u}_2^n)$. This scheme induces the following random p.m.f.
\begin{align}
\hat{P}(u^n,&u_1^n,u_2^n,x^n,y^n,m_0,f,m^*,f_1,b_1,f_2,b_2,\nonumber\\
&\hat{u}_1^n,\hat{u}_2^n,\hat{u}^n_{(1)},\hat{u}^n_{(2)})\nonumber\\
&\hspace{-1cm}=p^{\text{Unif}}(b_1)p^{\text{Unif}}(f_1)p^{\text{Unif}}(b_2)p^{\text{Unif}}(f_2)p^{\text{Unif}}(m_0)p^{\text{Unif}}(f)\nonumber\\
&\times P(u^n,u_1^n,u_2^n|b_1,b_2,f_1,f_2,m_0,f)P(m^*|u^n)\nonumber\\
&\times P^{SW}(\hat{u}^n_{(1)},\hat{u}_1^n|f_1,b_1,m_0,f,m^*)p(x^n|\hat{u}^n_{(1)},\hat{u}_1^n)\nonumber\\
&\times P^{SW}(\hat{u}^n_{(2)},\hat{u}_2^n|f_2,b_2,m_0,f,m^*)p(y^n|\hat{u}^n_{(2)},\hat{u}_2^n)\label{eqn1:protB}
\end{align}
Recall that we write $P_{X^{(n)}}\approx Q_{X^{(n)}}$ to denote that $\lim_{n\rightarrow \infty}\mathbbm{E} \norm{ P_{X^{(n)}}-Q_{X^{(n)}} }_1=0$. We use Yassaee et al.~\cite[Theorem~1]{YassaeeAG14}. By substituting $T=3, X_1=(U,U_1), X_2=(U,U_2), X_3=U,$ and $Z=\emptyset$ in their result we have
\begin{align*}
P(&b_1,b_2,f_1,f_2,m_0,f)\nonumber\\
&\approx p^{\text{Unif}}(b_1)p^{\text{Unif}}(b_2)p^{\text{Unif}}(f_1)p^{\text{Unif}}(f_2)p^{\text{Unif}}(m_0)p^{\text{Unif}}(f)
\end{align*}
if the following conditions hold:
\begin{align*}
 \tilde{R_1}+\hat{R_1}& < H(U_1,U),\\
  \tilde{R_2}+\hat{R_2}& < H(U_2,U),\\
  R_0+\hat{R_0}& < H(U),\\
  \tilde{R_1}+\hat{R_1}+\tilde{R_2}+\hat{R_2}& < H(U,U_1,U_2),\\
  \tilde{R_1}+\hat{R_1}+R_0+\hat{R_0}& < H(U,U_1),\\
  \tilde{R_2}+\hat{R_2}+R_0+\hat{R_0}& < H(U,U_2),\\
  \tilde{R_1}+\hat{R_1}+\tilde{R_2}+\hat{R_2}+R_0+\hat{R_0}& < H(U,U_1,U_2).
\end{align*}

Note that the first, second and fourth constraints above are redundant.
 \begin{align}
  R_0+\hat{R_0}& < H(U),\nonumber\\
  \tilde{R_1}+\hat{R_1}+R_0+\hat{R_0}& < H(U,U_1),\nonumber\\
  \tilde{R_2}+\hat{R_2}+R_0+\hat{R_0}& < H(U,U_2),\nonumber\\
  \tilde{R_1}+\hat{R_1}+\tilde{R_2}+\hat{R_2}+R_0+\hat{R_0}& < H(U,U_1,U_2)\label{eqn1:1}.
 \end{align}
Hence, when \eqref{eqn1:1} is satisfied,
\begin{align}\label{eqn1:2}
P(&u^n,u_1^n,u_2^n,m_0,f,m^*,f_1,b_1,f_2,b_2,\hat{u}_1^n,\hat{u}_2^n,\hat{u}^n_{(1)},\hat{u}^n_{(2)})\nonumber\\
&\hspace{-0.5cm}\approx\hat{P}(u^n,u_1^n,u_2^n,m_0,f,m^*,f_1,b_1,f_2,b_2,\hat{u}_1^n,\hat{u}_2^n,\hat{u}^n_{(1)},\hat{u}^n_{(2)}).
\end{align}
Now, for the Slepian-Wolf decoder at processor $P_1$ to succeed applying Lemma 1 of Yassaee et al.~ \cite{YassaeeAG14} (with $T=2, X_1=(U,U_1), X_2=U, Z=\emptyset$), it suffices if the following conditions hold: 
 \begin{align*}
 \tilde{R_1}+\hat{R_1}& \geq H(U,U_1|U)=H(U_1|U),\\
 R_0+\hat{R_0}+R^*& \geq H(U|U,U_1)=0,\\
 \tilde{R_1}+\hat{R_1}+R_0+\hat{R_0}+R^*& \geq H(U,U_1).\\
  \end{align*}
Note that the second constraint above is redundant. And similarly for the other decoder. 
 \begin{align}
 \tilde{R_1}+\hat{R_1}& \geq H(U,U_1|U)=H(U_1|U),\nonumber\\
 \tilde{R_1}+\hat{R_1}+R_0+\hat{R_0}+R^*& \geq H(U,U_1),\nonumber\\
 \tilde{R_2}+\hat{R_2}& \geq H(U,U_2|U)=H(U_2|U),\nonumber\\
 \tilde{R_2}+\hat{R_2}+R_0+\hat{R_0}+R^*& \geq H(U,U_2)\label{eqn1:3}.
 \end{align}
 Hence, when the conditions in \eqref{eqn1:3} are met, 
\begin{align}\label{eqn1:4}
P(&u^n,u_1^n,u_2^n,m_0,f,m^*,f_1,b_1,f_2,b_2,\hat{u}_1^n,\hat{u}_2^n,\hat{u}^n_{(1)},\hat{u}^n_{(2)})\nonumber\\
&\approx P(u^n,u_1^n,u_2^n,m_0,f,m^*,f_1,b_1,f_2,b_2)\nonumber\\
&\hspace{1cm}\times \mathbbm{1}\{\hat{u}^n_{(1)}=u^n=\hat{u}^n_{(2)},\hat{u}_1^n=u_1^n,\hat{u}_2^n=u_2^n\}.
\end{align}
Now, we have 
\begin{align}\hat{P}(u^n,&u_1^n,u_2^n,x^n,y^n,m_0,f,m^*,f_1,b_1,f_2,b_2,\nonumber\\
&\hat{u}_1^n,\hat{u}_2^n,\hat{u}^n_{(1)},\hat{u}^n_{(2)})\nonumber\\
&\hspace{-1cm}=\hat{P}(u^n,u_1^n,u_2^n,m_0,f,m^*,f_1,b_1,f_2,b_2,\hat{u}_1^n,\hat{u}_2^n,\hat{u}^n_{(1)},\hat{u}^n_{(2)})\nonumber\\
&\times  p(x^n|\hat{u}^n_{(1)},\hat{u}_1^n)p(y^n|\hat{u}^n_{(2)},\hat{u}_2^n)\label{eqn:theorem:achievability_slepian_1}\\
&\hspace{-1cm}\approx P(u^n,u_1^n,u_2^n,m_0,f,m^*,f_1,b_1,f_2,b_2)\nonumber\\
&\times\mathbbm{1}{\{\hat{u}^n_{(1)}=u^n=\hat{u}^n_{(2)},\hat{u}_1^n=u_1^n,\hat{u}_2^n=u_2^n\}}\nonumber\\
&\times p(x^n|\hat{u}^n_{(1)},\hat{u}_1^n)p(y^n|\hat{u}^n_{(2)},\hat{u}_2^n)\label{eqn:theorem:achievability_slepian_2}\\
&\hspace{-1cm}=P(u^n,u_1^n,u_2^n,m_0,f,m^*,f_1,b_1,f_2,b_2)\nonumber\\
&\times\mathbbm{1}{\{\hat{u}^n_{(1)}=u^n=\hat{u}^n_{(2)},\hat{u}_1^n=u_1^n,\hat{u}_2^n=u_2^n\}}\nonumber\\
&\times p(x^n|{u}^n,{u}_1^n)p(y^n|{u}^{n},{u}_2^n)\nonumber\\
&\hspace{-1cm}=P(u^n,u_1^n,u_2^n,x^n,y^n,m_0,f,m^*,f_1,b_1,f_2,b_2)\nonumber\\
&\times \mathbbm{1}{\{\hat{u}^n_{(1)}=u^n=\hat{u}^n_{(2)},\hat{u}_1^n=u_1^n,\hat{u}_2^n=u_2^n\}}\nonumber.
\end{align}
where \eqref{eqn:theorem:achievability_slepian_1} follows from \eqref{eqn1:protB}, \eqref{eqn:theorem:achievability_slepian_2} follows from \eqref{eqn1:2} $\&$ \eqref{eqn1:4}.

Thus, we have
\begin{align}\label{eqn1:5}
\hat{P}(u^n,&u_1^n,u_2^n,x^n,y^n,m_0,f,m^*,f_1,b_1,f_2,b_2,\nonumber\\
&\hat{u}_1^n,\hat{u}_2^n,\hat{u}^{n^1},\hat{u}^{n^2})\nonumber\\
&\hspace{-1cm}\approx P(u^n,u_1^n,u_2^n,x^n,y^n,m_0,f,m^*,f_1,b_1,f_2,b_2,)\nonumber\\
&\times\mathbbm{1}{\{\hat{u}^{n}_{(1)}=u^n=\hat{u}^{n}_{(2)} ,\hat{u}_1^n=u_1^n,\hat{u}_2^n=u_2^n\}}.
\end{align}
Marginalizing $(u^n,u_1^n,u_2^n,m_0,m^*,b_1,b_2,\hat{u}^n_{(1)},\hat{u}^n_{(2)},\hat{u}_1^n,\hat{u}_2^n)$ from \eqref{eqn1:5}, we have
\begin{equation}\label{eqn1:6}
\hat{P}(x^n,y^n,f_1,f_2,f)\approx P(x^n,y^n,f_1,f_2,f).
\end{equation}
We need $(X^n,Y^n)$ to be independent of the extra shared randomness $(F,F_1,F_2)$ to eliminate them without actually disturbing the desired i.i.d. distribution. For this, we again use \cite[Theorem~1]{YassaeeAG14} (with $T=3, X_1=(U,U_1), X_2=(U,U_2), X_3=U, Z=(X,Y)$) which states that,
\begin{equation}\label{eqn1:7}
P(x^n,y^n,f_1,f_2,f)\approx p^{\text{Unif}}(f_1)p^{\text{Unif}}(f_2)p^{\text{Unif}}(f)p(x^n,y^n),
\end{equation}
if the following conditions hold:
 \begin{align*}
 \hat{R_1}& < H(U,U_1|X,Y),\\
 \hat{R_2}& < H(U,U_2|X,Y),\\
 \hat{R_0}& < H(U|X,Y),\\
 \hat{R_1}+\hat{R_2}& < H(U,U_1,U_2|X,Y),\\
 \hat{R_1}+\hat{R_0}& < H(U,U_1|X,Y),\\
 \hat{R_2}+\hat{R_0}& < H(U,U_2|X,Y),\\
 \hat{R_1}+\hat{R_2}+\hat{R_0}& < H(U,U_1,U_2|X,Y).
 \end{align*}
Note that the first, second and fourth constraints above are redundant.
 \begin{align}
 \hat{R_0}& < H(U|X,Y),\nonumber\\
 \hat{R_1}+\hat{R_0}& < H(U,U_1|X,Y),\nonumber\\
 \hat{R_2}+\hat{R_0}& < H(U,U_2|X,Y),\nonumber\\
 \hat{R_1}+\hat{R_2}+\hat{R_0}& < H(U,U_1,U_2|X,Y)\label{eqn1:8}.
 \end{align}
 Now from \eqref{eqn1:6} \& \eqref{eqn1:7}, if the constraints in \eqref{eqn1:1}, \eqref{eqn1:3} and \eqref{eqn1:8} are satisfied,  
\begin{equation}\label{eqn1:9}
\hat{P}(x^n,y^n,f_1,f_2,f)\approx p^{\text{Unif}}(f_1)p^{\text{Unif}}(f_2)p^{\text{Unif}}(f)p(x^n,y^n). 
\end{equation}
Condition \eqref{eqn1:9} implies the existence of a particular realization of the random binning with corresponding p.m.f. $p$ so that we can replace $P$ with $p$ in \eqref{eqn1:protB} and denote the resulting p.m.f. by $\hat{p}$. Then \eqref{eqn1:9} implies
\begin{equation*}
\hat{p}(x^n,y^n,f_1,f_2,f)\approx p^{\text{Unif}}(f_1)p^{\text{Unif}}(f_2)p^{\text{Unif}}(f)p(x^n,y^n) 
\end{equation*}
which, by second part of \cite[Lemma 4]{YassaeeAG14}, implies that there exists instances $f^*, f_1^*, f_2^*$ of $F,F_1,F_2$ such that,
\begin{equation}\label{eqn1:10}
\hat{p}(x^n,y^n|f_1^*,f_2^*,f^*)\approx p(x^n,y^n).
\end{equation}
Note that the rate of common message $R$, and respective rates of shared randomness $R_1,R_2$ are given by,
\begin{align}
R&=\frac{R_0}{2}+R^*,\nonumber\\
R_1&=\tilde{R_1}+\frac{R_0}{2},\nonumber\\
R_2&=\tilde{R_2}+\frac{R_0}{2}\label{eqn1:11}.
\end{align}
We gather all the constraints from \eqref{eqn1:1}, \eqref{eqn1:3}, \eqref{eqn1:8} and \eqref{eqn1:11},
\begin{align}
  R_0+\hat{R_0}& < H(U),\nonumber\\
  \tilde{R_1}+\hat{R_1}+R_0+\hat{R_0}& < H(U,U_1),\nonumber\\
  \tilde{R_2}+\hat{R_2}+R_0+\hat{R_0}& < H(U,U_2),\nonumber\\
  \tilde{R_1}+\hat{R_1}+\tilde{R_2}+\hat{R_2}+R_0+\hat{R_0}& < H(U,U_1,U_2)\label{eqn:theorem:achievability_gather_1},
  \end{align}
  \begin{align}
 \tilde{R_1}+\hat{R_1}& \geq H(U_1|U),\nonumber\\
 \tilde{R_1}+\hat{R_1}+R_0+\hat{R_0}+R^*& \geq H(U,U_1),\nonumber\\
 \tilde{R_2}+\hat{R_2}& \geq H(U_2|U),\nonumber\\
 \tilde{R_2}+\hat{R_2}+R_0+\hat{R_0}+R^*& \geq H(U,U_2)\label{eqn:theorem:achievability_gather_2},
\end{align}
 \begin{align}
 \hat{R_0}& < H(U|X,Y),\nonumber\\
 \hat{R_1}+\hat{R_0}& < H(U,U_1|X,Y),\nonumber\\
 \hat{R_2}+\hat{R_0}& < H(U,U_2|X,Y),\nonumber\\
 \hat{R_1}+\hat{R_2}+\hat{R_0}& < H(U,U_1,U_2|X,Y)\label{eqn:theorem:achievability_gather_3},
 \end{align}
 \begin{align}
R&=\frac{R_0}{2}+R^*,\nonumber\\ 
R_1&=\tilde{R_1}+\frac{R_0}{2},\nonumber\\ 
R_2&=\tilde{R_2}+\frac{R_0}{2}\label{eqn1:full}.
\end{align}
In addition, we need to impose non-negativity constraints on all the rates to eliminate all but $R,R_1,R_2$. But it turns out that the non-negativity constraints on $\hat{R_0},\hat{R_1},\hat{R_2}$ are redundant. To see this, along similar lines as Yassaee et al.~\cite[Remark 4]{YassaeeAG14}, we show that if $\hat{R_0},\hat{R_1},\hat{R_2}$ (not necessarily all positive) along with the other rates satisfy \eqref{eqn:theorem:achievability_gather_1}-\eqref{eqn1:full} for some random variables $U,U_1,U_2$ with $X-(U,U_1)-(U,U_2)-Y$, then there exists random variables $U_{\text{new}},U_{1\text{new}},U_{2\text{new}}$ with $X-(U_{\text{new}},U_{1\text{new}})-(U_{\text{new}},U_{2\text{new}})-Y$ and $\hat{R}_{0\text{new}}\geq 0,\hat{R}_{1\text{new}}\geq 0,\hat{R}_{2\text{new}}\geq 0$, such that $\hat{R}_{0\text{new}},\hat{R}_{1\text{new}},\hat{R}_{2\text{new}}$ along with the same other rates satisfy \eqref{eqn:theorem:achievability_gather_1}-\eqref{eqn1:full} for $U_{\text{new}},U_{1\text{new}},U_{2\text{new}}$ instead of $U,U_1,U_2$. We consider an extreme case, i.e., when $\hat{R_0}<0,\hat{R_1}<0,\hat{R_2}<0$ (other cases can be dealt similarly). Let $W,W_1,W_2$ be random variables such that $H(W)>\lvert\hat{R_0}\rvert, H(W_1)>\lvert\hat{R_1}\rvert$ and $H(W_2)>\lvert\hat{R_2}\rvert$. Further, we assume that $W,W_1,W_2$ are independent of each other and independent of all other random variables. Let $\hat{R}_{0\text{new}}=\hat{R_0}+H(W), \hat{R}_{1\text{new}}=\hat{R_1}+H(W_1)$ and $\hat{R}_{2\text{new}}=\hat{R_2}+H(W_2)$ and $U_{\text{new}}=(U,W), U_{1\text{new}}=(U_1,W_1)$ and $U_{2\text{new}}=(U_2,W_2)$. Now clearly, $\hat{R}_{0\text{new}}\geq 0,\hat{R}_{1\text{new}}\geq 0,\hat{R}_{2\text{new}}\geq 0$ and it can be easily shown that $\hat{R}_{0\text{new}},\hat{R}_{1\text{new}},\hat{R}_{2\text{new}}$ along with other rates satisfy \eqref{eqn:theorem:achievability_gather_1}-\eqref{eqn1:full} for $U_{\text{new}},U_{1\text{new}},U_{2\text{new}}$ using the independence of each of  $W,W_1,W_2$ with all the other random variables and the fact that $\hat{R_0},\hat{R_1},\hat{R_2}$ along with other rates satisfy \eqref{eqn:theorem:achievability_gather_1}-\eqref{eqn1:full} (See Appendix~\ref{appreviewers2}). 

Notice that we can assume that the constraints in \eqref{eqn:theorem:achievability_gather_2} hold with equality, because we can reduce the rates $\hat{R_0},\hat{R_1},\hat{R_2}$ to get equalities in \eqref{eqn:theorem:achievability_gather_2} without disturbing the other constraints. Rate elimination becomes simpler with this observation. This leads to,
\begin{align}
\hat{R}_1&=H(U_1|U)-\tilde{R}_1,\nonumber\\
\hat{R}_0&=H(U)-R_0-R^*,\nonumber\\
\hat{R}_2&=H(U_2|U)-\tilde{R}_2\label{eqn:theorem:achievability_SW_simplified}.
\end{align} 
Substituting \eqref{eqn:theorem:achievability_SW_simplified} in \eqref{eqn:theorem:achievability_gather_1} and \eqref{eqn:theorem:achievability_gather_3} gives the following constraints after ignoring the redundant inequalities.
\begin{align}
R^*&>I(U_1;U_2|U),\nonumber\\
R_0+R^*&>I(X,Y;U),\nonumber\\
R_0+\tilde{R}_1+R^*&>I(X,Y;U,U_1),\nonumber\\
R_0+\tilde{R}_2+R^*&>I(X,Y;U,U_2),\nonumber\\
R_0+\tilde{R}_1+\tilde{R}_2+R^*&>I(U_1;U_2|U)+I(X,Y;U,U_1,U_2)\label{eqn:theorem:achievability_after_elim_extra}.
\end{align} 
Also, from \eqref{eqn1:full} we get
\begin{align}
R^*&=R-\frac{R_0}{2},\nonumber\\
\tilde{R}_1&=R_1-\frac{R_0}{2},\nonumber\\
\tilde{R}_2&=R_2-\frac{R_0}{2}.\label{eqn:theorem:achievability_RR_1R_2}
\end{align}
Non-negativity constraints on $R^*,\tilde{R}_1,\tilde{R}_2$ imply from \eqref{eqn:theorem:achievability_RR_1R_2} that
\begin{align}
R&\geq \frac{R_0}{2},\nonumber\\
R_1&\geq \frac{R_0}{2},\nonumber\\
R_1&\geq \frac{R_0}{2}.\label{eqn:theorem:achievability_nonneg}
\end{align}
Substituting \eqref{eqn:theorem:achievability_RR_1R_2} in \eqref{eqn:theorem:achievability_after_elim_extra} gives the following constraints on $R,R_1,R_2$ and $R_0$.
\begin{align}
R-0.5R_0&>H(U)+H(U_1|U)+H(U_2|U)\nonumber\\
&\hspace{01cm}-H(U,U_1,U_2),\nonumber\\    
R+0.5R_0&>I(X,Y;U),\nonumber\\
R+R_1&>I(X,Y;U,U_1),\nonumber\\
R+R_2&>I(X,Y;U,U_2),\nonumber\\
R+R_1+R_2-0.5R_0&> H(U)+H(U_1|U)+H(U_2|U)\nonumber\\
&\hspace{1cm}-H(U,U_1,U_2|X,Y).\label{eqn:theorem:achievability_before_elim}
\end{align}
Now, notice that $R_0$ is the only variable which needs to be eliminated from \eqref{eqn:theorem:achievability_nonneg}, \eqref{eqn:theorem:achievability_before_elim} along with a non-negativity constraint, $R_0\geq0$. We use Fourier-Motzkin elimination (FME) to eliminate $R_0$ to get the following:
\begin{align}
R+R_1&>I(X,Y;U,U_1),\nonumber\\
R+R_2&>I(X,Y;U,U_2),\nonumber\\
R&>I(U_1;U_2|U),\nonumber\\
R+R_1+R_2&>I(U_1;U_2|U)+I(X,Y;U,U_1,U_2),\nonumber\\
2R+R_1+R_2&>I(U_1;U_2|U)+I(X,Y;U)\nonumber\\
&\hspace{1cm}+I(X,Y;U,U_1,U_2),\nonumber\\
2R&>I(U_1;U_2|U)+I(X,Y;U)\label{eqn1:final}.
\end{align}  
Thus, when the conditions in \eqref{eqn1:final} are met, there exists a sequence of $(n,2^{nR},2^{nR_1},2^{nR_2})$ simulation codes with coordinator and processors as described in the random coding scheme above with the particular realization of random binning along with fixed instances of $f^*,f_1^*,f_2^*$ resulting in desired vanishing total variation distance. 
\end{IEEEproof}

\begin{IEEEproof}[{Proof of Theorem~\ref{theorem:unlimited_shared_randomness}}]
For achievability, when rates $R_1, R_2$ are large enough, Theorem~\ref{theorem:achievability} implies that a rate of 
$\max \{I(U_1;U_2|U)$, $\frac{1}{2}\big(I(U_1;U_2|U)+I(U;X,Y)\big)\}$
is achievable when $X-(U,U_1)-(U,U_2)-Y$. It is easy to see that $U_1=X, U_2=Y$ satisfy the Markov chain $X-(U,U_1)-(U,U_2)-Y$ for any $U$.
So, for any $p(u|x,y)$, if $R=\max\left\{I(X;Y|U),\frac{1}{2}\big(I(X,Y;U)+I(X;Y|U)\big)\right\}$, then there exists $R_1$ and $R_2$ so that $(R,R_1,R_2)\in\mathcal{R}$. Hence, $R_{\mrm{opt}} \leq \min \max \left\{I(X;Y|U), \frac{1}{2}\big(I(X,Y;U)+I(X;Y|U)\big)\right\}=:R_U$, where the minimum is over all conditional p.m.f.'s $p(u|x,y)$. 

For the converse, suppose a rate triplet $(R,R_1,R_2)$ is achievable for $q(x,y)$. Fix an $\epsilon\in(0,\frac{1}{4})$. Then there exists an $(n,2^{nR},2^{nR_1},2^{nR_2})$ simulation code such that 
\begin{equation}
\lVert p_{X^nY^n}-q_{XY}^{(n)}\rVert_1 < \epsilon\label{eqn:correctness1}
\end{equation}
for large enough $n$.
First, we show that there exists a p.m.f. $\gamma_{X,Y,U}$ with $|\mathcal{U}|\leq|\mathcal{X}||\mathcal{Y}|+2$ such that $\lVert \gamma_{X,Y}-q_{X,Y}\rVert_1<\epsilon$ and
\begin{align}
R&\geq I(X;Y|U),\label{eqn1:newthing}\\
R&\geq I(X,Y;U)-g(\epsilon)\label{eqn1:wyner},
\end{align} 
where $\lim_{\epsilon\downarrow 0} g(\epsilon)=0$. We will show \eqref{eqn1:wyner} along the lines of Wyner~\cite{Wyner75}. To obtain \eqref{eqn1:newthing}, we will first show that $nR\geq I(X^n;Y^n|M)$. In Wyner's model~\cite{Wyner75}, the term $I(X^n;Y^n|M)$ is precisely zero. This is not the case here, in general, because of the presence of shared random variables $W_1$ and $W_2$. We will further lower bound the term $I(X^n;Y^n|M)$ by a single-letter form to obtain \eqref{eqn1:newthing}:
\begin{align}
I(X^n;Y^n|M)&\leq I(X^n,W_1;Y^n,W_2|M)\nonumber\\
&=I(W_1;Y^n,W_2|M)\label{eqn1:new_1}\\
&=I(W_1;W_2|M)\label{eqn1:new_2}\\
&\leq I(W_1;M)+I(W_1;W_2|M)\nonumber\\
&\hspace{12pt}-I(W_1;W_2)\label{eqn1:new_3}\\
%&=I(W_1;M,W_2)-I(W_1;W_2)\nonumber\\
&=I(W_1;M|W_2)\nonumber\\
&\leq H(M|W_2)\nonumber\\
&\leq H(M)\nonumber\\
&\leq nR\label{eqn1:new},
\end{align}
where \eqref{eqn1:new_1} and \eqref{eqn1:new_2} follow from the Markov chain $X^n-(M,W_1)-(M,W_2)-Y^n$, \eqref{eqn1:new_3} follows because $W_1$ is independent of $W_2$ and $I(W_1;M)\geq0$. 

Let $T$ be a random variable uniformly distributed over $[1:n]$ and independent of all other variables. Then, by continuing \eqref{eqn1:new}, we have
\iftoggle{paper}
{\begin{align}
nR&\geq I(X^n;Y^n|M)\nonumber\\
&=\sum_{i=1}^n I(X_i;Y^n|M,X^{i-1})\nonumber\\
%&=\sum_{i=1}^n I(X_i;Y^{i-1}|M,X^{i-1})\nonumber\\
%&\hspace{2cm}+\sum_{i=1}^n I(X_i;Y_i^n|M,X^{i-1},Y^{i-1})\nonumber\\
%&\geq \sum_{i=1}^n I(X_i;Y_i^n|M,X^{i-1},Y^{i-1})\nonumber\\
&\geq \sum_{i=1}^n I(X_i;Y_i|M,X^{i-1},Y^{i-1})\nonumber\\
&=\sum_{i=1}^n I(X_i;Y_i|U_i)\label{eqn:converse_newthing_1}\\
&=nI(X_T;Y_T|U_T,T),\label{eqn:converse_newthing}
\end{align}}
{\begin{align}
nR&\geq I(X^n;Y^n|M)\nonumber\\
&=\sum_{i=1}^n I(X_i;Y^n|M,X^{i-1})\nonumber\\
&=\sum_{i=1}^n I(X_i;Y^{i-1}|M,X^{i-1})\nonumber\\
&\hspace{2cm}+\sum_{i=1}^n I(X_i;Y_i^n|M,X^{i-1},Y^{i-1})\nonumber\\
&\geq \sum_{i=1}^n I(X_i;Y_i^n|M,X^{i-1},Y^{i-1})\nonumber\\
&= \sum_{i=1}^n I(X_i;Y_i|M,X^{i-1},Y^{i-1}) \nonumber\\
&\hspace{2cm}+\sum_{i=1}^n I(X_i;Y_{i+1}^n |M,X^{i-1},Y^i)\nonumber\\
&\geq \sum_{i=1}^n I(X_i;Y_i|M,X^{i-1},Y^{i-1})\nonumber\\
&=\sum_{i=1}^n I(X_i;Y_i|U_i)\label{eqn:converse_newthing_1}\\
&=nI(X_T;Y_T|U_T,T),\label{eqn:converse_newthing}
\end{align}}
where \eqref{eqn:converse_newthing_1} follows by defining $U_i=(M,X^{i-1},Y^{i-1})$. Following Wyner~\cite{Wyner75}, we lower bound $R$ in another fashion,
\begin{align}
nR &\geq H(M)\nonumber\\
&\geq I(X^n,Y^n;M)\nonumber\\
&=H(X^n,Y^n)-H(X^n,Y^n|M)\nonumber\\
&\geq\sum_{i=1}^n [H(X_i,Y_i)-\epsilon^\prime]-\sum_{i=1}^n H(X_i,Y_i|M,X^{i-1},Y^{i-1})\label{eqn:converse_closeness_of_dist}\\
&\geq\sum_{i=1}^n[I(X_i,Y_i;M,X^{i-1},Y^{i-1})-\epsilon^\prime]\nonumber\\
&=\sum_{i=1}^n[I(X_i,Y_i;U_i)-\epsilon^\prime]\nonumber\\
&= n[I(X_T,Y_T;U_T|T)-\epsilon^\prime]\nonumber\\
&= n[I(X_T,Y_T;U_T,T)-I(X_T,Y_T;T)-\epsilon^\prime]\nonumber\\
&\geq n[I(X_T,Y_T;U_T,T)-\epsilon^\prime-\delta]\label{eqn:T_independent}\\
&=nI(X_T,Y_T;U_T,T)-ng(\epsilon).\label{eqn:converse_wyner}
\end{align}
In \eqref{eqn:converse_closeness_of_dist} and \eqref{eqn:T_independent}, $\epsilon^\prime,\delta\rightarrow 0$ as $\epsilon\rightarrow 0$. We show these steps using \eqref{eqn:correctness1} (details are in \iftoggle{paper}{the extended version \cite{arxiv})}{Appendix \ref{details_omitted})}. In \eqref{eqn:converse_wyner}, $g(\epsilon):=\epsilon^\prime+\delta$,  so $g(\epsilon)\rightarrow 0$ as $\epsilon \rightarrow 0$. Now, we claim that we can find a $\gamma_{X,Y,U}$ such that
\begin{align}
\gamma_{XY}&=p_{X_TY_T},\label{convex_cover_1}\\
I_{\gamma}(X;Y|U)&=I_p(X_T;Y_T|U_T,T),\label{convex_cover_2}\\
I_{\gamma}(X,Y;U)&=I_p(X_T,Y_T;U_T,T),\label{convex_cover_3}\\
|\mathcal{U}|&\leq|\mathcal{X}||\mathcal{Y}|+2\label{convex_cover_4}.
\end{align}
This directly follows from an application of Convex Cover Method~\cite[Appendix C]{GamalK12} along the same lines as Cuff~\cite[Lemma VI.1]{Cuff13}. 
Note that
\begin{align}
\lVert \gamma_{XY}-q_{XY}\rVert_1&=\lVert p_{X_TY_T}-q_{XY}\rVert_1\nonumber\\
&\leq\lVert p_{X^nY^n}-q_{XY}^{(n)}\rVert_1\label{eqn:cuff}\\
&<\epsilon,\nonumber
\end{align}
where \eqref{eqn:cuff} follows from~\cite[Lemma VI.2]{Cuff13}. Let $\mathcal{S}_\epsilon$, for $\epsilon\geq 0$ be defined as the set of all non-negative rates $R$ such that
\begin{align}
R&\geq I(X;Y|U),\\
R&\geq I(X,Y;U)-g^\prime(\epsilon),
\end{align}
for some p.m.f. $p(x,y,u)$ satisfying \eqref{convex_cover_4} and $\lVert p_{XY}-q_{XY}\rVert_1\leq\epsilon$ with $g^\prime(\epsilon)=g(\epsilon)$, for $\epsilon>0$ and $g^\prime(0)=0$.
Thus,  for every $\epsilon>0$, it follows from \eqref{eqn:converse_newthing}, \eqref{eqn:converse_wyner} and \eqref{convex_cover_1}-\eqref{eqn:cuff} that, $R\in \mathcal{S}_\epsilon$. Using the continuity of total variation distance and mutual information in the probability simplex, we can show that that $\underset{\epsilon>0}{\bigcap}\mathcal{S}_\epsilon=~\mathcal{S}_0$ along the same lines as Yassaee et al.~\cite[Lemma 6]{YassaeeGA15}. Hence $R_{\mrm{opt}}\geq \min \max \big\{I(X;Y|U),I(X,Y;U)\big\}=:R_L$, where the minimum is over all conditional p.m.f.'s $p(u|x,y)$ with $|\mathcal{U}|\leq |\mathcal{X}||\mathcal{Y}|+2$. So, achievability and converse give us $R_L\leq R_{\mrm{opt}}\leq R_U$. And since it is trivial to see that $R_L\geq R_U$, we have $R_{\mrm{opt}}=R_L=R_U$ ($R_L=R_U$ can be proved directly also as shown in Lemma~\ref{lemma:RL=RU} in Appendix~\ref{details_omitted}).
\end{IEEEproof}

\begin{IEEEproof}[Proof of Theorem~\ref{theorem:bounds}]
Consider the second expression for $R_{\mrm{opt}}$ in Theorem~\ref{theorem:unlimited_shared_randomness}. To see the lower bound, notice that
 \begin{align*}I(X,Y;U)+&I(X;Y|U)\\
 &=I(X;Y)+I(X;U|Y)+I(Y;U|X)\\
 &\geq I(X;Y).
 \end{align*} 
 
 For the upper bound, choosing $U$ to be a minimizer in \eqref{eqn:wyner_discussion} gives us $R_{\mrm{opt}}\leq 0.5C(X;Y)$. Choosing $U=\emptyset$ gives us $R_{\mrm{opt}} \leq I(X;Y)$. 
 
If $X$ and $Y$ are independent, it is easy to see that $R_{\mrm{opt}}=0$ by choosing $U=\emptyset$. Recall from Remark~\ref{gastpar} that $R_{\mrm{opt}}=C_{\gamma^*}$, where $\gamma^*$ is such that $C_{\gamma^*}=\gamma^*$, where $C_\gamma$ is relaxed Wyner's common information. For the other direction, suppose that $R_{\mrm{opt}}=I(X;Y)$. Then, $C_{\gamma^*}=\gamma^*=I(X;Y)$. However, from the definition of $C_\gamma$, if $\gamma=I(X;Y)$, it is easy to see that $C_{\gamma}=0$ by choosing $U=\emptyset$. Therefore, $C_{\gamma^*}=0=I(X;Y)$, which implies that $X$ is independent of $Y$. 

Notice that \begin{align*}I(X;Y|U)=\frac{I(X;Y)+I(U;X|Y)+I(U;Y|X)}{2}\\
+\frac{I(X;Y)-I(X;U)-I(Y;U)}{2},
\end{align*}
\begin{align*}
I(X,Y;U)+I(X;Y|U)\\
=I(X;Y)+I(U;X|Y)+I(U;Y|X).
\end{align*}
If there exists a $P_{U|XY}$ such that $U-X-Y$ and $U-Y-X$ are Markov chains and $I(X;Y)\leq I(X;U)+I(Y;U)$, then $R_{\mrm{opt}}=\frac{I(X;Y)}{2}$. For the other direction, suppose that for every $P_{U|XY}$, we have $I(U;Y|X)>0$ or $I(U;X|Y)>0$ or $I(X;Y)>I(X;U)+I(Y;U)$. In that case, it is easy to see that $R_{\mrm{opt}}>\frac{I(X;Y)}{2}$. 

For $X=(X^\prime,V)$ and $Y=(Y^\prime,V)$, where $X$ and $Y$ are conditionally independent given $V$, we have $C(X;Y)=I(X;Y)=H(V)$. Now since $\frac{I(X;Y)}{2}\leq R_{\mrm{opt}}\leq \frac{C(X;Y)}{2}$, we have $R_{\mrm{opt}}=\frac{C(X;Y)}{2}$. 
\end{IEEEproof}

\begin{IEEEproof}[{Proof of Theorem~\ref{theorem:noshared}}]
For the achievability, it is easy to see from Theorem \ref{theorem:achievability} that $\left(C(X;Y),0,0\right)\in\mathcal{R}$ by identifying that for any $U$ satisfying $X-U-Y$, we have $\left(I(X,Y;U),0,0\right)\in\mathcal{R}$ with the corresponding other auxiliary random variables defined by $U_1=\emptyset$, $U_2=\emptyset$. Hence, $\RNO\leq C(X;Y)$.

 For the converse, suppose $R$ is such that $(R,\epsilon,\epsilon)$ is achievable for every $\epsilon>0$. This implies that for a fixed $\epsilon>0$, there exists an $(n,2^{nR},2^{n\epsilon},2^{n\epsilon})$ simulation code such that  
\begin{align}
\lVert p_{X^nY^n}-q_{XY}^{(n)} \rVert_1 < \epsilon\label{eqn:arxiv_noshared_1},
\end{align}
for large enough $n$. $R$ can be bounded using \eqref{eqn:arxiv_noshared_1} along the similar lines as \eqref{eqn:converse_wyner}, which gives us 
\begin{align}
R\geq I(X_T,Y_T;U_T,T)-g(\epsilon),\label{eqn:arxiv_noshared_4}
\end{align}
where $\lim_{\epsilon\downarrow 0}g(\epsilon)=0$, $U_T=(M,X^{T-1},Y^{T-1})$ and $T$ is a random variable uniformly distributed over $[1:n]$ and independent of everything else.

 Next, we lower bound $n\epsilon$ in the following fashion.
 \begin{align}
 n\epsilon&\geq H(W_1)\nonumber\\
 &\geq I(W_1;W_2|M)\nonumber\\
 &=I(X^n,W_1;Y^n,W_2|M)\label{eqn:arxiv_noshared_2}\\
 &\geq I(X^n;Y^n|M)\nonumber\\
 &\geq nI(X_T;Y_T|U_T,T),\label{eqn:arxiv_noshared_3}
 \end{align}
 where \eqref{eqn:arxiv_noshared_2} follows from $X^n-(M,W_1)-(M,W_2)-Y^n$, and \eqref{eqn:arxiv_noshared_3} follows along similar lines as \eqref{eqn:converse_newthing}.
 
 Now from \eqref{eqn:arxiv_noshared_4} and \eqref{eqn:arxiv_noshared_3} and using arguments similar to \eqref{convex_cover_1}-\eqref{eqn:cuff} one can show that $R\in\mathcal{M}_\epsilon$, where $\mathcal{M}_\epsilon$ is defined to be the set of all rates $R$ such that
 \begin{align*}
 R&\geq I(X;Y|U)-g(\epsilon),\\
 \epsilon&\geq I(X;Y|U),
 \end{align*}
 where $\lim_{\epsilon\downarrow 0}g(\epsilon)=0$ for some p.m.f. $p(x,y,u)$ satisfying $|\mathcal{U}|\leq|\mathcal{X}||\mathcal{Y}|+2$ and $\lVert p_{XY}-q_{XY}\rVert_1\leq\epsilon$.
 
 Using the continuity of total variation distance and mutual information in the probability simplex, we can show that $\underset{\epsilon>0}{\bigcap}\mathcal{M}_\epsilon=\mathcal{M}$ along the same lines as~Yassaee et al.~\cite[Lemma~6]{YassaeeGA15}, where $\mathcal{M}$ is defined to be the set of all rates $R$ such that
\begin{align}
R&\geq I(X,Y;U),
\end{align}
for some conditional p.m.f. $p(u|x,y)$ satisfying $X-U-Y$ and $|\mathcal{U}|\leq|\mathcal{X}||\mathcal{Y}|+2$. Hence, $\RNO\geq C(X;Y)$.
\end{IEEEproof}

The proof of Theorem~\ref{theorem:X=Y=Z} is subsumed by the proof of Theorem~\ref{theorem_greedy} proved in Section~\ref{proofs_multi}.

% !TEX root = Generalization.tex

\section{Omniscient Coordinator Setting: Multiple processors}\label{section:multi_proc}

We now turn to the omniscient coordinator setting with $t > 2$ processors. The proof techniques for the multiple-processor setting use similar methods as for two processors. In this setting, however, the individually shared randomness model and the randomness-on-the-forehead model are not identical. We show a number of results for these two models as well as an achievable scheme for the general model where $\mc{V}_i$'s are arbitrary subsets of $[1:h]$.

\subsection{Results}

We provide results for the individually shared randomness model and randomness-on-the-forehead model. Results for the former and the latter models involve Watanabe's total correlation~\cite{Watanabe60} and Han's dual total correlation~\cite{Han75} measures, respectively.

\subsubsection{Individually Shared Randomness Model}

When the shared randomness rates are sufficiently large, we characterize the optimal rate of communication. Let $I(X_1;\dots;X_t|U)$ denote Watanabe's total correlation~\cite{Watanabe60}.
\begin{align}
I(X_1;\dots;X_t|U):=\left(\sum_{i=1}^tH(X_i|U)\right)-H(X_1,\dots,X_t|U).
\label{eq:watanabe}
\end{align}

\begin{thm}\label{theorem:generalize}
The optimal transmission rate for the individually shared randomness model is given by
	\begin{align*}
	&R_{\mrm{opt}}^{\mrm{Indv}}=\min \max \big\{I(X_1;\dots;X_t|U), I(X_1,\dots,X_t;U)\big\},
	%&=\min \max \bigg\{I(X_1;\dots;X_t|U), \nonumber\\
	%&\hspace{1.2cm}\frac{1}{t}\big((t-1)I(X_1,\dots,X_t;U)+I(X_1;\dots;X_t|U)\big)\bigg\},
	\end{align*}
where the minimum is over all probability mass functions
	\begin{align*}
	p(x_{[1:t]},u)=q(x_{[1:t]})p(u|x_{[1:t]})
	\end{align*}
with 
	\begin{align*}
	|\mathcal{U}|\leq\left(\prod_{i=1}^t |\mathcal{X}_i|\right)+t.
	\end{align*}
\end{thm}

For the case when all $X_i$ are equal, we can completely characterize the simulation rate region $\mathcal{R}^{\mrm{Indv}}$. 

\begin{thm}\label{theorem_greedy}
Suppose $q_{X_1\dots X_t}$ is such that $X_1=\dots=X_t = X$. Then the simulation rate region $\mathcal{R}^{\mrm{Indv}}$ for the individually shared randomness model is the set of all non-negative rate tuples $(R,R_1,\dots,R_h)$ such that
\begin{align}
R+\min\{R_1,\dots,R_t\}&\geq H(X)\label{eqn:indv:remarkequal_1},\\
R&\geq \frac{(t-1)H(X)}{t}\label{eqn:indv:remarkequal}.
\end{align}
\end{thm}

\subsubsection{Randomness-on-the-Forehead Model}

We give an upper bound on $R^{\mrm{Forehead}}_{\mrm{opt}}$. Let $\tilde{I}(X_1;\dots;X_t|U)$ denote Han's dual total correlation~\cite{Han75}:
\begin{align}
\tilde{I}(X_1;\dots;X_t|U):=H(X_{[1:t]}|U)-\sum_{i=1}^tH(X_i|U,X_{[1:t]\setminus\{i\}}).
\label{eq:han}
\end{align}

\begin{thm}\label{theorem:t=3}
The optimal transmission rate for the randomness-on-the-forehead model is upper bounded as follows:
	\begin{align}
	R^{\mrm{Forehead}}_{\mrm{opt}}\leq \min\max_{i\in[1:t]}\left\{\frac{r_i}{i}\right\}\label{eqn:forehead:bound},
	\end{align}
where
	\begin{align*}
	r_i=\max_{\{l_1,\dots,l_{i+1}\}\subseteq [1:t]}\tilde{I}(U_{l_1};\dots;U_{l_{i+1}}|U,U_{[1:t]\setminus \{l_1,\dots l_{i+1}\}}),
	\end{align*}
	 for $i\in[1:t-1]$, and $r_t=\tilde{I}(U_1;\dots;U_t|U)+I(X_{[1:t]};U)$. 
The minimum in \eqref{eqn:forehead:bound} is taken over all probability mass functions of the form 	
	\begin{align*}
	p(x_{[1:t]},u,u_{[1:t]})\linebreak=q(x_{[1:t]})p(u,u_{[1:t]}|x_{[1:t]})
	\end{align*}
such that
	\begin{align}
	p(u,u_{[1:t]},x_{[1:t]})=p(u,u_{[1:t]})\prod_{m=1}^t p(x_m|u,u_{[1:t]\setminus\{m\}})\label{eqn:t=3markov}.
	\end{align}
\end{thm}

\textbf{Special cases:} We identify several special cases of the above result to help illustrate the structure of the problem.
\begin{enumerate}[$(a)$]
\item When $X_1$ is independent of $(X_2,\dots,X_t)$, clearly a rate of zero is achievable because processor $P_1$ samples i.i.d. $X_1$ using $W_t$ and other processors sample i.i.d. $(X_2,\dots,X_t)$ using $W_1$. We recover this by taking $U=U_2=\dots=U_{t-1}=\emptyset,U_1=(X_2,\dots,X_t),U_t=X_1$ in Theorem~\ref{theorem:t=3}. So, $R^{\text{Forehead}}_{\text{opt}}=0$. 
\item When $q_{X_1,\dots,X_t}$ is such that $X_1=\dots=X_t=X$, a rate of $\frac{H(X)}{t}$ is achievable by taking $U_1=\dots=U_t=\emptyset$ and $U=X$ in Theorem~\ref{theorem:t=3}. The converse follows from the converse of Theorem~\ref{theorem:foreheadequal} (in particular, by substituting $i=t$ in \eqref{eqn:foreheadequal1}). So, $R^{\text{Forehead}}_{\text{opt}}=\frac{H(X)}{t}$. 
\end{enumerate}
For the case when all $X_i$ are equal, we can completely characterize the simulation rate region $\mathcal{R}^{\text{Forehead}}$.

\begin{thm}\label{theorem:foreheadequal}
Suppose $q_{X_1\dots X_t}$ is such that $X_1=\dots=X_t=X$. Then the simulation rate region $\mathcal{R}^{\emph{Forehead}}$ for the randomness-on-the-forehead model is given by the set of all non-negative rate tuples $(R,R_1,\dots,R_t)$ such that
\begin{align}
%(t-1)R+\min\{R_1,\dots,R_t\}&\geq H(X)\,\label{eqn:foreheadequal1}\\
%R+\sum_{i=1,i\neq j}^t R_i&\geq H(X), \ \emph{for} \ j\in[1:t],\label{eqn:foreheadequal2}\\
%R&\geq \frac{H(X)}{t}\label{eqn:foreheadequal3}.
&\hspace{0.4cm}iR+\sum_{j\in \mathcal{S}}R_j\geq H(X), \label{eqn:foreheadequal1}\\ 
&\hspace{-2.6cm}\text{for}\ i\in[1:t], \mathcal{S}\subsetneq[1:t] \ \text{s.t.}\ |\mathcal{S}|=t-i\nonumber.
\end{align}
\end{thm}

\subsubsection{The general case}\label{section:equal}

For the general model (i.e., when $\mathcal{V}_i$'s are arbitrary subsets of $[1:h]$),  we can completely characterize the simulation rate region $\mathcal{R}(\mathcal{V})$ when all $X_i$ are equal. 

\begin{thm}\label{theorem:equal}
Suppose $q_{X_1\dots X_t}$ is such that $X_1=\dots=X_t=X$. The simulation rate region $\mathcal{R}(\mathcal{V})$ for the omniscient coordinator setting is the set of all non-negative rate tuples $(R,R_1,\dots,R_h)$ s.t. there exists non-negative $r,r_1,\dots,r_h$ satisfying
\begin{align}
R&\geq \left(\sum_{j:j\notin\mathcal{V}_i}r_j\right)+r,\ i\in[1:t],\label{eqn:equal1}\\
H(X)&\leq r+\sum_{j=1}^h r_j,\label{eqn:equal2}\\
R_j&\geq r_j,\ j\in[1:h]\label{eqn:equal3}.
\end{align}
\end{thm}

\begin{remark}
Even though Theorems~\ref{theorem_greedy} \& \ref{theorem:foreheadequal} can be recovered from Theorem~\ref{theorem:equal} by eliminating $r$'s (see Section~\ref{section:proofs}), they are of independent interest because their proofs give an optimal choice of $r$'s for explicitly constructing an achievable scheme and the rate regions have nice closed form expressions in terms of rates $R,R_1,\dots,R_t$ (as in \eqref{eqn:indv:remarkequal_1} and \eqref{eqn:indv:remarkequal}, and \eqref{eqn:foreheadequal1}).
\end{remark}

%
% PROOFS
%
% !TEX root = Generalization.tex

\subsection{Proofs}\label{proofs_multi}
\subsubsection{Individually Shared Randomness Model}

\begin{IEEEproof}[Proof of Theorem~\ref{theorem:generalize}]
\if \arxive 0
The achievability proof is based on a generalization of the idea behind \cite[Theorem~1]{KurriPS18}. We present a rough outline here. Individually shared randomness is used by the processors in two different ways: some part as common randomness and the remaining part as private randomness. The common randomness is established via simple network coding in the following way. For all the $t$ processors to recover common randomness $m_0=(m_{01},\dots,m_{0t})$, the coordinator sends $(m_{01}\oplus m_{02},\dots, m_{01}\oplus m_{0t})$ as a part of the common message, i.e., $m_{0i}$ is the part of individually shared randomness that is converted to common randomness. The role of private randomness is along the same lines as \cite[Theorem~1]{KurriPS18}. Rate analysis and total variation analysis, employing the proof technique of output statistics of random binning (OSRB)~\cite{YassaeeAG14} along the same lines as \cite[Theorem~1]{KurriPS18}, gives us achievability. For more details, see an extended version~\cite{Full}.
\fi
\if \arxive 1
The achievability proof is based on generalization of the idea behind the proof of Theorem~\ref{theorem:achievability}. The intuition is as follows. Fix a conditional p.m.f. $p{(u|x_1,\dots,x_t)}$, generate a binned codebook $\{u^n(j,k)\}_{j,k}\sim p_u\ \text{i.i.d.}$, where $j\in[1:2^{nR_0}]$ denotes the number of bins and $k\in[1:2^{nR^*}]$ specifies a particular $u^n$ sequence inside a bin. For each $u^n(j,k)$, generate an $X_i-$conditional codebook~$\sim p(x_i | u)$ i.i.d., where each codeword is represented as $x_i^n(i,j,b_i)$ (here $b_i$ is assumed to be of sufficiently large rate), for $i\in[1:t]$. We treat shared randomness as bit strings. Indices $m_0, b_i,i\in[1:t]$ are determined by the shared randomness in the following way: Index $m_0$ which is uniformly distributed on $[1:2^{nR_0}]$ is a concatenation of `$t$' $\frac{nR_0}{t}-$length bit strings $m_{0i},i\in[1:t]$, where $m_{0i}$ is obtained from shared randomness $w_i$. Index $b_i$ which is independent of $m_0$ is also obtained from shared randomness $w_i$, for $i\in[1:t]$. Note that $m_0,b_i,i\in[1:t]$ are mutually independent of each other. The coordinator finds an $m^*$ inside the bin indexed by $m_0$ such that $(u^n(m_0,m^*),x_1^n(m_0,m^*,b_1),\dots,x_t^n(m_0,m^*,b_t))$ is consistent with high probability. Loosely, $R^*>I(X_1;\dots;X_t|U)$ ensures that there exists such an $m^*$. The coordinator then sends $(m_{01}\oplus m_{02},\dots, m_{01}\oplus m_{0t},m^*)$ as a common message to the processors at a rate $R=\frac{t-1}{t}R_0+R^*$. Note that processor $P_i$ has access to $m_{0i}$ and recovers $m_0$.Then, the processors $P_i,i\in[1:t]$ output $x_i^n(m_0,m^*,b_i),i\in[1:t]$, respectively. Roughly, $R_0+R^*>I(X_1,\dots,X_t;U)$ ensures that the output is according to the desired distribution. Since $R=\frac{t-1}{t}R_0+R^*$, the above rate constraints imply that $\max \Big\{I(X_1;\dots;X_t|U), \frac{1}{t}\big((t-1)I(X_1,\dots,X_t;U)+I(X_1;\dots;X_t|U)\big)\Big\}$ is achievable. A formal proof can be written down along similar lines as that of Theorem~1 employing the proof technique of OSRB framework~\cite{YassaeeAG14} (the proof is outlined in Appendix~\ref{appendix:wynerstyle}). 
\fi

The converse argument is broadly along the lines of the converse in Theorem~\ref{theorem:unlimited_shared_randomness}. The key step is to show that $nR\geq I(X_1^n;\dots;X_t^n|M)$, where $I(X_1^n;\dots;X_t^n|M)$ is the Watanabe total correlation in \eqref{eq:watanabe}. Notice that the notion of multivariate mutual information in the R.H.S. of this inequality can be viewed as a generalization of a corresponding mutual information term in the converse of Theorem~\ref{theorem:unlimited_shared_randomness}. Following the chain of inequalities: 
\begin{align}
I(X_1&^n;\dots;X_t^n|M)\nonumber\\
&=\left(\sum_{i=1}^t H(X_i^n|M)\right)-H(X_1^n,\dots,X_t^n|M)\nonumber\\
&=\sum_{i=1}^t\left[H(X_i^n|M)-H(X_i^n|M,X_1^n,\dots,X_{i-1}^n) \right]\nonumber\\
&=\sum_{i=2}^{t} I(X_1^n,\dots,X_{i-1}^n;X_{i}^n|M)\nonumber\\
&\leq \sum_{i=2}^{t} I(W_1,\dots,W_{i-1};W_{i}|M)\label{eqn:dataprocessin_mutual}\\
&=\left(\sum_{i=1}^t H(W_i|M)\right)-H(W_1,\dots,W_t|M)\nonumber\\
&=\sum_{i=1}^t \big[H(W_i|M)-H(W_i)\big]-H(W_1,\dots,W_t|M)\nonumber\\
&\hspace{2cm}+H(W_1,\dots,W_t)\label{eqn:genindependence}\\
&=I(M;W_1,\dots,W_t)-\sum_{i=1}^n I(M;W_i)\nonumber\\
&\leq H(M)\nonumber\\
&\leq nR \label{eqn:generalcontinuition},
\end{align}
where \eqref{eqn:dataprocessin_mutual} follows from the Markov chains $X_{i}^n-(M,W_{i})-(M,W_1,\dots,W_{i-1})-(X_1^n,\dots,X_{i-1}^n)$, for $i\in [2:t]$, \eqref{eqn:genindependence} follows because $W_1,\dots,W_t$ are mutually independent random variables.

Let $Q$ be a random variable uniformly distributed over $[1:n]$ and independent of all other random variables. 
\if \arxive 1 
Then, by continuing \eqref{eqn:generalcontinuition}, we have
\begin{align}
nR&\geq I(X_1^n;\dots;X_t^n|M)\nonumber\\
&= \left(\sum_{i=1}^t H(X_i^n|M)\right)-H(X_1^n,\dots,X_t^n|M)\nonumber\\
&= \left(\sum_{i=1}^t\sum_{j=1}^n H(X_{ij}|M,X_i^{1:j-1})\right)\nonumber\\
&\hspace{1cm}-\sum_{j=1}^nH(X_{1j},\dots,X_{tj}|M,X_1^{1:j-1},\dots,X_t^{1:j-1})\nonumber\\
&= \sum_{j=1}^n\bigg[\sum_{i=1}^t H(X_{ij}|M,X_i^{1:j-1})\nonumber\\
&\hspace{1cm}-H(X_{1j},\dots,X_{tj}|M,X_1^{1:j-1},\dots,X_t^{1:j-1})\bigg]\nonumber\\
&\geq \sum_{j=1}^n \bigg[\sum_{i=1}^t H(X_{ij}|M,X_1^{1:j-1},\dots,X_t^{1:j-1})\nonumber\\
&\hspace{1cm}-H(X_{1j},\dots,X_{tj}|M,X_1^{1:j-1},\dots,X_t^{1:j-1})\bigg]\nonumber\\
&=\sum_{j=1}^n I(X_{1j};\dots;X_{tj}|M,X_1^{1:j-1},\dots,X_t^{1:j-1})\nonumber\\
&=\sum_{j=1}^n I(X_{1j};\dots;X_{tj}|U_j)\label{eqn:generalauxil}\\
&=nI(X_{1Q};\dots;X_{tQ}|U_Q,Q)\nonumber,
\end{align}
where \eqref{eqn:generalauxil} follows by defining $U_j=(M,X_1^{1:j-1},\dots,X_t^{1:j-1})$.
\fi
\if \arxive 0
Single-letterizing $I(X_1^n;\dots;X_t^n|M)$ (details are in the extended version~\cite{Full}) and using \eqref{eqn:generalcontinuition}, we get $R\geq  I(X_{1Q};\dots;X_{tQ}|U_Q,Q)$, where $U_j=(M,X_1^{1:j-1},\dots,X_t^{1:j-1})$.
\fi 
 Following Wyner~\cite{Wyner75}, we lower bound $R$ in another fashion as $R\geq I(X_{1Q},\dots,X_{tQ};U_Q,Q)-g(\epsilon)$, where $g(\epsilon)\rightarrow 0$ as $\epsilon \rightarrow 0$  (details are in\if \arxive 1 Appendix~\ref{appendix:wynerstyle}\fi \if \arxive 0 the extended version~\cite{Full}\fi). Note that $\lVert p_{X_{1Q}\dots X_{tQ}}-q_{X_1\dots X_t} \rVert<\epsilon$, which follows from Cuff~\cite[Lemma~VI.2]{Cuff13}. Using the continuity of total variation distance and mutual information in the probability simplex, it follows along the same lines as Theorem~\ref{theorem:unlimited_shared_randomness} and Yassaee et al.~\cite[Lemma~6]{YassaeeGA15} that $R^{\mrm{Indv}}_{\mrm{opt}}\geq \min \max \big\{I(X_1;\dots;X_t|U), I(X_1,\dots,X_t;U)\big\}$, where the minimum is over all conditional p.m.f.'s $p(u|x_1,\dots,x_t)$ with $|\mc{U}|\leq\left(\prod_{i=1}^t |\mc{X}_i|\right)+t$. Note that this cardinality bound on $\mc{U}$ follows from an application of Convex Cover Method~\cite[Appendix~C]{GamalK12}. This completes the proof of Theorem~\ref{theorem:generalize}.
\end{IEEEproof}

\begin{IEEEproof}[Proof of Theorem~\ref{theorem_greedy}]
This is a special case of Theorem~\ref{theorem:equal} where $\mc{V}_i=\{i\}$. The rate region $\mc{R}^{\text{Indv}}$ is given by 
the set of all non-negative rate tuples $(R,R_1,\dots,R_t)$ such that there exist non-negative $r,r_1,\dots,r_t$ satisfying
\begin{align}
R&\geq \left(\sum_{j:j\in[1:t]\setminus\{i\}}r_j\right)+r,\ i\in[1:t],\label{eqn:thm2_1}\\
H(X)&\leq r+\sum_{i=1}^tr_i,\label{eqn:thm2_2}\\
R_i&\geq r_i,\ i\in[1:t].\label{eqn:thm2_3}
\end{align}
Let $\mc{R}^\prime$ be the set of all non-negative rate tuples $(R,R_1,\dots,R_t)$ satisfying \eqref{eqn:indv:remarkequal_1} and \eqref{eqn:indv:remarkequal}, the region given in the theorem. To show that $\mc{R}^\prime\subseteq \mc{R}^{\text{Indv}}$, let $(R,R_1,\dots,R_t)\in\mc{R}^\prime$. Without loss of generality, let $R_1\leq\dots\leq R_t$. Consider two cases.

\underline{\emph{Case (i)}} $\big(R_1\geq \frac{H(X)}{t}\big)$: 
Choose $r_i=\frac{H(X)}{t}$ for $i\in[1:t]$ and $r=0$. 

\underline{\emph{Case (ii)}} $\big(R_1<\frac{H(X)}{t}\big)$:
Choose $r_i=R_1$ for $i\in[1:t]$ and $r=H(X)-tR_1$ (note that $r>0$ since $tR_1<H(X)$).

In both the cases, it is easy to see that the choice of $r,r_1,\dots,r_t$ ensures that $(R,R_1,\dots,R_t)\in\mc{R}^{\text{Indv}}$.

To show that $\mc{R}^{\text{Indv}}\subseteq\mc{R}^\prime$, let $(R,R_1,\dots,R_t)\in\mc{R}^{\text{Indv}}$. For any $i\in[1:t]$, adding \eqref{eqn:thm2_1} and \eqref{eqn:thm2_3} gives $R+R_i\geq \sum_{j=1}^tr_j+r\geq H(X)$, where the last inequality follows from \eqref{eqn:thm2_2}. This gives \eqref{eqn:indv:remarkequal_1}. Adding \eqref{eqn:thm2_1} over respective $i\in[1:t]$ gives $tR\geq (t-1)\left(\sum_{i=1}^tr_j+r\right)+r\geq (t-1)H(X)$, where the last inequality follows from \eqref{eqn:thm2_2} and the fact that $r\geq0$. This gives \eqref{eqn:indv:remarkequal} and thus $(R,R_1,\dots,R_t)\in\mc{R}^\prime$. This completes the proof of Theorem~\ref{theorem_greedy}.
\end{IEEEproof}

\subsubsection{Randomness-on-the-Forehead Model}

\begin{IEEEproof}[Proof sketch of Theorem~\ref{theorem:t=3}]
Here we give a proof sketch for $t=3$ (a detailed proof can be found in\if \arxive 1 Appendix~\ref{appendix_foreheadproof}\fi \if \arxive 0 the extended version~\cite{Full}\fi). The proof employs the OSRB framework~\cite{YassaeeAG14}. 
Let $(U^n,U_1^n,U_2^n,U_3^n,X_1^n,X_2^n,X_3^n)$ be i.i.d. with distribution $p(u,u_{[1:3]},x_{[1:3]})=q(x_{[1:3]})p(u,u_{[1:3]}|x_{[1:3]})$ satisfying \eqref{eqn:t=3markov}. Bin indices $f,m^*,b_1,b_2,b_3$ with respective rates $\hat{R}_0,R^*,\tilde{R}_1,\tilde{R}_2,\tilde{R}_3$ are created from $(U^n,U_1^n,U_2^n,U_3^n)$ in a way that can be understood from the following joint probability distribution:
\begin{align}
&P(u^n,u_1^n,u_2^n,u_3^n,x_1^n,x_2^n,x_3^n,f,m^*,b_1,b_2,b_3)\nonumber\\
&=p(u^n,u_1^n,u_2^n,u_3^n)P(f|u^n)P(m^*|u^n)P(b_1|u^n,u_1^n)\nonumber\\
&\hspace{0.6cm}\times P(b_2|u^n,u_2^n)P(b_3|u^n,u_3^n)p(x_1^n|u^n,u_2^n,u_3^n)\nonumber\\
&\hspace{0.6cm}\times p(x_2^n|u^n,u_1^n,u_3^n)p(x_3^n|u^n,u_1^n,u_2^n)\nonumber\\
&=P(b_1,b_2,b_3,f)P(u^n,u_1^n,u_2^n,u_3^n|b_1,b_2,b_3,f)P(m^*|u^n)\nonumber\\
&\hspace{0.6cm}\times p(x_1^n|u^n,u_2^n,u_3^n)p(x_2^n|u^n,u_1^n,u_3^n)p(x_3^n|u^n,u_1^n,u_2^n)\label{eqn:forehead:pmf1_5page}.
\end{align}
Further, we use Slepian-Wolf decoders to estimate $(u^n,u_{(i)_3+1})$ from $b_{(i+2)_3+1},f,m^*$, $i=0,1,2$, where $(i)_3=i \mod 3$. Now we impose a series of constraints on the rates (for details see\if \arxive 0 the extended version~\cite{Full}\fi \if \arxive 1 Appendix~\ref{appendix_foreheadproof}\fi). The first set of constraints ensure that $b_1,b_2,b_3,f$ are approximately (i.e., with vanishing total variation distance) uniformly distributed and mutually independent of each other~\cite[Theorem~1]{YassaeeAG14}. The second set of constraints guarantees the success of Slepian-Wolf decoders with high probability~\cite[Lemma~1]{YassaeeAG14}. Under these two sets of rate constraints, the above p.m.f. becomes approximately close to the p.m.f. described below, which is related to our original problem.
 We generate $b_1,b_2,b_3,f$ independently and uniformly from the respective alphabets. For $i\in[1:3]$, we treat $b_i$ as the shared randomness $w_i$ that is not available to processor $P_i$. In addition, we have extra shared randomness $f$ (to be eliminated later), which is shared among coordinator and all the three processors. The coordinator on observing $b_1,b_2,b_3,f$ produces $u^n,u_1^n,u_2^n,u_3^n$ according to the random p.m.f. $P(u^n,u_1^n,u_2^n,u_3^n|b_1,b_2,b_3,f)$ of \eqref{eqn:forehead:pmf1_5page} and sends $(m^*(u^n))$ as a common message $m$ to the processors, where $m^*(u^n)$ is produced according to $P(m^*|u^n)$ of \eqref{eqn:forehead:pmf1_5page}. The processors use (random) Slepian-Wolf decoders mentioned below \eqref{eqn:forehead:pmf1_5page} to produce their respective estimates. Then they generate $x_1^n,x_2^n,x_3^n$ according to respective p.m.f.'s mentioned in the last line of \eqref{eqn:forehead:pmf1_5page}. We need a third set of rate constraints so that $X_i^n, i\in[1:3]$ becomes approximately independent of $F$ (for details see\if \arxive 0 the extended version~\cite{Full}\fi \if \arxive 1 Appendix~\ref{appendix_foreheadproof}\fi). All these three sets of rate constraints ensures the correctness of the output distribution with a particular realization of the binning. Noting that $R=R^*$ and eliminating all the other rates gives us \eqref{eqn:forehead:bound} for $t=3$.
\end{IEEEproof}

%%%
%%%
\begin{IEEEproof}[Proof of Theorem~\ref{theorem:foreheadequal}]
This is a special case of Theorem~\ref{theorem:equal} where $\mc{V}_i=[1:t]\setminus\{i\}$. The rate region $\mc{R}^{\text{Forehead}}$ is given by 
the set of all non-negative rate tuples $(R,R_1,\dots,R_t)$ such that there exists non-negative $r,r_1,\dots,r_t$ satisfying
\begin{align}
R&\geq r_i+r,\ i\in[1:t],\label{eqn:thm4_1}\\
H(X)&\leq r+\sum_{i=1}^tr_i,\label{eqn:thm4_2}\\
R_i&\geq r_i,\ i\in[1:t].\label{eqn:thm4_3}
\end{align}
Let $\mc{R}^\prime$ be the set of all non-negative rate tuples $(R,R_1,\dots,R_t)$ satisfying \eqref{eqn:foreheadequal1}, the region given in the theorem. To show that $\mc{R}^\prime\subseteq \mc{R}^{\text{Forehead}}$, let $(R,R_1,\dots,R_t)\in\mc{R}^\prime$. Without loss of generality, let $R_1\leq\dots\leq R_t$. We consider two cases.

\underline{\emph{Case (i)}} $\big(H(X) \leq \sum_{i=1}^tR_i\big)$: 
Let $i\in[1:t]$ such that $\sum_{j=1}^{i-1}R_j<H(X)\leq \sum_{j=1}^{i}R_j$. Choose $r_j=R_j$, $j\in[1:i-1]$, $r_j=\frac{(H(X)-\sum_{k=1}^{i-1}R_k)}{(t-(i-1))},j\in[i:t]$ and $r=0$.

\underline{\emph{Case (i)}} $\big(H(X) > \sum_{i=1}^tR_i\big)$: Choose $r_j=R_j,i\in[1:t]$ and $r=H(X)-\sum_{i=1}^tR_i$.

 It is easy to see that, in both the cases, the choice of $r,r_1,\dots,r_t$ ensures that $(R,R_1,\dots,R_t)\in\mc{R}^{\text{Forehead}}$.

To show that $\mc{R}^{\text{Forehead}}\subseteq\mc{R}^\prime$, let $(R,R_1,\dots,R_t)\in\mc{R}^{\text{Forehead}}$. Fix a set $\mc{S}\subsetneq [1:t]$. Adding \eqref{eqn:thm4_1} over $i\in[1:t]\setminus\mc{S}$ and \eqref{eqn:thm4_3} over $i\in\mc{S}$ gives $(t-|\mc{S}|)R+\sum_{j\in\mc{S}}R_i\geq (t-|\mc{S}|-1)r+r+\sum_{i=1}^tr_i\geq H(X)$, where the last inequality follows from \eqref{eqn:thm4_2} and the facts that $|\mc{S}|\leq t-1$ and $r\geq0$. Considering this over all the possible sets $\mc{S}$ gives us $(R,R_1,\dots,R_t)\in\mc{R}^\prime$.
This completes the proof of Theorem~\ref{theorem:foreheadequal}.
\end{IEEEproof}

\subsubsection{The general case}

In the general case we can only prove results when all of the output variables are equal.

\begin{IEEEproof}[Proof of Theorem~\ref{theorem:equal}]
Suppose for a rate tuple $(R,R_1,\dots,R_h)$, there exist $r,r_1,\dots,r_h$ such that \eqref{eqn:equal1}, \eqref{eqn:equal2}  and \eqref{eqn:equal3} hold. For each $i\in[1:h]$, from randomness $W_i$ only a randomness of rate $r_i$ is utilised in the achievability. For $i\in[1:h]$, since $r_i\leq R_i$, without loss of generality, assume that $W_i$ is of rate $R_i$. The coordinator sends a message which consists of two parts. Since coordinator has access to all $W_i$'s, by network coding~\cite{Yeung2008,fragouli2007}, a multicast message of rate at least $\max_i \sum_{j:j\notin\mc{V}_i}r_j$ can be used to deliver all the sources of randomness to all the processors. This constitutes the first part of the common message. The second part of the message is a uniform randomness of rate $r$ which might be required additionally so that a common randomness of rate of atleast $H(X)$ is available to all the processors. This gives an achievable scheme for the rate tuple $(R,R_1,\dots,R_h)$ satisfying \eqref{eqn:equal1}, \eqref{eqn:equal2}  and \eqref{eqn:equal3}, since a common randomness of rate $H(X)$ is sufficient for sampling the same i.i.d. sequence approximately according to $q_X$ by all the processors.

For the converse, suppose a rate tuple $(R,R_1,\dots,R_h)$ is achievable. Consider
\begin{align}
nR&\geq H(M)\nonumber\\
&\geq H(M|W_{\mc{V}_i})\nonumber\\
&\geq I(M;X_i^n|W_{\mc{V}_i})\nonumber\\
&=I(M,W_{[1:h]\setminus\mc{V}_i};X_i^n|W_{\mc{V}_i})\label{eqn:thm5_markov}\\
&=I(W_{[1:h]\setminus\mc{V}_i};X_i^n|W_{\mc{V}_i})+I(M;X_i^n|W_{[1:h]})\label{eqn:thm5_exp},
\end{align}
where \eqref{eqn:thm5_markov} follows from the Markov chain $X_i^n-(M,W_{\mc{V}_i})-W_{[1:h]\setminus\mc{V}_i}$. For the first term in \eqref{eqn:thm5_exp} with $i=1$, note that
\begin{align}
I(W_{[1:h]\setminus\mc{V}_1}&;X_1^n|W_{\mc{V}_1})\nonumber\\
&=\sum_{j:j\notin\mc{V}_1}I(W_j;X_1^n|W_{[1:j-1]\cap\left([1:h]\setminus\mc{V}_1\right)},W_{\mc{V}_1})\label{eqn:thm5_chainrule}\\
&=\sum_{j:j\notin\mc{V}_1}I(W_j;X_1^n,W_{[1:j-1]\cap\left([1:h]\setminus\mc{V}_1\right)},W_{\mc{V}_1})\label{eqn:thm5_indp}\\
&\geq \sum_{j:j\notin\mc{V}_1}I(W_j;X_1^n,W^{j-1})\nonumber,
\end{align}
where \eqref{eqn:thm5_chainrule} follows from chain rule of mutual information, \eqref{eqn:thm5_indp} follows because, for $j\notin\mc{V}_1$, $W_j$ is independent of $\left(W_{[1:j-1]\cap\left([1:h]\setminus\mc{V}_1\right)},W_{\mc{V}_1}\right)$. For $j\in[1:h]$, let $r_j=n^{-1}I(W_j;X_1^n,W^{j-1})$.

For $i>1$, note that
\begin{align}
I(&W_{[1:h]\setminus\mc{V}_i};X_i^n|W_{\mc{V}_i})&\nonumber\\
&=I(W_{[1:h]\setminus\mc{V}_i};X_1^n,X_i^n|W_{\mc{V}_i})-I(W_{[1:h]\setminus\mc{V}_i};X_1^n|X_i^n,W_{\mc{V}_i})\nonumber\\
&\geq I(W_{[1:h]\setminus\mc{V}_i};X_1^n|W_{\mc{V}_i})-H(X_1^n|X_i^n)\nonumber\\
&\geq I(W_{[1:h]\setminus\mc{V}_i};X_1^n|W_{\mc{V}_i})-n\epsilon_1\label{thm5_correct},
\end{align}
where \eqref{thm5_correct} follows from the correctness of the output distribution with $\epsilon_1\rightarrow 0$ as $\epsilon\rightarrow 0$. For the second term in \eqref{eqn:thm5_exp}, $I(M;X_i^n|W_{[1:h]})=I(M;X_i^n,X_1^n|W_{[1:h]})-I(M;X_1^n|X_i^n,W_{[1:h]})\geq I(M;X_1^n|W_{[1:h]})-n\epsilon^\prime$, where $\epsilon^\prime\rightarrow 0$ as $\epsilon\rightarrow 0$.  Let $r=n^{-1}I(M;X_1^n|W_{[1:h]})$. This gives \eqref{eqn:equal1}. Consider
\begin{align}
nr+\sum_{j=1}^hnr_j&=I(M;X_1^n|W_{[1:h]})+\sum_{j=1}^hI(W_j;X_1^n,W^{j-1})\nonumber\\
&=I(M;X_1^n|W_{[1:h]})+\sum_{j=1}^hI(W_j;X_1^n|W^{j-1})\nonumber\\
&=I(M;X_1^n|W_{[1:h]})+I(W_{[1:h]},X_1^n)\nonumber\\
&=I(M,W_{[1:h]};X_1^n)\nonumber\\
&\geq I(M,W_{\mc{V}_1};X_1^n)\nonumber\\
&\geq I(X_2^n;X_1^n)\label{eqn:thm5:markov7}\\
&\geq nH(X)-n\epsilon_2\label{eqn:thm5_correctness2},
\end{align}
where \eqref{eqn:thm5:markov7} follows from the Markov chain $X_1^n-(M,W_{\mc{V}_1})-X_2^n$, and \eqref{eqn:thm5_correctness2} follow from the correctness of the output distribution with $\epsilon_2\rightarrow 0$ as $\epsilon\rightarrow 0$. This gives \eqref{eqn:equal2}. For \eqref{eqn:equal3}, note that $r_j=n^{-1}I(W_j;X_1^n,W^{j-1})\leq n^{-1}H(W_j)=R_j$. This completes the proof of Theorem~\ref{theorem:equal}.
\end{IEEEproof}
\subsection{An Achievable Strategy for General Model}
We remark that the idea behind the achievabilities of Theorems~\ref{theorem:achievability}, \ref{theorem:unlimited_shared_randomness}, \ref{theorem:generalize} and \ref{theorem:t=3} is not confined only to either individually shared randomness model or randomness-on-the-forehead model. A similar achievable strategy can be written down along the same lines for the general model where $\mc{V}_i$'s are arbitrary subsets of $[1:h]$ even though its not direct to attain closed form expression(s). We outline this achievable strategy here. 

Let $(U^n,U_1^n,\dots,U_h^n,X_1^n,\dots,X_t^n)$ be i.i.d. with distribution $p(u,u_{[1:h]},x_{[1:t]})=q(x_{[1:t]})p(u,u_{[1:h]}|x_{[1:t]})$ satisfying $$p(u,u_{[1:h]},x_{[1:t]})=p(u,u_{[1:h]})\prod_{i=1}^t p(x_i|u,u_{\mc{V}_i}).$$
Bin indices $f,m^*,b_{[1:h]}$ with respective rates $\hat{R}_0,R^*,\tilde{R}_1,\dots,\tilde{R}_h$ are created from $(U^n,U_1^n,\dots,U_h^n)$ in a way that can be understood from the following joint probability distribution:
\begin{align}
&P(u^n,u_1^n,\dots,u_h^n,x_1^n,\dots,x_t^n,f,m^*,b_1,\dots,b_h)\nonumber\\
&=p(u^n,u_1^n,\dots,u_h^n)P(f|u^n)P(m^*|u^n)\left(\prod_{j=1}^hP(b_j|u^n,u_j^n)\right)\nonumber\\
&\hspace{0.6cm}\times \left(\prod_{i=1}^tp(x_i^n|u^n,u_{\mc{V}_i}^n)\right)\nonumber\\
&=P(b_1,\dots,b_h,f)P(u^n,u_1^n,\dots,u_h^n|b_1,\dots,b_h,f)P(m^*|u^n)\nonumber\\
&\hspace{0.6cm}\times \left(\prod_{i=1}^tp(x_i^n|u^n,u_{\mc{V}_i}^n)\right)\label{eqn:general:strategy}.
\end{align}
Further, we use Slepian-Wolf decoders to estimate $(u^n,u_{\mc{V}_i})$ from $b_{\mc{V}_i},f,m^*$, $i\in[1:t]$. Now we impose a series of constraints on the rates as in the proofs of Theorems~\ref{theorem:achievability}, \ref{theorem:unlimited_shared_randomness}, \ref{theorem:generalize} and \ref{theorem:t=3}. The first set of constraints can be written down as in \eqref{eqn:forehead:indep_1} so that $b_1,\dots,b_h,f$ are approximately (i.e., with vanishing total variation distance) uniformly distributed and mutually independent of each other~\cite[Theorem~1]{YassaeeAG14}. The second set of constraints can be written down as in \eqref{eqn:forehead:sw} to guarantee the success of Slepian-Wolf decoders with high probability~\cite[Lemma~1]{YassaeeAG14}. Under these two sets of rate constraints, the above p.m.f. becomes approximately close to the p.m.f. described below, which is related to our original problem.
 We generate $b_1,\dots,b_h,f$ independently and uniformly from the respective alphabets. For $j\in[1:h]$, we treat $b_j$ as $j^{\text{th}}$ shared randomness. In addition, we have extra shared randomness $f$ (to be eliminated later), which is shared among coordinator and all the $n$ processors. The coordinator on observing $b_1,\dots,b_h,f$ produces $u^n,u_1^n,\dots,u_h^n$ according to the random p.m.f. $P(u^n,u_1^n,\dots,u_h^n|b_1,\dots,b_h,f)$ of \eqref{eqn:general:strategy} and sends $(m^*(u^n))$ as a common message $m$ to the processors, where $m^*(u^n)$ is produced according to $P(m^*|u^n)$ of \eqref{eqn:general:strategy}. The processors use (random) Slepian-Wolf decoders mentioned below \eqref{eqn:general:strategy} to produce their respective estimates. Then they generate $x_1^n,\dots,x_t^n$ according to respective p.m.f.'s mentioned in the last line of \eqref{eqn:general:strategy}. We need a third set of rate constraints so that $X_i^n, i\in[1:t]$ becomes approximately independent of $F$ as in \eqref{eqn:forehead:indep_2}. All these three sets of rate constraints ensures the correctness of the output distribution with a particular realization of the binning. Noting that $R=R^*$ and eliminating all the other rates will give us an achievable rate.

% !TEX root = Generalization.tex

\section{Omniscient Coordinator with Correlated Shared Randomness Model}\label{section:assisted}

\begin{figure}[htbp]
 \begin{center}
\includegraphics{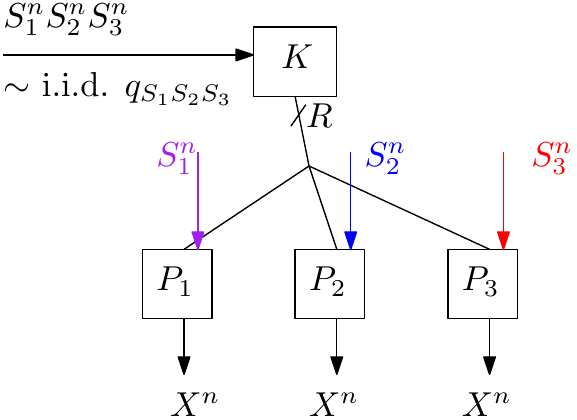}
\end{center}
\caption[Omniscient coordinator with correlated shared randomness model.]{Omniscient coordinator with correlated shared randomness model. $t=3$ case is shown. Coordinator $K$ having access to $(S_1^n,S_2^n,S_3^n)\sim $ i.i.d. $q_{S_1S_2S_3}$ sends a common message of rate $R$ to the processors, where processor $P_i$ has access to $S_i^n$, for $i\in[1:3]$, so that each of them outputs (approximately) the same random sequence $X^n\sim$ i.i.d. $ q_X$.}\label{figure:prabh}
\end{figure}

In this section, we study the model where the shared random variables are arbitrarily correlated instead of being independent as assumed in previous sections. In particular, the coordinator has access to $(S_1^n,\dots,S_t^n)$, where $(S_{1i},\dots,S_{ti})$, $i=1,\dots,n$, are i.i.d with distribution $q_{S_1,\dots,S_t}$, and processor $P_i$ has access to $S_i^n$, for $i\in[1:t]$ (see Figure~\ref{figure:prabh}). A simulation code and an achievable rate (note that in this setting, there is only one rate involved, the rate of message communicated from the coordinator to all the processors) can be defined analogously to Definitions~\ref{defn1} and \ref{defn2}. We are interested in characterizing the infimum of all the achievable rates, i.e., the {optimal communication rate}. In this section, we prove results for the case when $q_{X_1\dots X_t}$ is such that $X_1=\dots=X_t$. The following theorem characterizes the optimal communication rate for this model.

\begin{thm}\label{theorem:assisted}
Suppose $q_{X_1\dots X_t}$ is such that $X_1=\dots=X_t=X$. Then the optimal communication rate for the correlated shared randomness model is given by 
\begin{align}
\min_{\substack{r\geq 0,p_{U|S_1,\dots,S_t}:\\I(U;S_{[1:t]})+r\geq H(X)}}\max_i\left(I(U;S_{[1:t]\setminus\{i\}}|S_i)+r\right)\label{eqn:corr:ach}.
\end{align}
\end{thm} 

\begin{IEEEproof}
Fix a conditional p.m.f. $p_{U|S_1,\dots,S_t}$. For the achievability, it suffices to show that an uniform common randomness of rate $I(U;S_1,\dots,S_t)$ is recovered at all the processors if $R>I(U;S_{[1:t]\setminus\{i\}}|S_i)$, for $i\in[1:t]$. This is because if in case $I(U;S_{1},\dots,S_t)<H(X)$, then an additional randomness of rate $r$ such that $I(U;S_1\dots,S_t)+r\geq H(X)$ can be sent by the coordinator giving us \eqref{eqn:corr:ach}, as atleast uniform randomness of rate  $H(X)$ is sufficient to produce i.i.d. $X^n$ at all the processors~\cite{Wyner75}. Let $(U^n,S_1^n,\dots,S_t^n)$ be i.i.d. with distribution $p(u,s_1,\dots,s_t)=q(s_1\dots,s_t)\times p(u|s_1\dots,s_t)$. To each $u^n$ sequence, assign uniformly and independently three bin indices $m\in[1:2^{nR}]$, $m^\prime\in[1:2^{nR^\prime}]$ and $f\in[1:2^{n\hat{R}}]$. The induced random p.m.f. will be
\begin{align}
&P(u^n,s_{[1:t]}^n,m,m^\prime,f)=p(u^n,s_{[1:t]}^n)P(m,m^\prime,f|u^n)\nonumber\\
&=P(s_{[1:t]}^n,f)P(u^n|s_{[1:t]}^n,f)P(m|u^n)P(m^\prime|u^n)\label{eqn:corr:prot1}.
\end{align}
Also, for each $i\in[1:n]$, there is a Slepian-Wolf decoder to reconstruct $u^n$ from $(m,f,s_i^n)$. Now, using  \cite[Theorem~1]{YassaeeAG14}, if 
\begin{align}
\hat{R}<H(U|S_1\dots,S_t)\label{eqn:corr_achv1},
\end{align}
 we have
\begin{align}
P(f,s_{[1:t]}^n)\approx p^{\text{Unif}}(f)p(s_{[1:t]}^n)\label{eqn:corr:achvcons}.
\end{align}
For the success of Slepian-Wolf decoders with high probability, using \cite[Lemma~1]{YassaeeAG14} we need
\begin{align}
R+\hat{R}>H(U|S_i)\label{eqn:corr:SW},
\end{align}
for $i\in[1:t]$. Now, the p.m.f. in \eqref{eqn:corr:prot1} becomes approximately close to the protocol corresponding to the main problem with additional shared randomness $F$, i.e., the coordinator produces message $m$ according to $(P(u^n|s_{[1:t]}^n,f)\times P(m|u^n))$ and the processors implement the Slepian-Wolf decoders mentioned before. After all the processors recover $u^n$ correctly with high probability, they find index $m^\prime$ according to $P(m^\prime|u^n)$. Using \cite[Theorem~1]{YassaeeAG14}, if 
\begin{align}
R^\prime+\hat{R}<H(U)\label{eqn:corr:achv2},
\end{align}
we have 
\begin{align}
P(m^\prime,f)\approx p^{\text{Unif}}(m^\prime)p^{\text{Unif}}(f).\label{eqn:corr:achv3}
\end{align}
Conditions \eqref{eqn:corr:achvcons} and \eqref{eqn:corr:achv3} imply the existence of a particular realization of the random binning with corresponding p.m.f. $p$ so that we can replace $P$ with $p$. This implies
\begin{align}
p(f,s_{[1:t]}^n)\approx p^{\text{Unif}}(f)p(s_{[1:t]}^n)\nonumber\\
p(m^\prime,f)\approx p^{\text{Unif}}(m^\prime)p^{\text{Unif}}(f)\label{eqn:corr;achv4}.
\end{align}
The conditions in \eqref{eqn:corr;achv4} implies that there exists an instance $f^*$ such that
\begin{align}
p(s^n_{[1:t]}|f^*)\approx p(s_{[1:t]}^n),\nonumber\\
p(m^\prime|f)\approx p^{\text{Unif}}(m^\prime).
\end{align}
This ensures that after fixing the instance $f^*$, the shared random sequences are according to i.i.d. with the distribution $q_{S_1,\dots,S_t}$ and all the processors are able to recover uniform randomness of rate $R^\prime$ under the conditions \eqref{eqn:corr_achv1}, \eqref{eqn:corr:SW} and \eqref{eqn:corr:achv2}. Conditions \eqref{eqn:corr_achv1} and  \eqref{eqn:corr:SW} imply that 
\begin{align}\label{eqn:corr:final1}
R>I(U;S_{[1:i]\setminus\{i\}}|S_i),
\end{align}
 for $i\in[1:t]$. Conditions \eqref{eqn:corr:SW} and \eqref{eqn:corr:achv2} imply that
 \begin{align}
R^\prime-R<I(U;S_i),\label{eqn:corr:final2}
\end{align}
for $i\in[1:t]$. Choosing $R^\prime$ and $R$ such that $R^\prime-R=I(U;S_i)-\epsilon$, where $\epsilon$ is arbitrarily small in \eqref{eqn:corr:final2} and using \eqref{eqn:corr:final1}, it can be seen that an uniform common randomness of rate $R^\prime> I(U;S_{[1:t]})-\epsilon$ is available to all the processors if $R>\max\limits_iI(U;S_{[1:t]\setminus\{i\}}|S_i),$ for $i\in[1:t]$. This completes the achievability.  

For the converse, suppose a rate $R$ is achievable. For $i>1$, consider
\begin{align}
nR&\geq H(M)\nonumber\\
&\geq H(M|S_1^n)\nonumber\\
&=I(M;S_{[1:t]\setminus\{i\}}^n|S_i^n)+H(M|S_1^n,\dots,S_t^n)\nonumber\\
&=I(M,X_i^n;S_{[1:t]\setminus\{i\}}^n|S_i^n)+H(M|S_1^n,\dots,S_t^n)\label{eqn:correlated:conv1}\\
&\geq I(X_i^n;S_{[1:t]\setminus\{i\}}^n|S_i^n)+H(M|S_1^n,\dots,S_t^n)\label{eqn:correlated:conv2}
\end{align}
where \eqref{eqn:correlated:conv1} follows from the Markov chain $X_i^n-(M,S_i^n)-S_{[1:t]\setminus\{i\}}^n$. Let $Q$ be a random variable uniformly distributed over $[1:n]$ and independent of all other random variables. For the first term in \eqref{eqn:correlated:conv2} with $i=1$, note that
\begin{align}
&I(X_1^n;S_2^n,\dots,S_t^n|S_1^n)\\
&=\sum_{j=1}^nI(X_1^n;S_{2j},\dots,S_{tj}|S_{2}^{1:j-1},\dots,S_t^{1:j-1},S_1^n)\nonumber\\
&=\sum_{j=1}^nI(X_1^n,S_1^{1:j-1},S_1^{j+1:n},S_{[2:t]}^{1:j-1};S_{2j},\dots,S_{tj}|S_{1j})\label{eqn:correlated:conv3}\\
&\geq\sum_{j=1}^nI(X_1^n,S_{[1:t]}^{1:j-1};S_{2j},\dots,S_{tj}|S_{1j})\nonumber\\
&=nI(X_1^n,S_{[1:t]}^{1:Q-1};S_{2Q},\dots,S_{tQ}|S_{1Q},Q)\nonumber\\
&\geq n[I(X_1^n,S_{[1:t]}^{1:Q-1},Q;S_{2Q},\dots,S_{tQ}|S_{1Q})-\epsilon_1]\label{eqn:correlated:conv4}\\
&=n[I(U;S_{2Q},\dots,S_{tQ}|S_{1Q})-\epsilon_1]\label{eqn:correlated:conv41},
\end{align}
where \eqref{eqn:correlated:conv3} follows since $(S_1^n,\dots,S_t^n)$ are i.i.d., \eqref{eqn:correlated:conv4} follows along similar lines as \eqref{eqn:T_independent} with $\epsilon_1\rightarrow 0$ as $\epsilon\rightarrow 0$, \eqref{eqn:correlated:conv4} follows by defining $U=(X_1^n,S_{[1:t]}^{1:Q-1},Q)$. For $i>1$, note that
\begin{align}
&I(X_i^n;S_{[1:t]\setminus\{i\}}^n|S_i^n)\nonumber\\
& =I(X_i^n,X_1^n;S_{[1:t]\setminus\{i\}}^n|S_i^n)-I(X_1^n;S_{[1:t]\setminus\{i\}}^n|S_i^n,X_i^n)\nonumber\\
&\geq I(X_i^n,X_1^n;S_{[1:t]\setminus\{i\}}^n|S_i^n)-H(X_1^n|X_i^n)\nonumber\\
&\geq I(X_1^n;S_{[1:t]\setminus\{i\}}^n|S_i^n)-n\epsilon_2\label{eqn:correlated:conv5}
\end{align}
where \eqref{eqn:correlated:conv5} follows from the correctness of the output distribution with $\epsilon_2\rightarrow 0$ as $\epsilon\rightarrow 0$. For the second term in \eqref{eqn:correlated:conv2}, note that
\begin{align}
H(M|S_1^n,&\dots,S_t^n)\nonumber\\
&\geq I(M;X_1^n|S_1^n,\dots,S_t^n)\nonumber\\
&=H(X_1^n|S_1^n,\dots,S_t^n)-H(X_1^n|M,S_1^n,\dots,S_t^n)\nonumber\\
&\geq H(X_1^n|S_1^n,\dots,S_t^n)-H(X_1^n|M,S_1^n)\nonumber\\
&\geq H(X_1^n|S_1^n,\dots,S_t^n)-H(X_1^n|X_2^n)\label{eqn:correlated:conv6}\\
&\geq H(X_1^n|S_1^n,\dots,S_t^n)-n\epsilon_3\label{eqn:correlated:cpnv7},
\end{align}
where \eqref{eqn:correlated:conv6} follows from the Markov chain $X_1^n-(M,S_1^n)-X_2^n$, \eqref{eqn:correlated:cpnv7} follows from the correctness of the output distribution with $\epsilon_3\rightarrow 0$ as $\epsilon\rightarrow 0$. By defining $r=n^{-1}H(X_1^n|S_1^n,\dots,S_t^n)$, consider 
\begin{align}
&nH(X)\nonumber\\
&\leq H(X_1^n)+n\epsilon_4\label{eqn:correlated:conv8}\\
&= I(X_1^n;S_1^n,\dots,S_t^n)+H(X_1^n|S_1^n,\dots,S_t^n)+n\epsilon_4\nonumber\\
&=\sum_{j=1}^nI(X_1^n;S_{1j},\dots,S_{tj}|S_1^{1:j-1},\dots,S_t^{1:j-1})+nr+n\epsilon_4\nonumber\\
&\leq\sum_{j=1}^nI(X_1^n,S_{[1:t]}^{1:j-1};S_{1j},\dots,S_{tj})+nr+n\epsilon_4\nonumber\\
&\leq nI(U;S_{1Q},\dots,S_{tQ})+nr+n\epsilon_4\label{eqn:correlated:conv9}
\end{align}
where \eqref{eqn:correlated:conv8} follows from the correctness of the output distribution with $\epsilon_4\rightarrow 0$ as $\epsilon\rightarrow 0$.  From \eqref{eqn:correlated:conv2}, \eqref{eqn:correlated:conv41}, \eqref{eqn:correlated:conv5}, \eqref{eqn:correlated:cpnv7} and \eqref{eqn:correlated:conv9}, using the continuity of total variation distance and mutual information in the probability simplex, it follows along the same lines as Theorem~\ref{theorem:unlimited_shared_randomness}, \cite[Lemma~6]{YassaeeGA15} that 
\begin{align*}
R\geq I(U;S_{[1:t]\setminus\{i\}})+r, \ i\in[1:t]\\
\end{align*}
where $I(U;S_{[1:t]})+r\geq H(X),$
for some $r\geq 0$ and p.m.f. $p_{U|S_1,\dots,S_t}$. This completes the converse.
\end{IEEEproof}

% !TEX root = Generalization.tex

\section{Oblivious Coordinator Setting}\label{section:oblivious}
In this section we study a variant of our problem where instead of having access to all shared random variables, the coordinator does not have access to any shared random variables (See Figure~\ref{fig:modelobv}).
We call this the oblivious coordinator setting. A simulation code, an achievable rate tuple, and simulation rate region can be defined analogously to the omniscient coordinator setting. Note that the common message $M$ sent by the coordinator is independent of the shared randomness here. We treat it as a uniformly distributed random variable on $[1:2^{nR}]$ and denote it by $W$ for the oblivious coordinator setting. Notice that this problem is similar to Wyner's common information problem~\cite{Wyner75}, whose multi-user generalization, among other things, was studied by Xu et al.~\cite{Liu2010}. Even though the coordinator sends uniformly distributed common random message to all the processors in both the problems, the main difference here is that the processors have access to some of the shared random variables, which can potentially reduce the rate of common message. Thus, this problem reduces to Wyner's common information problem in the absence of shared random variables. In this model, when $\mathcal{V}_i$'s are arbitrary subsets of $[1:h]$, we completely characterize the simulation rate-region, i.e., the trade-off region between shared randomness rates and the rate of uniform message communicated from coordinator to all the processors. We first present and prove the rate-region for the case when $t=h=3$ and $\mathcal{V}_i=[1:3]\setminus\{i\}$ which essentially illustrates the proof idea behind more general rate region where there are $t$ processors and $\mathcal{V}_i$'s are arbitrary subsets of $[1:h]$.

\begin{thm}\label{theorem_oblv_3users}
For the oblivious coordinator setting, when $t=h=3$ and $\mathcal{V}_i=[1:3]\setminus\{i\}$, the simulation rate region is given by the set of all non-negative rate tuples $(R,R_1,R_2,R_3)$ such that
\begin{align}
R+R_{\mathcal{S}}\geq I(X_1,X_2,X_3;U,U_{\mathcal{S}}), \ \mathcal{S}\subseteq [1:3],
\end{align}
for some probability mass function 
	\begin{align*}
	p(u,u_{[1:3]},x_{[1:3]})=q(x_{[1:3]})p(u,u_{[1:3]}|x_{[1:3]})
	\end{align*}
such that
	\begin{align*}
	p(u,u_{[1:3]},x_{[1:3]})&=p(u)\left(\prod_{i=1}^3 p(u_i)\right) \\
	&\qquad\qquad\qquad
	\left(\prod_{i=1}^3p(x_i|u,u_{[1:3]\setminus\{i\}})\right).
	\end{align*}
\end{thm}
\begin{IEEEproof}
The proof of achievability is in the spirit of versions of channel resolvability that appear in recent works \cite{Hayashi,Cuff13,BlochK13}. Fix a p.m.f. $p(u,u_{[1:3]},x_{[1:3]})$ as given in the theorem. We generate four codebooks randomly in the following way.
\begin{itemize}
\item Randomly and independently generate $2^{nR}$ sequences $u^n(m)$, $m\in[1:2^{nR}]$,  each according to i.i.d. $p_U$. 
\item For each $u^n(w)$, randomly and independently generate $2^{nR_i}$ sequences $u_i^n(w,w_i)$, $w_i\in[1:2^{nR_i}]$, each according to i.i.d. $p_{U_i}$, for $i\in[1:3]$. 
\end{itemize}
Processor $P_1$ on observing $w,w_2,w_3$ produces $x_1^n$ according to $P(x_1^n|U^n(w),U_2^n(w,w_2),U_3^n(w,w_3))$, which is a random p.m.f. as $U^n(w),U_2^n(w,w_2),U_3^n(w,w_3)$ are random codewords. In a similar manner, processors $P_2$ and $P_3$ also produce $x_2^n$ and $x_3^n$, respectively. In the sequel of this proof, whenever we need not treat $x_1^n,x_2^n,x_3^n$ separately,  we denote $a^n:=(x_1^n,x_2^n,x_3^n)$. The induced output random p.m.f. can be written as
\begin{align}
&P(a^n)=2^{-n(R+R_1+R_2+R_3)}\times\nonumber\\
&\sum_{w,w_1,w_2,w_3}P(a^n|U^n(w),U_1^n(w,w_1),U_2^n(w,w_2),U_3^n(w,w_3))\label{eqn:ach:oblv1}.
\end{align} 
We denote by $\mathcal{T}_1^\epsilon$ and $\mathcal{T}_2^\epsilon$ the $\epsilon$-typical sets with distributions $p_A$ and $p_{UU_{[1:3]}A}$, respectively. Note that $P(a^n)$ can be written as
\begin{align}
P(a^n)=P_1(a^n)+P_2(a^n)\nonumber
\end{align}
where
\begin{align}
&P_1(a^n)=2^{-n(R+R_1+R_2+R_3)}\times\nonumber\\
&\sum_{w,w_{[1:3]}}\Big[ P(a^n|U^n(w),U_1^n(w,w_1),U_2^n(w,w_2),U_3^n(w,w_3))\nonumber\\
&\mathbbm{1}\{ (U^n(w),U_1^n(w,w_1),U_2^n(w,w_2),U_3^n(w,w_3),a^n)\in\mathcal{T}_2^\epsilon\}\Big],\nonumber\\
&P_2(a^n)=2^{-n(R+R_1+R_2+R_3)}\times\nonumber\\
&\sum_{w,w_{[1:3]}}\Big[ P(a^n|U^n(w),U_1^n(w,w_1),U_2^n(w,w_2),U_3^n(w,w_3))\nonumber\\
&\mathbbm{1}\{ (U^n(w),U_1^n(w,w_i),U_2^n(w,w_2),U_3^n(w,w_3),a^n)\notin\mathcal{T}_2^\epsilon\}\Big].\nonumber
\end{align}
Notice that $\mathbbm{E}P(a^n)=q(a^n)$, where the expectation is over the randomness of codebooks. Now, we analyse the total variation distance. Using the triangle inequality, we have 
\begin{align}
\mathbbm{E}\lVert P(a^n)-\mathbbm{E}P(a^n) \rVert_1&\leq \sum_{a^n\in\mathcal{T}_1^\epsilon}\mathbbm{E}\lvert P_1(a^n)-\mathbbm{E}P_1(a^n) \rvert\nonumber\\
& +\sum_{a^n\in\mathcal{T}_1^\epsilon}\mathbbm{E}\lvert P_2(a^n)-\mathbbm{E}P_2(a^n) \rvert\nonumber\\
&+ \sum_{a^n\notin\mathcal{T}_1^\epsilon}\mathbbm{E}\lvert P(a^n)-\mathbbm{E}P(a^n) \rvert\nonumber
\end{align}
It can be easily seen that the second and third terms vanishes asymptotically as shown below.
\begin{align}
&\sum_{a^n\in\mathcal{T}_1^\epsilon}\mathbbm{E}\lvert P_2(a^n)-\mathbbm{E}P_2(a^n) \rvert\nonumber\\
&\leq \sum_{a^n}2\mathbbm{E}P_2(a^n)\nonumber\\
& =2\sum_{a^n}2^{-n(R+R_1+R_2+R_3)}\times\nonumber\\
&\sum_{w,w_{[1:3]}}\mathbbm{E}\Big[ P(a^n|U^n(w),U_1^n(w,w_1),U_2^n(w,w_2),U_3^n(w,w_3))\nonumber\\
&\mathbbm{1}\{ (U^n(w),U_1^n(w,w_i),U_2^n(w,w_2),U_3^n(w,w_3),a^n)\notin\mathcal{T}_2^\epsilon\}\Big]\nonumber\\
& =2\sum_{a^n}\mathbbm{E}\Big[P(a^n|U^n(1),U_1^n(1,1),U_2^n(1,1),U_3^n(1,1))\nonumber\\
&\mathbbm{1}\{ (U^n(1),U_1^n(1,1),U_2^n(1,1),U_3^n(1,1),a^n)\notin\mathcal{T}_2^\epsilon\}\Big]\label{eqn:achv:oblvcode}\\
&=2\sum_{(u^n,u_1^n,u_2^n,u_3^n,a^n)\notin\mathcal{T}_2^\epsilon}p(u^n,u_1^n,u_2^n,u_3^n,a^n)\nonumber\\
&\rightarrow 0\  \text{as}  \ n\rightarrow \infty\nonumber. 
\end{align}
where \eqref{eqn:achv:oblvcode} follows from the symmetry of the codebook construction.
\begin{align}
\sum_{a^n\notin\mathcal{T}_1^\epsilon}\mathbbm{E}\lvert P(a^n)-\mathbbm{E}P(a^n) \rvert&\leq 2\sum_{a^n\notin\mathcal{T}_1^\epsilon}\mathbbm{E}P(a^n)\nonumber\\
&=2\sum_{a^n\notin\mathcal{T}_1^\epsilon}q(a^n)\nonumber\\
&\rightarrow 0\  \text{as}  \ n\rightarrow \infty\nonumber. 
\end{align}
Using Jensen's inequality, the first term can be upper bounded as
\begin{align}
&\sum_{a^n\in\mathcal{T}_1^\epsilon}\mathbbm{E}\lvert P_1(a^n)-\mathbbm{E}P_1(a^n) \rvert\nonumber\\
&\leq \sum_{a^n\in\mathcal{T}_1^\epsilon}\sqrt{\mathbbm{E}{\left( P_1(a^n)-\mathbbm{E}P_1(a^n) \right)}^2}\nonumber\\
&= \sum_{a^n\in\mathcal{T}_1^\epsilon}\sqrt{\mathbbm{E}{\left( P_1(a^n)\right)}^2-{\left(\mathbbm{E}P_1(a^n)\right)}^2}\label{eqn:ach_oblv_final}
\end{align}
$\mathbbm{E}{\left( P_1(a^n)\right)}^2$ can be precisely written as
\begin{align*}
\mathbbm{E}{\left( P_1(a^n)\right)}^2=\sum_{\substack{w,w_1,w_2,w_3,\\w^\prime,w_1^\prime,w_2^\prime,w_3^\prime}}T_{{w,w_1,w_2,w_3,w^\prime,w_1^\prime,w_2^\prime,w_3^\prime}},
\end{align*}
where 
\begin{align}
&T_{{w,w_1,w_2,w_3,w^\prime,w_1^\prime,w_2^\prime,w_3^\prime}}=2^{-2n(R+R_1+R_2+R_3)}\times\nonumber\\
&\mathbbm{E}\Big(P(a^n|U^n(w),U_1^n(w,w_1),U_2^n(w,w_2),U_3^n(w,w_3))\nonumber\\
&P(a^n|U^n(w^\prime),U_1^n(w^\prime,w_1^\prime),U_2^n(w^\prime,w_2^\prime),U_3^n(w^\prime,w_3^\prime))\nonumber\\
&\mathbbm{1}\{ (U^n(w),U_1^n(w,w_1),U_2^n(w,w_2),U_3^n(w,w_3),a^n)\in\mathcal{T}_2^\epsilon\}\nonumber\\
&\mathbbm{1}\{ (U^n(w^\prime),U_1^n(w^\prime,w_1^\prime),U_2^n(w^\prime,w_2^\prime),U_3^n(w^\prime,w_3^\prime),a^n)\in\mathcal{T}_2^\epsilon\}\Big).\nonumber
\end{align}
We divide the above summation into 9 parts each part specified by a case as below.

\noindent Case $(1)$: $w\neq w^\prime$\\
Case $(2)$: $w=w^\prime,w_1=w_1^\prime,w_2=w_2^\prime,w_3= w_3^\prime$\\ 
Case $(3)$: $w=w^\prime,w_1=w_1^\prime,w_2=w_2^\prime,w_3\neq w_3^\prime$\\
Case $(4)$: $w=w^\prime,w_1=w_1^\prime,w_2\neq w_2^\prime,w_3= w_3^\prime$\\
Case $(5)$: $w=w^\prime,w_1\neq w_1^\prime,w_2=w_2^\prime,w_3= w_3^\prime$\\
Case $(6)$: $w=w^\prime,w_1\neq w_1^\prime,w_2\neq w_2^\prime,w_3= w_3^\prime$\\
Case $(7)$: $w=w^\prime,w_1=w_1^\prime,w_2\neq w_2^\prime,w_3\neq  w_3^\prime$\\
Case $(8)$: $w=w^\prime,w_1\neq w_1^\prime,w_2=w_2^\prime,w_3\neq w_3^\prime$\\
Case $(9)$: $w=w^\prime,w_1\neq w_1^\prime,w_2\neq w_2^\prime,w_3\neq w_3^\prime$\\

\noindent Consider case $(1)$. It can be seen that 
\begin{align}
\sum_{\substack{w=w^\prime,w_1,w_2,w_3,\\w_1^\prime,w_2^\prime,w_3^\prime}}T_{w,w_1,w_2,w_3,w^\prime,w_1^\prime,w_2^\prime,w_3^\prime}\leq {\left(\mathbbm{E}P_1(a^n)\right)}^2\label{eqn:newoblv11}.
\end{align}
Consider case $(2)$. The corresponding part equals
\begin{align}
&2^{-2n(R+R_1+R_2+R_3)}\sum_{w,w_1,w_2,w_3}\nonumber\\
&\mathbbm{E}\Big(P^2(a^n|U^n(w),U_1^n(w,w_1),U_2^n(w,w_2),U_3^n(w,w_3))\nonumber\\
&\mathbbm{1}\{ (U^n(w),U_1^n(w,w_i),U_2^n(w,w_2),U_3^n(w,w_3),a^n)\in\mathcal{T}_2^\epsilon\}\Big)\nonumber\\
&=2^{-n(R+R_1+R_2+R_3)}\times\nonumber\\
&\mathbbm{E}\Big(P^2(a^n|U^n(1),U_1^n(1,1),U_2^n(1,1),U_3^n(1,1))\nonumber\\
&\mathbbm{1}\{ (U^n(1),U_1^n(1,1),U_2^n(1,1),U_3^n(1,1),a^n)\in\mathcal{T}_2^\epsilon\}\Big)\label{eqn:oblv_achv:case2}\\
&=2^{-n(R+R_1+R_2+R_3)}\times\nonumber\\
&\sum_{\substack{(u^n,u_1^n,u_2^n,u_3^n):\\(u^n,u_1^n,u_2^n,u_3^n,a^n)\in\mathcal{T}^\epsilon_2}}p ^2(a^n|u^n,u_1^n,u_2^n,u_3^n)p(u^n,u_1^n,u_2^n,u_3^n)\nonumber\\
&\leq 2^{-n(R+R_1+R_2+R_3)} 2^{-n(H(A|U,U_1,U_2,U_3)-\delta_1(\epsilon))}\times\nonumber\\
&\sum_{(u^n,u_1^n,u_2^n,u_3^n)}p(a^n|u^n,u_1^n,u_2^n,u_3^n)p(u^n,u_1^n,u_2^n,u_3^n)\label{eqn:oblv_achv:case21}\\
&= 2^{-n(R+R_1+R_2+R_3)} 2^{-n(H(A|U,U_1,U_2,U_3)-\delta_1(\epsilon))}\times p(a^n)\nonumber\\
&\leq 2^{-n(R+R_1+R_2+R_3+H(A|U,U_1,U_2,U_3)+H(A)-\delta_1(\epsilon)-\delta_2(\epsilon))}\label{eqn:oblv_achv:case22}
\end{align}
where \eqref{eqn:oblv_achv:case2} follows from the symmetry of the codebook construction, \eqref{eqn:oblv_achv:case21} and \eqref{eqn:oblv_achv:case22} follow from the properties of the typical sequences with $\delta_1(\epsilon),\delta_2(\epsilon)\rightarrow 0$ as $\epsilon\rightarrow 0$ (note that \eqref{eqn:oblv_achv:case22} holds only for typical $a^n$ sequences). Now, consider case $(3)$. Using the symmetry of the codebook construction and noting that the corresponding part contains $2^{n(R+R_1+R_2+2R_3)}$ number of terms, it equals 
\begin{align}
&2^{-n(R+R_1+R_2)}\times\nonumber\\
&\mathbbm{E}\Big(P(a^n|U^n(1),U_1^n(1,1),U_2^n(1,1),U_3^n(1,1))\nonumber\\
&P(a^n|U^n(1),U_1^n(1,1),U_2^n(1,1),U_3^n(1,2))\nonumber\\
&\mathbbm{1}\{ (U^n(1),U_1^n(1,1),U_2^n(1,1),U_3^n(1,1),a^n)\in\mathcal{T}_2^\epsilon\}\nonumber\\
&\mathbbm{1}\{ (U^n(1),U_1^n(1,1),U_2^n(1,1),U_3^n(1,2),a^n)\in\mathcal{T}_2^\epsilon\}\Big).\nonumber\\
&=2^{-n(R+R_1+R_2)}\sum_{\substack{(u^n,u_1^n,u_2^n,u_3^n,\bar{u}_3^n):\\(u^n,u_1^n,u_2^n,u_3^n,a^n)\in\mathcal{T}^\epsilon_2\\ (u^n,u_1^n,u_2^n,\bar{u}_3^n,a^n)\in\mathcal{T}^\epsilon_2}}\Big[p(a^n|u^n,u_1^n,u_2^n,u_3^n)\nonumber\\
&p(a^n|u^n,u_1^n,u_2^n,\bar{u}_3^n)p(u^n,u_1^n,u_2^n)p(u_3^n)p(\bar{u}_3^n)\Big]\nonumber\\
&\leq2^{-n(R+R_1+R_2)}\sum_{\substack{(u^n,u_1^n,u_2^n):\\(u^n,u_1^n,u_2^n,a^n)\in\mathcal{T}^\epsilon_3}}\Bigg[ \nonumber\\
&\Big(\sum_{u_3^n}p(a^n|u^n,u_1^n,u_2^n,u_3^n)p(u_3^n)\Big)\nonumber\\
&\Big(\sum_{\bar{u}_3^n}p(a^n|u^n,u_1^n,u_2^n,\bar{u}_3^n)p(\bar{u}_3^n)\Big)p(u^n,u_1^n,u_2^n)\Bigg]\label{eqn:oblv_achv_case2newt}\\
&=2^{-n(R+R_1+R_2)}\times\nonumber\\
&\Big[\sum_{\substack{(u^n,u_1^n,u_2^n):\\(u^n,u_1^n,u_2^n,a^n)\in\mathcal{T}^\epsilon_3}}p^2(a^n|u^n,u_1^n,u_2^n)p(u^nu_1^n,u_2^n)\Big]\nonumber\\
&\leq 2^{-n(R+R_1+R_2)} 2^{-n(H(A|U,U_1,U_2)-\delta_3(\epsilon))}\times\nonumber\\
&\sum_{(u^n,u_1^n,u_2^n)}p(a^n|u^n,u_1^n,u_2^n)p(u^n,u_1^n,u_2^n)\label{eqn:oblv_achv:case23}\\
&= 2^{-n(R+R_1+R_2)} 2^{-n(H(A|U,U_1,U_2)-\delta_3(\epsilon))}\times p(a^n)\nonumber\\
&\leq 2^{-n(R+R_1+R_2+H(A|U,U_1,U_2)+H(A)-\delta_3(\epsilon)-\delta_4(\epsilon))},\label{eqn:oblv_achv:case24}
\end{align}
where \eqref{eqn:oblv_achv_case2newt} follows by defining $\mathcal{T}^\epsilon_3$ as the $\epsilon$-typical set with distribution $p_{UU_1U_2A}$, \eqref{eqn:oblv_achv:case23} and \eqref{eqn:oblv_achv:case24} follow from the properties of typical sequences with $\delta_3(\epsilon),\delta_4(\epsilon)\rightarrow 0$ as $\epsilon\rightarrow 0$ (note that \eqref{eqn:oblv_achv:case24} hold only for typical $a^n$ sequences). Other cases can also be dealt similarly giving us that the parts corresponding to cases $(4), (5), (6), (7), (8), (9)$ are respectively less than or equal to
\begin{align}
&2^{-n(R+R_1+R_3+H(A|U,U_1,U_3)+H(A)-\delta(\epsilon))},\label{eqn:oblv_ach_case4}\\
&2^{-n(R+R_2+R_3+H(A|U,U_2,U_3)+H(A)-\delta(\epsilon))},\label{eqn:oblv_ach_case5}\\
&2^{-n(R+R_1+H(A|U,U_1)+H(A)-\delta(\epsilon))},\label{eqn:oblv_ach_case6}\\
&2^{-n(R+R_2+H(A|U,U_2)+H(A)-\delta(\epsilon))},\label{eqn:oblv_ach_case7}\\
&2^{-n(R+R_3+H(A|U,U_3)+H(A)-\delta(\epsilon))}\ \text{and}\label{eqn:oblv_ach_case8}\\
&2^{-n(R+H(A|U)+H(A)-\delta(\epsilon))}\label{eqn:oblv_ach_case9}
\end{align}
with $\delta(\epsilon)\rightarrow 0$ as $\epsilon\rightarrow 0$. Now, substituting \eqref{eqn:newoblv11}, \eqref{eqn:oblv_achv:case22}, \eqref{eqn:oblv_achv:case24}-\eqref{eqn:oblv_ach_case9} in \eqref{eqn:ach_oblv_final} and using the bounds $|\mathcal{T}^\epsilon_1|\leq 2^{n(H(A)+\delta^\prime(\epsilon))}$ and $\sqrt{x+y}\leq \sqrt{x}+\sqrt{y}$, it can be seen that if
\begin{align}
R+R_{\mathcal{S}}> I(A;U,U_{\mathcal{S}}),\ \mathcal{S}\subseteq [1:3],
\end{align} 
then $ \sum_{a^n\in\mathcal{T}_1^\epsilon}\mathbbm{E}\lvert P_1(a^n)-\mathbbm{E}P_1(a^n) \rvert_1\rightarrow 0$ as $n\rightarrow 0$.

For the converse, suppose a rate tuple $(R,R_1,R_2,R_3)$ is achievable for $q_{X_1X_2X_3}$. For any $\mathcal{S}\subseteq[1:3]$, consider
\begin{align}
&n(R+R_{\mathcal{S}})\nonumber\\
&\geq H(W,W_{\mathcal{S}})\nonumber\\
&\geq I(W,W_{\mathcal{S}};X_1^n,X_2^n,X_3^n)\nonumber\\
&=H(X_1^n,X_2^n,X_3^n)-H(X_1^n,X_2^n,X_3^n|W,W_{\mathcal{S}})\nonumber\\
&\geq \sum_{i=1}^n\left[H(X_{1i},X_{2i},X_{3i})-\epsilon^\prime \right]\nonumber\\ 
&\hspace{12pt}-\sum_{i=1}^n H(X_{1i},X_{2i},X_{3i}|W,W_{\mathcal{S}},X_1^{1:i-1},X_2^{1:i-1},X_3^{1:i-1})\label{oblv1}\\
&= \sum_{i=1}^n\left[I(X_{1i},X_{2i},X_{3i};W,W_{\mathcal{S}},X_1^{1:i-1},X_2^{1:i-1},X_3^{1:i-1})\right]\nonumber\\
&\hspace{12pt}-n\epsilon^\prime\nonumber\\
&\geq \sum_{i=1}^n\left[I(X_{1i},X_{2i},X_{3i};W,W_{\mathcal{S}})-\epsilon^\prime\right]\nonumber\\
&= n\left[I(X_{1Q},X_{2Q},X_{3Q};W,W_{\mathcal{S}}|Q)-\epsilon^\prime\right]\nonumber\\
&\geq n\left[I(X_{1Q},X_{2Q},X_{3Q};W,W_{\mathcal{S}},Q)-\epsilon^\prime-\epsilon^{\prime\prime} \right] \label{oblv2}\\
& =n[I(X_{1Q},X_{2Q}.X_{3Q};U,U_{\mathcal{S}})-\epsilon^\prime-\epsilon^{\prime\prime}]\label{eqn:oblv_defn}
\end{align}
where \eqref{oblv1} and \eqref{oblv2} follow from the correctness of the output distribution with $\epsilon^\prime ,\epsilon^{\prime\prime}\rightarrow 0$ as $\epsilon\rightarrow 0$ along similar lines as \eqref{eqn:converse_closeness_of_dist} and \eqref{eqn:T_independent}, respectively, and \eqref{eqn:oblv_defn} follows by defining $U=(W,Q), U_i=W_i$, for $i\in[1:3]$. Note that $\lVert p_{X_{1Q}X_{2Q}X_{3Q}}-q_{X_1X_2X_3} \rVert<\epsilon$, which follows from Cuff~\cite[Lemma~VI.2]{Cuff13}. Using the structure of the problem (i.e., oblivious coordinator and that $\mathcal{V}_i=[1:3]\setminus\{i\}$) and the continuity of total variation distance and mutual information in the probability simplex, it follows along the same lines as Theorem~\ref{theorem:unlimited_shared_randomness} and Yassaee et al.~\cite[Lemma~6]{YassaeeGA15} that
\begin{align}
R+R_{\mathcal{S}}\geq I(X_1,X_2,X_3;U,U_{\mathcal{S}}), \ \mathcal{S}\subseteq [1:3] \nonumber
\end{align} 
for some p.m.f. $$p(u,u_{[1:3]},x_{[1:3]})=q(x_{[1:3]})p(u,u_{[1:3]}|x_{[1:3]})$$ s.t.
$$p(u,u_{[1:3]},x_{[1:3]})=p(u)\left(\prod_{i=1}^3 p(u_i)\right)\left(\prod_{i=1}^np(x_i|u,u_{[1:3]\setminus\{i\}})\right) \label{eqn:oblv_struc}.$$
This completes the proof.
\end{IEEEproof}
 In the above proof, we remark that, the analysis of total variation distance does not depend on how the processors share random variables, i.e., the same part of the proof works even for an arbitrary $\mathcal{V}=(\mathcal{V}_i)_{i\in[1:3]}$ as long as $h=3$. In fact, the above theorem can be readily extended to $t>3, h>3$ and arbitrary $\mathcal{V}$ as follows.
 \begin{thm}\label{theorem_oblv_general}
 For the oblivious coordinator setting, the simulation rate region is given by the set of all non-negative rate tuples $(R,R_1,\dots,R_t)$ such that
 \begin{align}
 R+R_{\mathcal{S}}\geq I(X_{[1:t]};U,U_{\mathcal{S}}),\ \mathcal{S}\subseteq [1:h],\label{eqn:oblv_genthm}
 \end{align}
 for some p.m.f. $$p(u,u_{[1:h]},x_{[1:t]})=q(x_{[1:t]})p(u,u_{[1:h]}|x_{[1:t]})$$
 s.t.
 \begin{align*}
 p(u,u_{[1:h]},x_{[1:t]})=p(u)\left(\prod_{i=1}^hp(u_i) \right)\left(\prod_{i=1}^tp(x_i|u,u_{\mathcal{V}_i})\right).
 \end{align*}
  \end{thm} 
 \begin{remark}
  Theorem~\ref{theorem_oblv_general} recovers multi-user Wyner's common information~\cite{Liu2010} in the absence of shared randomness.
  \end{remark}
  
  The proof of Theorem~\ref{theorem_oblv_general} is similar to the proof of Theorem~\ref{theorem_oblv_3users}. Appendix~\ref{appendix:oblv:general} contains a proof outline.

\section{Conclusion}\label{conclusion}
We studied the role of shared randomness in coordination. We considered various coordination problems involving shared randomness and obtained tight expressions for optimal communication and shared randomness rates. We confined our attention only to the distributed sampling problem which is a special case of the more general setting where some of the users have inputs and all the users want to output samples from a desired distribution conditioned on the inputs~\cite{Cuff13}. The two main resources that aid users in achieving this coordination are the shared randomness and the underlying communication network.
For the most part, we restricted our attention to independent sources of shared randomness. Instead, it is of interest to study settings with correlated sources of shared randomness. In Section~\ref{section:assisted}, we studied one such model. Coming to the communication network, we considered models where only one user (i.e., coordinator) transmits a message to other users (star topology). More generally, it might be interesting to study generic network topologies (e.g., combination networks, hybrid networks). Even more generally, we might consider coordination/distributed computation over multiple-input and multiple-output (MIMO) channels. With this generality, it might be quite challenging to study coordination as is already evident from the fact that source-channel separation does not necessarily hold (see, e.g., Nazer and Gastpar~\cite{NazerG07a}).

In the omniscient coordinator setting, we confined our attention mainly to the individually shared randomness model and the randomness-on-the-forehead model. It would be interesting to study if there are any other models in the omniscient coordinator setting for which closed form rate expressions can be obtained. One more limitation of our study of the omniscient coordinator setting is that whenever the processors have to output dependent random variables, we have assumed that shared randomness rates are sufficiently large and analyzed only the communication rates. Characterizing the trade-off between communication and shared randomness rates, as done in the oblivious coordinator setting (Theorem~\ref{theorem_oblv_general}),  remains open. In the omniscient coordinator with correlated shared randomness model, we have studied only the scenario when all the processors output equal random variables. It might be more challenging to study these settings in the general scenario where the processors may output dependent random variables instead of equal random variables. 

Also, it might be interesting to solve the optimization problem for the optimal transmission rate in the individually shared randomness model in Theorem~\ref{theorem:unlimited_shared_randomness} at least for the $\DSBS$. To that end, proving/disproving the conjecture presented at the end of Example~\ref{open} might give some insights. Even though the upper bound on the optimal communication rate for the randomness-on-the-forehead model turns out to be tight for some special case, it remains open whether it is tight in general. Furthermore, obtaining closed form expressions for optimal transmission rate in omniscient coordinator setting with general shared randomness access structures may also be of interest.

\if \arxive 1
\begin{appendices}
\section{Details Omitted From Proof of Theorem~\ref{theorem:unlimited_shared_randomness}}\label{details_omitted}
\textbf{Explanation for \eqref{eqn:converse_closeness_of_dist}}.
\begin{align}
&H(X^n,Y^n)-H(X^n,Y^n|M)\nonumber\\
&=H_p(X^n,Y^n)-H_p(X^n,Y^n|M)\nonumber\\
&\geq H_q(X^n,Y^n)-n\epsilon_1-H_p(X^n,Y^n|M)\label{eqn:converse_closeness_of_dist_1}\\
&=\sum_{i=1}^n [H_q(X_i,Y_i)-\epsilon_1]-\sum_{i=1}^n H_p(X_i,Y_i|M,X^{i-1},Y^{i-1})\nonumber\\
&\geq\sum_{i=1}^n [H_p(X_i,Y_i)-\epsilon_1-\epsilon_2]\nonumber\\
&\hspace{1cm}-\sum_{i=1}^n H_p(X_i,Y_i|M,X^{i-1},Y^{i-1})\label{eqn:converse_closeness_of_dist_2}\\
&=\sum_{i=1}^n [H(X_i,Y_i)-\epsilon_1-\epsilon_2]\nonumber\\
&\hspace{1cm}-\sum_{i=1}^n H(X_i,Y_i|M,X^{i-1},Y^{i-1})\nonumber\\
&=\sum_{i=1}^n [H(X_i,Y_i)-\epsilon^\prime]-\sum_{i=1}^n H(X_i,Y_i|M,X^{i-1},Y^{i-1})\label{eqn:converse_closeness_of_dist_3}.
\end{align}
We used the following fact in \eqref{eqn:converse_closeness_of_dist_1} and \eqref{eqn:converse_closeness_of_dist_2}: if two random variables $A$ and $A'$ with 
  same support set $\mathcal{A}$  satisfy $||p_{A} -  p_{A'}||_{1} \leq \epsilon \leq 1/4 $, then it follows from from standard results~\cite[Theorem 17.3.3]{Cover} that $|H(A) - H(A')|\leq \eta \log |\mathcal{A}|$, where $\eta \rightarrow 0$ as $\epsilon \rightarrow 0$. Now \eqref{eqn:correctness} implies \eqref{eqn:converse_closeness_of_dist_1}, where $\epsilon_1\rightarrow 0$ as $\epsilon\rightarrow 0$. Also, note that \eqref{eqn:correctness} implies $\lVert p_{X_i,Y_i}-q_{X,Y} \rVert_1\leq \epsilon$, $\forall i\in [1:n]$, which implies \eqref{eqn:converse_closeness_of_dist_2}, where $\epsilon_2\rightarrow 0$ as $\epsilon\rightarrow 0$. In \eqref{eqn:converse_closeness_of_dist_3}, $\epsilon^\prime:=\epsilon_1+\epsilon_2$.

\textbf{Explanation for \eqref{eqn:T_independent}}. 
\begin{align}
I(X_T,Y_T;T)&=H_p(X_T,Y_T)-H_p(X_T,Y_T|T)\nonumber\\
&\leq H_q(X_T,Y_T)+\delta_1-\frac{1}{n}\sum_{i=1}^n H_p(X_i,Y_i|T=i)\label{eqn:T_independent_1}\\
&= H_q(X_T,Y_T)+\delta_1-\frac{1}{n}\sum_{i=1}^n H_p(X_i,Y_i)\nonumber\\
&\leq H_q(X_T,Y_T)+\delta_1-\frac{1}{n}\sum_{i=1}^n[H_q(X_i,Y_i)-\delta_2]\label{eqn:T_independent_2}\\
&=H_q(X_T,Y_T)-H_q(X_T,Y_T)+\delta_1+\delta_2\nonumber\\
&=H_q(X_T,Y_T)-H_q(X_T,Y_T)+\delta\label{eqn:T_independent_3}\\
&\leq \delta\nonumber.
\end{align}
We used the following fact in \eqref{eqn:T_independent_1} and \eqref{eqn:T_independent_2}: if two random variables $A$ and $A'$ with 
  same support set $\mathcal{A}$  satisfy $||p_{A} -  p_{A'}||_{1} \leq \epsilon \leq 1/4 $, then it follows from standard results~\cite[Theorem 17.3.3]{Cover} that $|H(A) - H(A')|\leq \eta \log |\mathcal{A}|$, where $\eta \rightarrow 0$ as $\epsilon \rightarrow 0$. Now using Cuff~\cite[Lemma VI.2]{Cuff13}, \eqref{eqn:correctness} implies $\lVert p_{X_T,Y_T}-q_{X,Y}\rVert_1\leq\epsilon$, which implies \eqref{eqn:T_independent_1} and \eqref{eqn:T_independent_2}. In \eqref{eqn:T_independent_3}, we defined $\delta:=\delta_1+\delta_2$, where $\delta\rightarrow 0$ as $\epsilon\rightarrow 0$.
  
 \begin{lemma}\label{lemma:RL=RU}
\begin{align*}
&\min \max \left\{I(X;Y|U), \frac{1}{2}\big(I(X,Y;U)+I(X;Y|U)\big)\right\}\\
&\hspace{0pt}=\min \max \big\{I(X;Y|U),I(X,Y;U)\big\},
\end{align*}  
where the minimum is over all conditional p.m.f.'s $p(u|x,y)$ with $|\mathcal{U}|\leq |\mathcal{X}||\mathcal{Y}|+2$ in both the L.H.S and R.H.S.
\end{lemma}
\begin{IEEEproof}
Firstly, we define
\begin{multline}
R_U:=\\
\min_{p(u|x,y)}\max\left\{I(X;Y|U),\frac{1}{2}\big(I(X,Y;U)+I(X;Y|U)\big)\right\},\label{eqn:RU_opt_pblm}
\end{multline}
\begin{align}
R_L:&=\min_{{p(u|x,y)}}\max\big\{I(X;Y|U),I(X,Y;U)\big\}\label{eqn:RL_opt_pblm}.
\end{align}
It is trivial to see that $R_L\geq R_U$. Since $R_L\leq R_{\text{opt}}\leq R_U$, we have $R_L\leq R_U$ also and hence $R_L=R_U$. We can see the inequality $R_L\leq R_U$ directly also in the following way. For simplicity, we abbreviate $p(u|x,y)$ by $p$, $I(X;Y|U)$ by $f_1(p)$ and $I(X,Y;U)$ by $f_2(p)$ in the following.

Assume without loss of generality that $I(X;Y)\neq 0$, since otherwise $R_L=R_U=0$. Notice that there always exists a minimizer for the minimization problem in \eqref{eqn:RU_opt_pblm} since a continuous function (objective function of the minimization problem in this case) attains its minimum on a compact set (the set of all conditional p.m.f.s $p(u|x,y)$ with $|\mathcal{U}|\leq |\mathcal{X}||\mathcal{Y}|+2$ in this case). We argue that there must exist a minimizer $p_{\text{min}}:=p(u_{\text{min}}|x,y)$ for \eqref{eqn:RU_opt_pblm} such that
\begin{align}
f_1(p_{\text{min}})\geq \frac{f_1(p_{\text{min}})+f_2(p_{\text{min}})}{2}\label{eqn:lemma_RL=RU_opt}.
\end{align} 
Once we have such a minimizer, it follows that
\begin{align}
R_U&=f_1(p_{\text{min}})\label{eqn:lemma_RL=RU_opt_1}\\
&=\max\left\{f_1(p_{\text{min}}),f_2(p_{\text{min}})\right\}\label{eqn:lemma_RL=RU_opt_2}\\
&\geq R_L\label{eqn:lemma_RL=RU_opt_3},
\end{align}
where \eqref{eqn:lemma_RL=RU_opt_1}-\eqref{eqn:lemma_RL=RU_opt_2} follow from \eqref{eqn:lemma_RL=RU_opt} and \eqref{eqn:lemma_RL=RU_opt_3} follows from definition of $R_L$ in \eqref{eqn:RL_opt_pblm}.

To prove the claim made in \eqref{eqn:lemma_RL=RU_opt} we start with a minimizer, $p^*:=p(u^*|x,y)$ of \eqref{eqn:RU_opt_pblm}. Let $Q$ be a binary random variable independent of $(X,Y,U^*)$ with pmf $p(Q=1)=\theta=1-p(Q=0)$, where $\theta\in(0,1]$ will be fixed later. Let $U_0=U^*$ and $U_1=k$ (a constant random variable) and $U^{\prime}:=(U_Q,Q)$ and denote $p(u^\prime|x,y)$ by $p^\prime$. Note that $|\mathcal{U}^\prime|$ may be greater than $|\mathcal{X}||\mathcal{Y}|+2$ but, an application of Convex Cover Method \cite[Appendix C]{GamalK12} guarantees us another such pmf with $|\mathcal{U}^\prime|\leq|\mathcal{X}||\mathcal{Y}|+2$ preserving $I(X;Y|U^\prime)$ and $I(X,Y;U^\prime)$. Now we consider two cases.

\underline{\emph{Case (i)}}  $\left(I(X;U^*|Y)+I(Y;U^*|X)\neq 0\right):$

In this case, we show that $p^*$ itself must satisfy the condition, $f_1(p^*)\geq \frac{f_1(p^*)+f_2(p^*)}{2}$. We prove this by a contradiction. Suppose $f_1(p^*)<\frac{f_1(p^*)+f_2(p^*)}{2}$. We have
\begin{align}
&\frac{f_1(p^\prime)+f_2(p^\prime)}{2}\\
&=\frac{1}{2}\left[I(X;Y|U^\prime)+I(X,Y;U^\prime)\right]\nonumber\\
&=\frac{1}{2}\left[I(X;Y)+I(X;U^\prime|Y)+I(Y;U^\prime|X)\right]\label{eqn:lemma:RL=RU_case1_1}\\
&=\frac{1}{2}\left[I(X;Y)+I(X;U_Q,Q|Y)+I(Y;U_Q,Q|X)\right]\nonumber\\
&=\frac{1}{2}\left[I(X;Y)+I(X;U_Q|Q,Y)+I(Y;U_Q|Q,X)\right]\label{eqn:lemma:RL=RU_case1_2}\\
&=\frac{1}{2}\left[I(X;Y)+(1-\theta)\{I(X;U^*|Y)+I(Y;U^*|X)\}\right]\label{eqn:lemma:RL=RU_case1_5}\\
&<\frac{1}{2}\left[I(X;Y)+I(X;U^*|Y)+I(Y;U^*|X)\right]\label{eqn:lemma:RL=RU_case1_3}\\
&=\frac{1}{2}\left[I(X;Y|U^*)+I(X,Y;U^*)\right]\label{eqn:lemma:RL=RU_case1_4}\\
&=\frac{f_1(p^*)+f_2(p^*)}{2}\nonumber,
\end{align}
where \eqref{eqn:lemma:RL=RU_case1_1} and \eqref{eqn:lemma:RL=RU_case1_4} follow from the fact that $I(X;Y|W)+I(X,Y;W)=I(X;Y)+I(X;W|Y)+I(Y;W|X)$, \eqref{eqn:lemma:RL=RU_case1_2} and \eqref{eqn:lemma:RL=RU_case1_5} follow since $Q$ is independent of $(X,Y,U^*)$, \eqref{eqn:lemma:RL=RU_case1_3} follows since $\theta>0$ and $I(X;U^*|Y)+I(Y;U^*|X)\neq 0$. 

Also,
\begin{align}
f_1(p^\prime)&=I(X;Y|U_Q,Q)\nonumber\\
&=\theta I(X;Y)+(1-\theta)f_1(p^*)\label{eqn1:f1(p`)}.
\end{align}
Let $\frac{f_1(p^*)+f_2(p^*)}{2}-f_1(p^*)=\Delta>0$ and we set $\theta=\min\left\{\frac{\Delta}{2I(X;Y},1\right\}$. Now, \eqref{eqn1:f1(p`)} implies that
\begin{align}
f_1(p^\prime)&\leq\frac{\Delta}{2}+f_1(p^*)\nonumber\\
&=\frac{f_1(p^*)+f_2(p^*)}{2}-\frac{\Delta}{2}\label{eqn:lemma:RL=RU_delta}\\
&<\frac{f_1(p^*)+f_2(p^*)}{2},\label{eqn1:f1}
\end{align}
where \eqref{eqn:lemma:RL=RU_delta} follows since $\frac{f_1(p^*)+f_2(p^*)}{2}-f_1(p^*)=\Delta$, \eqref{eqn1:f1} follows since $\Delta>0$.

Now, \eqref{eqn:lemma:RL=RU_case1_4} and \eqref{eqn1:f1} imply that,
\begin{align*}
\max \left\{f_1(p^\prime),\frac{f_1(p^\prime)+f_2(p^\prime)}{2}\right\}<\frac{f_1(p^*)+f_2(p^*)}{2},
\end{align*}
which is a contradiction since $p^*$ is assumed to be minimizer for \eqref{eqn:RU_opt_pblm} such that $f_1(p^*)<\frac{f_1(p^*)+f_2(p^*)}{2}$. Hence, $f_1(p^*)\geq \frac{f_1(p^*)+f_2(p^*)}{2}$.

\underline{\emph{Case(ii)}}  $\left(I(X;U^*|Y)=0=I(Y;U^*|X)\right):$

If  $f_1(p^*)\geq \frac{f_1(p^*)+f_2(p^*)}{2}$, there is nothing to prove. Suppose $f_1(p^*)<\frac{f_1(p^*)+f_2(p^*)}{2}$. Then, we show that $p^\prime$ defined before is also a minimizer and satisfies the condition, $f_1(p^\prime)\geq \frac{f_1(p^\prime)+f_2(p^\prime)}{2}$. We have 
\begin{align}
I(X;Y|U^*)&=f_1(p^*)\nonumber\\
&<\frac{f_1(p^*)+f_2(p^*)}{2}\nonumber\\
&=\frac{1}{2}\left[I(X;Y|U^*)+I(X,Y;U^*)\right]\nonumber\\
&=\frac{1}{2}\left[I(X;Y)+I(X;U^*|Y)+I(Y;U^*|X)\right]\label{eqn:lemma:RL=RU_case2_1}\\
&=\frac{I(X;Y)}{2}\label{eqn:lemma:RL=RU_case2_2},
\end{align}
where \eqref{eqn:lemma:RL=RU_case2_1} follows from the fact that $I(X;Y|W)+I(X,Y;W)=I(X;Y)+I(X;W|Y)+I(Y;W|X)$, \eqref{eqn:lemma:RL=RU_case2_2} follows since $I(X;U^*|Y)=0=I(Y;U^*|X)$.

Let $I(X;Y|U^*)=\alpha\frac{I(X;Y)}{2}$, where $\alpha\in[0,1)$.
We select $\theta$ such that $f_1(p^\prime)=\frac{I(X;Y)}{2}$, i.e.,
\begin{align*}
&I(X;Y|U^\prime)=\frac{I(X;Y)}{2}\\
&\Rightarrow I(X;Y|U_Q,Q)=\frac{I(X;Y)}{2}\\
&\Rightarrow \theta I(X;Y)+(1-\theta)I(X;Y|U^*)=\frac{I(X;Y)}{2}\\
&\Rightarrow \theta I(X;Y)+(1-\theta)\alpha\frac{I(X;Y)}{2}=\frac{I(X;Y)}{2},\\
&\Rightarrow \theta=\frac{1-\alpha}{2-\alpha}.
\end{align*} 
Now, we have
\begin{align}
&f_1(p^\prime)+f_2(p^\prime)\\
&=\left[I(X;Y|U_Q,Q)+I(X,Y;U_Q,Q)\right]\nonumber\\
&=\left[I(X;Y)+I(X;U_Q,Q|Y)+I(Y;U_Q,Q|X)\right]\label{eqn_case_2_4}\\
&=I(X;Y)\label{eqn:case2_3}.
\end{align}
where \eqref{eqn_case_2_4} follows from the fact that $I(X;Y|W)+I(X,Y;W)=I(X;Y)+I(X;W|Y)+I(Y;W|X)$, \eqref{eqn:case2_3} follows since $I(X;U_Q,Q|Y)=0=I(Y;U_Q,Q|X)$.

Since $f_1(p^\prime)=\frac{I(X;Y)}{2}$, \eqref{eqn:case2_3} implies that $f_2(p^\prime)=\frac{I(X;Y)}{2}$. So, $p^\prime$ is also a minimzer since
\begin{align}
\max\left\{f_1(p^\prime),\frac{f_1(p^\prime)+f_2(p^\prime)}{2}\right\}&=\max\left\{\frac{I(X;Y)}{2},\frac{I(X;Y)}{2}\right\}\nonumber\\
&=\frac{I(X;Y)}{2}\nonumber\\
&=R_U\label{eqn:lemma:RL=RU_case2_3},
\end{align}
where \eqref{eqn:lemma:RL=RU_case2_3} holds because $\frac{f_1(p^*)+f_2(p^*)}{2}>f_1(p*)$ and $\frac{f_1(p^*)+f_2(p^*)}{2}=\frac{I(X;Y)}{2}$.

Hence, $p^\prime$ is also a minimizer and $f_1(p^\prime)=\frac{f_1(p^\prime)+f_2(p^\prime)}{2}$.
This concludes the proof.
\end{IEEEproof}  
\section{Details Omitted From the Proof of Theorem~\ref{theorem:generalize}}\label{appendix:wynerstyle}
We first outline the achievability proof of Theorem~\ref{theorem:generalize}. It generalizes the idea behind the proof of Theorem~1.
Let $(U^n,U_1^n,\dots,U_t^n$, $X_1^n,\dots,X_t^n)$ be i.i.d. with distribution $p(u,u_{[1:t]},x_{[1:t]})=q(x_{[1:t]})p(u,u_{[1:h]}|x_{[1:t]})$ satisfying 
\begin{align}
p(u,u_{[1:t]},x_{[1:t]})=p(u,u_{[1:h]})\prod_{i=1}^t p(x_i|u,u_{i})\label{eqn:thm5_finalpmf}.
\end{align}
Bin indices $f,m^*,b_{[1:t]}$ with respective rates $\hat{R}_0,R^*,\tilde{R}_1,\dots,\tilde{R}_t$ are created from $(U^n,U_1^n,\dots,U_t^n)$ in a way that can be understood from the following joint probability distribution:
\begin{align}
&P(u^n,u_1^n,\dots,u_t^n,x_1^n,\dots,x_t^n,f,m^*,b_1,\dots,b_t)\nonumber\\
&=p(u^n,u_1^n,\dots,u_t^n)P(f|u^n)P(m^*|u^n)\left(\prod_{j=1}^hP(b_j|u^n,u_j^n)\right)\nonumber\\
&\hspace{0.6cm}\times \left(\prod_{i=1}^tp(x_i^n|u^n,u_{i}^n)\right)\nonumber\\
&=P(b_1,\dots,b_t,f)P(u^n,u_1^n,\dots,u_h^n|b_1,\dots,b_t,f)P(m^*|u^n)\nonumber\\
&\hspace{0.6cm}\times \left(\prod_{i=1}^tp(x_i^n|u^n,u_{i}^n)\right)\label{eqn:thm5_pmf}.
\end{align}
Further, we use Slepian-Wolf decoders to estimate $(u^n,u_{i})$ from $b_{i},f,m^*$, $i\in[1:t]$. This can be seen as a generalization of the random binning scheme in the proof of Theorem~\ref{theorem:achievability} to multiple processors\footnote{Notice that the bin indices corresponding to $f_1,f_2,m_0$ of Theorem~\ref{theorem:achievability} does not show up here. This is because it turns out that in the setting of unlimited shared randomness, the bound on transmission rate does not get affected in the absence of these bin indices.}. Now we impose a series of constraints on the rates. 
\begin{align}
\hat{R_0}+\tilde{R_\mathcal{S}}< H(U,U_\mathcal{S}), \ \mathcal{S}\subseteq [1:t]\label{eqn:thm5_appendix_1}, 
\end{align}
\begin{align}
 \tilde{R_i}& \geq H(U_i|U),\nonumber\\
 \tilde{R_i}+\hat{R_0}+R^*& \geq H(U,U_i)\ \text{for}\  i\in[1:t],\label{eqn:thm5_appendix_2}
\end{align}
\begin{align}
\hat{R_0}<H(U|X_1,\dots,X_t)\label{eqn:thm5_appendix_3},
\end{align}
The first set of constraints \eqref{eqn:thm5_appendix_1} (analogous to \eqref{eqn1:1} in the proof of Theorem~\ref{theorem:achievability}) ensure that $b_1,\dots,b_h,f$ are approximately (i.e., with vanishing total variation distance) uniformly distributed and mutually independent of each other~\cite[Theorem~1]{YassaeeAG14}. The second set of constraints~\eqref{eqn:thm5_appendix_2} (analogous to \eqref{eqn1:3}) guarantees the success of Slepian-Wolf decoders with high probability~\cite[Lemma~1]{YassaeeAG14}. The third set of constraints \eqref{eqn:thm5_appendix_3} (analogous to \eqref{eqn1:8}) implies that $(X_1^n,\dots,X_t^n)$ is approximately independent of $F$. All these three sets of rate constraints ensures the existence of a sequence of simulation codes with a particular realization of the binning resulting in desired vanishing total variation distance as in the proof of Theorem~\ref{theorem:achievability}. Now we eliminate the rates $\hat{R_0},\tilde{R_i}$, for $i\in[1:t]$.
Notice that we can assume that the constraints in \eqref{eqn:thm5_appendix_2} hold with equality, because we can reduce the rates $\hat{R_0},\tilde{R_1}$, for $i\in[1:t]$, to get equalities in \eqref{eqn:theorem:achievability_gather_2} without disturbing the other constraints. This leads to  
\begin{align}
\tilde{R_i}&=H(U_i|U),\ \text{for} \ i\in[1:t]\nonumber,\\
R^*+\hat{R_0}&=H(U)\label{eqn:thm5_appendix_4}.
\end{align}
Substituting \eqref{eqn:thm5_appendix_4} in \eqref{eqn:thm5_appendix_1} and \eqref{eqn:thm5_appendix_3} gives the following constraints after ignoring the redundant inequalities.
\begin{align}
R^*&>I(U_1;\dots;U_t|U),\nonumber\\
R^*&>I(X_1,\dots,X_t;U)\label{eqn:thm5_ref},
\end{align}
where $I(U_1;\dots,U_t|U)$ is the Watanabe's total correlation in \eqref{eq:watanabe}. Noticing that $U_i=X_i$ for $i\in[1:t]$ satisfies the condition \eqref{eqn:thm5_finalpmf} for any p.m.f. $p_{U|X_1,\dots,X_t}$ and using $R=R^*$, \eqref{eqn:thm5_ref} gives us that $R_{\mrm{opt}}^{\mrm{Indv}}\leq \min \max \big\{I(X_1;\dots;X_t|U), I(X_1,\dots,X_t;U)\big\}$, where the minimum is over all p.m.f.'s $p(u|x_1,\dots,x_t)$. This completes the achievability.

Now we show that $R\geq I(X_{1Q},\dots,X_{tQ};U_Q,Q)-g(\epsilon)$, whose proof is omitted from the converse.
\begin{align}
nR&\geq H(M)\nonumber\\
&\geq I(X_1^n,\dots,X_t^n;M)\nonumber\\
&=H(X_1^n,\dots,X_t^n)-H(X_1^n,\dots,X_t^n|M)\nonumber\\
&\geq H_{q^{(n)}}(X_1^n,\dots,X_t^n)-n\epsilon_1-H(X_1^n,\dots,X_t^n|M)\label{eqn:appendix:wyner1}\\
&=\sum_{j=1}^n[H_{q^{(n)}}(X_{1j},\dots,X_{tj})-\epsilon_1]-\nonumber\\
&\hspace{1cm}\sum_{j=1}^nH(X_{1j},\dots,X_{tj}|M,X_1^{1:j-1},\dots,X_t^{1:j-1})\nonumber\\
&\geq \sum_{j=1}^n[H(X_{1j},\dots,X_{tj})-\epsilon_1-\epsilon_2]\nonumber\\
&\hspace{1cm}-\sum_{j=1}^nH(X_{1j},\dots,X_{tj}|M,X_1^{1:j-1},\dots,X_t^{1:j-1})\label{eqn:appendix_wyner2}\\
&=\sum_{j=1}^n [I(X_{1j},\dots,X_{tj};M,X_1^{1:j-1},\dots,X_t^{1:j-1})-\epsilon^\prime]\label{eqn:appendix:wynerdefine1}\\
&=\sum_{j=1}^n[I(X_{1j},\dots,X_{tj};U_i)-\epsilon^\prime]\nonumber\\
&=n[I(X_{1Q}\dots,X_{tQ};U_Q|Q)-\epsilon^\prime]\nonumber\\
&=n[I(X_{1Q}\dots,X_{tQ};U_Q,Q)\nonumber\\
&\hspace{4cm}-I(X_{1Q},\dots,X_{tQ};Q)-\epsilon^\prime]\nonumber\\
&\geq n[I(X_{1Q}\dots,X_{tQ};U_Q,Q)-\delta-\epsilon^\prime]\label{eqn:appendix:wyner3}\\
&=nI(X_{1Q}\dots,X_{tQ};U_Q,Q)-ng(\epsilon)\label{eqn:appendix:wynerdefine2}.
\end{align}
We have used the following fact in \eqref{eqn:appendix:wyner1}-\eqref{eqn:appendix:wyner3}: if two random variables $A$ and $A'$ with 
  same support set $\mathcal{A}$  satisfy $||p_{A} -  p_{A'}||_{1} \leq \epsilon \leq 1/4 $, then it follows from standard results~\cite[Theorem 17.3.3]{Cover} that $|H(A) - H(A')|\leq \eta \log |\mathcal{A}|$, where $\eta \rightarrow 0$ as $\epsilon \rightarrow 0$. Now \eqref{eqn:correctness} implies \eqref{eqn:appendix:wyner1} and \eqref{eqn:appendix_wyner2}, where $\epsilon_1,\epsilon_2\rightarrow 0$ as $\epsilon\rightarrow 0$. Also, using Cuff~\cite[Lemma~VI.2]{Cuff13}, \eqref{eqn:correctness} implies $\lVert p_{X_{1Q},\dots,X_{tQ}}-q_{X_1,\dots,X_t} \rVert\leq \epsilon$, which in turn implies \eqref{eqn:appendix:wyner3}, where $\delta\rightarrow 0$ as $\epsilon\rightarrow 0$. In \eqref{eqn:appendix:wynerdefine1} and \eqref{eqn:appendix:wynerdefine2}, we defined $\epsilon^\prime=\epsilon_1+\epsilon_2$ and $g(\epsilon)=\delta+\epsilon^\prime$, respectively.

\section{Proof of Theorem~\ref{theorem:t=3}}\label{appendix_foreheadproof}
The proof employs the OSRB framework~\cite{YassaeeAG14}. We give a proof for $t=3$, but a similar proof can be written down for any $t$.

Let $(U^n,U_1^n,U_2^n,U_3^n,X_1^n,X_2^n,X_3^n)$ be i.i.d. with distribution $p(u,u_{[1:3]},x_{[1:3]})=q(x_{[1:3]})p(u,u_{[1:3]}|x_{[1:3]})$ satisfying \eqref{eqn:t=3markov}. Bin indices $f,m^*,b_1,b_2,b_3$ with respective rates $\hat{R}_0,R^*,\tilde{R}_1,\tilde{R}_2,\tilde{R}_3$ are created from $(U^n,U_1^n,U_2^n,U_3^n)$ in a way that can be understood from the following joint probability distribution:
\begin{align}
&P(u^n,u_1^n,u_2^n,u_3^n,x_1^n,x_2^n,x_3^n,f,m^*,b_1,b_2,b_3)\nonumber\\
&=p(u^n,u_1^n,u_2^n,u_3^n)P(m_0,f|u^n)P(m^*|u^n)P(b_1|u^n,u_1^n)\nonumber\\
&\hspace{0.6cm}\times P(b_2|u^n,u_2^n)P(b_3|u^n,u_3^n)p(x_1^n|u^n,u_2^n,u_3^n)\nonumber\\
&\hspace{0.6cm}\times p(x_2^n|u^n,u_1^n,u_3^n)p(x_3^n|u^n,u_1^n,u_2^n)\nonumber\\
&=p(b_1,b_2,b_3,f)P(u^n,u_1^n,u_2^n,u_3^n|b_1,b_2,b_3,f)P(m^*|u^n)\nonumber\\
&\hspace{0.6cm}\times p(x_1^n|u^n,u_2^n,u_3^n)p(x_2^n|u^n,u_1^n,u_3^n)p(x_3^n|u^n,u_1^n,u_2^n)\label{eqn:forehead:pmf1}.
\end{align}
Further, we use Slepian-Wolf decoders to estimate $(u^n,u_{(i)_3+1})$ from $b_{(i+2)_3+1},f,m^*$, $i=0,1,2$, where $(i)_3=i \mod 3$. Now we impose a series of constraints on the rates.
\begin{align}
\hat{R}_0&<H(U)\nonumber\\
\tilde{R}_1+\hat{R}_0&<H(U,U_1)\nonumber\\
\tilde{R}_2+\hat{R}_0&<H(U,U_2)\nonumber\\
\tilde{R}_3+\hat{R}_0&<H(U,U_3)\nonumber\\
\tilde{R}_1+\tilde{R}_2+\hat{R}_0&<H(U,U_1,U_2)\nonumber\\
\tilde{R}_1+\tilde{R}_3+\hat{R}_0&<H(U,U_1,U_3)\nonumber\\
\tilde{R}_3+\tilde{R}_2+\hat{R}_0&<H(U,U_2,U_3)\nonumber\\
\tilde{R}_1+\tilde{R}_2+\tilde{R}_3+\hat{R}_0&<H(U,U_1,U_2,U_3)\label{eqn:forehead:indep_1}
\end{align}

\begin{align}
\tilde{R}_2&>H(U_2|U,U_3)\nonumber\\
\tilde{R}_3&>H(U_3|U,U_2)\nonumber\\
\tilde{R}_2+\tilde{R}_3&>H(U_2,U_3|U)\nonumber\\
R^*+\hat{R}_0+\tilde{R}_2+\tilde{R}_3&>H(U,U_2,U_3)\nonumber\\
\tilde{R}_1&>H(U_1|U,U_3)\nonumber\\
\tilde{R}_3&>H(U_3|U,U_1)\nonumber\\
\tilde{R}_1+\tilde{R}_3&>H(U_1,U_3|U)\nonumber\\
R^*+\hat{R}_0+\tilde{R}_1+\tilde{R}_3&>H(U,U_1,U_3)\nonumber\\
\tilde{R}_1&>H(U_1|U,U_2)\nonumber\\
\tilde{R}_2&>H(U_2|U,U_1)\nonumber\\
\tilde{R}_1+\tilde{R}_2&>H(U_1,U_2|U)\nonumber\\
R^*+\hat{R}_0+\tilde{R}_1+\tilde{R}_2&>H(U,U_1,U_2)\label{eqn:forehead:sw}
\end{align}
The first set of constraints \eqref{eqn:forehead:indep_1} ensure that $b_1,b_2,b_3,f$ are approximately (i.e., with vanishing total variation distance) uniformly distributed and mutually independent of each other~\cite[Theorem~1]{YassaeeAG14}. The second set of constraints \eqref{eqn:forehead:sw} guarantees the success of Slepian-Wolf decoders with high probability~\cite[Lemma~1]{YassaeeAG14}. Thus, under these two sets of rate constraints \eqref{eqn:forehead:indep_1} and \eqref{eqn:forehead:sw}, the random p.m.f. comprising \eqref{eqn:forehead:pmf1} and Slepian-Wolf decoders approximately close to the p.m.f. below.
\begin{align}
&P(u^n,u_1^n,u_2^n,u_3^n,x_1^n,x_2^n,x_3^n,f,m^*,b_1,b_2,b_3,\nonumber\\
&\hspace{1cm}\hat{u}_{(1)2}^n,\hat{u}_{(1)3}^n,\hat{u}_{(2)1}^n,\hat{u}_{(2)3}^n,\hat{u}_{(3)1}^n,\hat{u}_{(3)2}^n,\hat{u}_{(1)}^n,\hat{u}_{(2)}^n,\hat{u}_{(3)}^n)\nonumber\\
&=p^{\text{Unif}}(b_1)p^{\text{Unif}}(b_2)p^{\text{Unif}}(b_3)p^{\text{Unif}}(f)\nonumber\\ 
&\hspace{0.2cm}\times P(u^n,u_1^n,u_2^n,u_3^n|b_1,b_2,b_3,f)P(m^*|u^n)\nonumber\\
&\hspace{0.2cm}\times P^{SW}(\hat{u}_{(1)}^n,\hat{u}_{(1)2}^n,\hat{u}_{(1)3}^n|b_2,b_3,f,m^*)\nonumber\\
&\hspace{0.2cm}\times P^{SW}(\hat{u}_{(2)}^n,\hat{u}_{(2)1}^n,\hat{u}_{(2)3}^n|b_1,b_3,f,m^*)\nonumber\\
&\hspace{0.2cm}\times P^{SW}(\hat{u}_{(3)}^n,\hat{u}_{(3)1}^n,\hat{u}_{(3)2}^n|b_1,b_2,f,m^*)\nonumber\\
&\hspace{0.2cm}\times p(x_1^n|\hat{u}_{(1)}^n,\hat{u}_{(1)2}^n,\hat{u}_{(1)3}^n)p(x_2^n|\hat{u}_{(2)}^n,\hat{u}_{(2)1}^n,\hat{u}_{(2)3}^n)\nonumber\\
&\hspace{0.2cm}\times p(x_3^n|\hat{u}_{(3)}^n,\hat{u}_{(3)1}^n,\hat{u}_{(3)2}^n)\label{eqn:forehead:originalpmf}
\end{align}
The connection between above p.m.f. and the original problem is described below. In p.m.f. \eqref{eqn:forehead:originalpmf} we generate $b_1,b_2,b_3,f$ independently and uniformly from the respective alphabets. For $i\in[1:3]$, we treat $b_i$ as the shared randomness $w_i$ that is not available to processor $P_i$. In addition, we have extra shared randomness $f$ (to be eliminated later), which is shared among coordinator and all the three processors. The coordinator on observing $b_1,b_2,b_3,f$ produces $u^n,u_1^n,u_2^n,u_3^n$ according to the random p.m.f. $P(u^n,u_1^n,u_2^n,u_3^n|b_1,b_2,b_3,f)$ of \eqref{eqn:forehead:pmf1} and sends $(m^*(u^n))$ as a common message $m$ to the processors, where $m^*(u^n)$ is produced according to $P(m^*|u^n)$ of \eqref{eqn:forehead:pmf1}. The processors use (random) Slepian-Wolf decoders mentioned below \eqref{eqn:forehead:pmf1} to produce their respective estimates. Then the processors produce $x_1^n,x_2^n$ and $x_3^n$ according to $p(x_1^n|\hat{u}_{(1)}^n,\hat{u}_{(1)2}^n,\hat{u}_{(1)3}^n), p(x_2^n|\hat{u}_{(2)}^n,\hat{u}_{(2)1}^n,\hat{u}_{(2)3}^n)$ and $p(x_3^n|\hat{u}_{(3)}^n,\hat{u}_{(3)1}^n,\hat{u}_{(3)2}^n)$, respectively.

To eliminate the extra shared randomness without disturbing the desired i.i.d. distribution on $X_1,X_2,X_3$, we need a third set of constraints on rates. Under these constraints below, $(X_1^n,X_2^n,X_3^n)$ and $F$ are approximately independent~\cite[Theorem~1]{YassaeeAG14}.
\begin{align}
\hat{R}_0&<H(U|X_1,X_2,X_3)\label{eqn:forehead:indep_2}
%\hat{R}_1+\hat{R}_0&<H(U,U_1|X_1,X_2,X_3)\nonumber\\
%\hat{R}_2+\hat{R}_0&<H(U,U_2|X_1,X_2,X_3)\nonumber\\
%\hat{R}_3+\hat{R}_0&<H(U,U_3|X_1,X_2,X_3)\nonumber\\
%\hat{R}_1+\hat{R}_2+\hat{R}_0&<H(U,U_1,U_2|X_1,X_2,X_3)\nonumber\\
%\hat{R}_1+\hat{R}_3+\hat{R}_0&<H(U,U_1,U_3|X_1,X_2,X_3)\nonumber\\
%\hat{R}_3+\hat{R}_2+\hat{R}_0&<H(U,U_2,U_3|X_1,X_2,X_3)\nonumber\\
%\hat{R}_1+\hat{R}_2+\hat{R}_3+\hat{R}_0&<H(U,U_1,U_2,U_3|X_1,X_2,X_3)
\end{align}
All these three sets of rate constraints \eqref{eqn:forehead:indep_1}, \eqref{eqn:forehead:sw} and \eqref{eqn:forehead:indep_2} guarantee the existence of a particular realization of random binning (so that we can replace $P$ with $p$ in \eqref{eqn:forehead:originalpmf} and denote the resulting p.m.f. by $\hat{p}$) such that
\begin{align*}
\hat{p}(x_1^n,x_2^n,x_3^n,f)&\approx p^{\text{Unif}}(f)p(x_1^n,x_2^n,x_3^n),
\end{align*}
which further implies that there exists instance $f^*$  of $F$ such that
\begin{align*}
\hat{p}(x_1^n,x_2^n,x_3^n|f^*)\approx p(x_1^n,x_2^n,x_3^n).
\end{align*}
Note that the above equation is the required correctness condition. 
Noting that the transmission rate $R=R^*$ and eliminating all the other rates from \eqref{eqn:forehead:indep_1}, \eqref{eqn:forehead:sw} and \eqref{eqn:forehead:indep_2} gives us \eqref{eqn:forehead:bound} for $t=3$.

\section{Proof Outline of Theorem~\ref{theorem_oblv_general}}\label{appendix:oblv:general}
The proof follows along the same lines as that of Theorem~\ref{theorem_oblv_3users}.   Fix a p.m.f. $p(u,u_{[1:h]},x_{[1:t]})$ as given in the theorem. We generate $(h+1)$ number of codebooks randomly in the following way.
\begin{itemize}
\item Randomly and independently generate $2^{nR}$ sequences $u^n(m)$, $m\in[1:2^{nR}]$,  each according to i.i.d. $p_U$. 
\item For each $u^n(w)$, randomly and independently generate $2^{nR_i}$ sequences $u_i^n(w,w_i)$, $w_i\in[1:2^{nR_i}]$, each according to i.i.d. $p_{U_i}$, for $i\in[1:h]$. 
\end{itemize}
For $i\in[1:t]$, processor $P_i$ on observing $w,w_{\mathcal{V}_i}$ produces $x_i^n$ according to a (random) p.m.f. analogous to the proof of Theorem~\ref{theorem_oblv_3users}. We denote $a^n:=(x_1^n,\dots,x_t^n)$. Let $P(a^n)$ be the induced random p.m.f. on $A^n$. It can be easily checked that $\mathbbm{E}P(a^n)=q(a^n)$. The analysis of total variation distance follows the same steps as that of Theorem~\ref{theorem_oblv_3users} by defining $P_1(a^n)$ and $P_2(a^n)$ analogously. $\mathbbm{E}{\left( P_1(a^n)\right)}^2$ is divided in to $2^{h}+1$ parts, each part corresponding to a case as in the proof of Theorem~\ref{theorem_oblv_3users}. Case $(1)$ is when $w\neq w^\prime$. Each of the remaining $2^h$ cases is specified by a set $\mathcal{S}\subseteq [1:h]$, in particular, by $w=w^\prime,w_i\neq w_i^\prime,i\in\mathcal{S}, w_i=w_i^\prime, i\in[1:h]\setminus\mathcal{S}$. The part corresponding to case $(1)$ is dealt similar to that of Theorem~\ref{theorem_oblv_3users}. Using the bound $\sqrt{\sum_{i=1}^lx_i}\leq \sum_{i=1}^l\sqrt{x_i}$, the expression $\sum_{a^n\in\mathcal{T}}\sqrt{\mathbbm{E}{\left( P_1(a^n)\right)}^2-{\left(\mathbbm{E}P_1(a^n)\right)}^2}$ is upper bounded by the summation of $2^h$ corresponding parts as in the proof of Theorem~\ref{theorem_oblv_3users}. The part corresponding to the set $\mathcal{S}\subseteq[1:h]$ asymptotically vanishes if $R+R_{\mathcal{S}}> I(A;U,U_{\mathcal{S}})$. This leads to asymptotically vanishing total variation distance $\mathbbm{E}\lVert P(a^n)-\mathbbm{E}P(a^n) \rVert_1$ using the properties of typicality and Jensen's inequality as in the proof of Theorem~\ref{theorem_oblv_3users}. This completes the achievability.

For the converse, suppose a rate tuple $(R,R_1,\dots,R_h)$ is achievable for $q_{X_1\dots X_t}$. For any $\mathcal{S}\subseteq[1:h]$, consider
\begin{align}
&n(R+R_{\mathcal{S}})\nonumber\\
&\geq H(W,W_{\mathcal{S}})\nonumber\\
&\geq I(W,W_{\mathcal{S}};X_1^n,\dots,X_t^n)\nonumber\\
&=H(X_1^n,\dots,X_t^n)-H(X_1^n,\dots,X_t^n|W,W_{\mathcal{S}})\nonumber\\
&\geq \sum_{i=1}^n\left[H(X_{1i},\dots,X_{ti})-\epsilon^\prime \right]\nonumber\\ 
&\hspace{1cm}-\sum_{i=1}^n H(X_{1i},\dots,X_{ti}|W,W_{\mathcal{S}},X_1^{1:i-1},\dots,X_t^{1:i-1})\label{general:oblv1}\\
&= \sum_{i=1}^n\left[I(X_{1i},\dots,,X_{ti};W,W_{\mathcal{S}},X_1^{1:i-1},\dots,X_t^{1:i-1})-
\epsilon^\prime\right]\nonumber\\
&\geq \sum_{i=1}^n\left[I(X_{1i},\dots,X_{ti};W,W_{\mathcal{S}})-\epsilon^\prime\right]\nonumber\\
&= n\left[I(X_{1Q},\dots,X_{tQ};W,W_{\mathcal{S}}|Q)-\epsilon^\prime\right]\nonumber\\
&\geq n\left[I(X_{1Q},\dots,X_{tQ};W,W_{\mathcal{S}},Q)-\epsilon^\prime-\epsilon^{\prime\prime} \right] \label{general:oblv2}\\
& =n[I(X_{1Q},\dots.X_{tQ};U,U_{\mathcal{S}})-\epsilon^\prime-\epsilon^{\prime\prime}]\label{eqn:oblv_defn:general}
\end{align}
where \eqref{general:oblv1} and \eqref{general:oblv2} follow from the correctness of the output distribution with $\epsilon^\prime ,\epsilon^{\prime\prime}\rightarrow 0$ as $\epsilon\rightarrow 0$ along similar lines as \eqref{eqn:converse_closeness_of_dist} and \eqref{eqn:T_independent}, respectively, and \eqref{eqn:oblv_defn:general} follows by defining $U=(W,Q), U_i=W_i$, for $i\in[1:h]$. Note that $\lVert p_{X_{1Q}\dots X_{tQ}}-q_{X_1\dots X_t} \rVert<\epsilon$, which follows from~\cite[Lemma~VI.2]{Cuff13}. Using the structure of the problem and the continuity of total variation distance and mutual information in the probability simplex, it follows along the same lines as Theorem~\ref{theorem:unlimited_shared_randomness},~\cite[Lemma~6]{YassaeeGA15} that
\begin{align}
R+R_{\mathcal{S}}\geq I(X_1,\dots,X_t;U,U_{\mathcal{S}}), \ \mathcal{S}\subseteq [1:h] \nonumber
\end{align} 
for some p.m.f. $$p(u,u_{[1:h]},x_{[1:t]})=q(x_{[1t3]})p(u,u_{[1:h]}|x_{[1:t]})$$ s.t.
$$p(u,u_{[1:h]},x_{[1:t]})=p(u)\left(\prod_{i=1}^h p(u_i)\right)\left(\prod_{i=1}^tp(x_i|u,u_{\mathcal{V}_i})\right) \label{eqn:oblv_struc:general}.$$
This completes the proof.
\section{Some Other Omitted Details}
\subsection{Details Omitted from Example~\ref{open}}\label{appreviewers1}
Here we show that $f(0)=0.5C(X;Y)$ and $f(1)=I(X;Y)$.
\begin{align*}
f(0)&=\max\{I_{p^*}(X;Y|U),\frac{1}{2}\left(I_{p^*}(X,Y;U)+I_{p^*}(X;Y|U)\right)\}\\
&=\max\{0,0.5C(X;Y)\}\\
&=0.5C(X;Y).
\end{align*}
\begin{align*}
f(1)&=\max\{I_{p^\bot}(X;Y|U),\frac{1}{2}\left(I_{p^\bot}(X,Y;U)+I_{p^\bot}(X;Y|U)\right)\}\\
&=\max\{I(X;Y),\frac{1}{2}(0+I(X;Y))\}\\
&=I(X;Y).
\end{align*}

\subsection{Details Omitted in Relaxing the Implicit Non-Negativity Constraints on Rates from the Proof of Theorem~\ref{theorem:achievability} }\label{appreviewers2}
Here we argue that the new non-negative rates and the auxiliary random variables defined in the achievability proof of Theorem~\ref{theorem:achievability} satisfy \eqref{eqn:theorem:achievability_gather_1}-\eqref{eqn:theorem:achievability_gather_3}. We argue this for one constraint each from \eqref{eqn:theorem:achievability_gather_1}, \eqref{eqn:theorem:achievability_gather_2}, and \eqref{eqn:theorem:achievability_gather_3}. The other constraints can be argued similarly. From \eqref{eqn:theorem:achievability_gather_1}, consider
\begin{align*}
 &\tilde{R_1}+\hat{R_1}+R_0+\hat{R_0} < H(U,U_1),\nonumber\\
& \Rightarrow  \tilde{R_1}+\hat{R_1}+H(W_1)+R_0+\hat{R_0}+H(W)\\
 &\hspace{50pt} < H(U,U_1)+H(W,W_1),\nonumber\\
& \Rightarrow \tilde{R_1}+\hat{R}_{1\text{new}}+R_0+\hat{R}_{0\text{new}}<H(U_{\text{new}},U_{1\text{new}}).
\end{align*}
From \eqref{eqn:theorem:achievability_gather_2}, consider
\begin{align*}
&\tilde{R}_1+\hat{R}_1\geq H(U_1|U)\\
&\Rightarrow \tilde{R}_1+\hat{R}_1+H(W_1)\geq H(U_1|U,W)+H(W_1)\\
&\Rightarrow \tilde{R}_1+\hat{R}_{1\text{new}}\geq H(U_1,W_1|U,W)\\
&\Rightarrow \tilde{R}_1+\hat{R}_{1\text{new}}\geq H(U_{1\text{new}}|U_{\text{new}}).
\end{align*}
From \eqref{eqn:theorem:achievability_gather_3}, consider
\begin{align*}
&\hat{R}_0+\hat{R}_1<H(U,U_1|X.Y)\\
&\Rightarrow\hat{R}_0+H(W)+\hat{R}_1+H(W_1)\\
&\hspace{50pt}<H(U,U_1|X.Y)+H(W,W_1)\\
&\Rightarrow\hat{R}_{0\text{new}}+\hat{R}_{1\text{new}}<H(U_{\text{new}},U_{1\text{new}}).
\end{align*}
\end{appendices}

\fi%
\bibliographystyle{IEEEtran}
\balance
\bibliography{Bibliography}
\begin{IEEEbiographynophoto}
{Gowtham R. Kurri} (Member, IEEE) graduated from the International Institute of Information Technology, Hyderabad, India, with a B.\ Tech.\ degree in Electronics and Communication Engineering, in 2011. He received his M.Sc. and Ph.D. degrees from the Tata Institute of Fundamental Research, Mumbai, India in 2020. He is currently a Post-Doctoral Researcher at the School of Electrical, Computer and Energy Engineering at Arizona State University.

From 2011-2012, he worked as an Associate Engineer at Qualcomm India Private Limited, Hyderabad, India. From July to October, 2019, he was a Research Intern in the Blockchain Technology Group at IBM Research, Bangalore, India. 
\end{IEEEbiographynophoto}
\begin{IEEEbiographynophoto}
{Vinod M. Prabhakaran} (Member, IEEE) received the M.E. degree from the Indian Institute of Science in 2001 and the Ph.D. degree from the University of California, Berkeley in 2007. He was a Post-Doctoral Researcher at the Coordinated Science Laboratory, University of Illinois, Urbana-Champaign from 2008 to 2010 and at Ecole Polytechnique Fédérale de Lausanne, Switzerland in 2011. Since 2011, he has been at the School of Technology and Computer Science at the Tata Institute of Fundamental Research, Mumbai. His research interests are in information theory, communication, cryptography, and signal processing.

He has received the Tong Leong Lim Pre-Doctoral Prize and the Demetri Angelakos Memorial Achievement Award from the EECS Department, University of California, Berkeley, and the Ramanujan Fellowship from the Department of Science and Technology, Government of India. He was an Associate Editor for IEEE Transactions on Information Theory during 2016-19.
\end{IEEEbiographynophoto}

\begin{IEEEbiographynophoto}
{Anand D. Sarwate}(Senior Member, IEEE) received the B.S. degrees in electrical engineering and computer science and mathematics from the Massachusetts Institute of Technology, Cambridge, MA, USA, in 2002, and the M.S. and Ph.D. degrees in electrical engineering from the Department of Electrical Engineering and Computer Sciences (EECS), University of California, Berkeley (U.C. Berkeley), Berkeley, CA, USA. He is a currently an Assistant Professor with the Department of Electrical and Computer Engineering, The State University of New Jersey, New Brunswick, NJ, USA, since January 2014. He was previously a Research Assistant Professor from 2011 to 2013 with the Toyota Technological Institute at Chicago; prior to this, he was a Postdoctoral Researcher from 2008 to 2011 with the University of California, San Diego, CA. His research interests include information theory, machine learning, signal processing, optimization, and privacy and security. 

Dr. Sarwate received the Rutgers Board of Trustees Research Fellowship for Scholarly Excellence in 2020, the A. Walter Tyson Assistant Professor Award from the Rutgers School of Engineering in 2018, and the NSF CAREER award in 2015. He was awarded the National Defense Science and Engineering Graduate Fellowship from 2002 to 2005. He is a member of Phi Beta Kappa and Eta Kappa Nu.
\end{IEEEbiographynophoto}
\end{document}